\documentclass[a4paper,11pt]{article}
\pdfoutput=1
\usepackage{jcappub}

\usepackage[demo,export]{adjustbox}

\usepackage{amsmath}
\usepackage{hhline}
\usepackage{bm}
\usepackage{xcolor,color,colortbl}
\usepackage[utf8]{inputenc}
\usepackage{hyperref}
\usepackage{url}
\usepackage{graphicx}
\usepackage{multirow,bigdelim}
\usepackage{amssymb}
\usepackage[amssymb]{SIunits}
\usepackage{multirow}
\usepackage{bbold}
\usepackage{verbatim}
\usepackage{fontawesome5}
\usepackage{orcidlink}

\definecolor{darkgreen}{HTML}{339933}

\title{
Cosmological constraints from the cross-correlation of DESI Luminous Red Galaxies with CMB lensing from Planck PR4 and ACT DR6
}

\author[1,2]{Noah Sailer\,\orcidlink{0000-0002-5333-8983}}
\author[3]{Joshua Kim\,\orcidlink{0000-0002-0935-3270}}
\author[2,1]{Simone Ferraro\,\orcidlink{0000-0003-4992-7854}}
\author[3]{Mathew S. Madhavacheril}
\author[1,2]{Martin White}
\author[4,5]{Irene Abril-Cabezas\,\orcidlink{0000-0003-3230-4589}}
\author[2]{Jessica Nicole Aguilar}
\author[6]{Steven Ahlen\,\orcidlink{0000-0001-6098-7247}}
\author[7]{J. Richard Bond}
\author[8]{David Brooks}
\author[9]{Etienne Burtin}
\author[10]{Erminia Calabrese}
\author[11]{Shi-Fan Chen}
\author[12]{Steve K. Choi\,\orcidlink{0000-0002-9113-7058}}
\author[2]{Todd Claybaugh}
\author[13]{Kyle Dawson}
\author[14]{Axel de la Macorra\,\orcidlink{0000-0002-1769-1640}}
\author[2]{Joseph DeRose}
\author[15]{Arjun Dey\,\orcidlink{0000-0002-4928-4003}}
\author[16]{Biprateep Dey\,\orcidlink{0000-0002-5665-7912}}
\author[8]{Peter Doel}
\author[17,18]{Jo Dunkley\,\orcidlink{0000-0002-7450-2586}}
\author[4]{Carmen Embil-Villagra\,\orcidlink{0009-0001-3987-7104}}
\author[4,5]{Gerrit S. Farren\,\orcidlink{0000-0001-5704-1127}}
\author[8,19]{Andreu Font-Ribera\,\orcidlink{0000-0002-3033-7312}}
\author[20,21]{Jaime E. Forero-Romero\,\orcidlink{0000-0002-2890-3725}}
\author[22,23,24]{Enrique Gaztañaga}
\author[25]{Vera Gluscevic\,\orcidlink{0000-0002-3589-8637}}
\author[2]{Satya Gontcho A Gontcho\,\orcidlink{0000-0003-3142-233X}}
\author[26,27,28]{Klaus Honscheid}
\author[29]{Cullan Howlett\,\orcidlink{0000-0002-1081-9410}}
\author[15]{Stephanie Juneau}
\author[30]{David Kirkby\,\orcidlink{0000-0002-8828-5463}}
\author[2]{Theodore Kisner\,\orcidlink{0000-0003-3510-7134}}
\author[2]{Anthony Kremin\,\orcidlink{0000-0001-6356-7424}}
\author[2]{Martin Landriau\,\orcidlink{0000-0003-1838-8528}}
\author[31]{Laurent Le Guillou\,\orcidlink{0000-0001-7178-8868}}
\author[2]{Michael Levi\,\orcidlink{0000-0003-1887-1018}}
\author[32,19]{Marc Manera\,\orcidlink{0000-0003-4962-8934}}
\author[15]{Aaron Meisner\,\orcidlink{0000-0002-1125-7384}}
\author[33,19]{Ramon Miquel}
\author[34,35]{Kavilan Moodley\,\orcidlink{000-0001-6606-7142}}
\author[36]{John Moustakas\,\orcidlink{0000-0002-2733-4559}}
\author[37,38]{Michael D. Niemack\,\orcidlink{0000-0001-7125-3580}}
\author[39,40]{Gustavo Niz\,\orcidlink{0000-0002-1544-8946}}
\author[9,2]{Nathalie Palanque-Delabrouille\,\orcidlink{0000-0003-3188-784X}}
\author[41,42,43]{Will Percival\,\orcidlink{0000-0002-0644-5727}}
\author[44]{Francisco Prada\,\orcidlink{0000-0001-7145-8674}}
\author[4,5,45]{Frank J. Qu\,\orcidlink{0000-0001-7805-1068}}
\author[46]{Graziano Rossi}
\author[47]{Eusebio Sanchez\,\orcidlink{0000-0002-9646-8198}}
\author[48,49]{Emmanuel Schaan}
\author[50]{Edward Schlafly\,\orcidlink{0000-0002-3569-7421}}
\author[2]{David Schlegel}
\author[51,52]{Michael Schubnell}
\author[53]{Neelima Sehgal\,\orcidlink{0000-0002-9674-4527}}
\author[54]{Hee-Jong Seo\,\orcidlink{0000-0002-6588-3508}}
\author[4,5]{Blake Sherwin}
\author[55]{Crist\'obal Sif\'on\,\orcidlink{0000-0002-8149-1352}}
\author[15]{David Sprayberry}
\author[17]{Suzanne T. Staggs\,\orcidlink{0000-0002-7020-7301}}
\author[52]{Gregory Tarlé\,\orcidlink{0000-0003-1704-0781}}
\author[15]{Benjamin Alan Weaver}
\author[9]{Christophe Yèche\,\orcidlink{0000-0001-5146-8533}}
\author[2]{Rongpu Zhou\,\orcidlink{0000-0001-5381-4372}}
\author[56]{Hu Zou\,\orcidlink{0000-0002-6684-3997}}

\affiliation[1]{Department of Physics, University of California, Berkeley, CA 94720, USA}
\affiliation[2]{Lawrence Berkeley National Laboratory, 1 Cyclotron Road, Berkeley, CA 94720, USA}
\affiliation[3]{Department of Physics and Astronomy, University of Pennsylvania, 209 South 33rd Street, Philadelphia, PA 19104, USA}
\affiliation[4]{DAMTP, Centre for Mathematical Sciences, University of Cambridge, Wilberforce Road, Cambridge CB3 OWA, UK}
\affiliation[5]{Kavli Institute for Cosmology Cambridge, Madingley Road, Cambridge CB3 0HA, UK}
\affiliation[6]{Physics Dept., Boston University, 590 Commonwealth Avenue, Boston, MA 02215, USA}
\affiliation[7]{Canadian Institute for Theoretical Astrophysics, University of Toronto, Toronto, ON, Canada M5S 3H8}
\affiliation[8]{Department of Physics \& Astronomy, University College London, Gower Street, London, WC1E 6BT, UK}
\affiliation[9]{IRFU, CEA, Universit\'{e} Paris-Saclay, F-91191 Gif-sur-Yvette, France}
\affiliation[10]{School of Physics and Astronomy, Cardiff University, The Parade, Cardiff, Wales CF24 3AA, UK}
\affiliation[11]{School of Natural Sciences, Institute for Advanced Study, 1 Einstein Drive, Princeton, NJ 08540, USA}
\affiliation[12]{Department of Physics and Astronomy, University of California, Riverside, CA 92521, USA}
\affiliation[13]{Department of Physics and Astronomy, The University of Utah, 115 South 1400 East, Salt Lake City, UT 84112, USA}
\affiliation[14]{Instituto de F\'{\i}sica, Universidad Nacional Aut\'{o}noma de M\'{e}xico,  Cd. de M\'{e}xico  C.P. 04510,  M\'{e}xico}
\affiliation[15]{NSF NOIRLab, 950 N. Cherry Ave., Tucson, AZ 85719, USA}
\affiliation[16]{Department of Physics \& Astronomy and Pittsburgh Particle Physics, Astrophysics, and Cosmology Center (PITT PACC), University of Pittsburgh, 3941 O'Hara Street, Pittsburgh, PA 15260, USA}
\affiliation[17]{Joseph Henry Laboratories of Physics, Jadwin Hall, Princeton University, Princeton, NJ 08544, USA}
\affiliation[18]{Department of Astrophysical Sciences, Peyton Hall, Princeton University, Princeton, NJ 08544, USA}
\affiliation[19]{Institut de F\'{i}sica d’Altes Energies (IFAE), The Barcelona Institute of Science and Technology, Campus UAB, 08193 Bellaterra Barcelona, Spain}
\affiliation[20]{Departamento de F\'isica, Universidad de los Andes, Cra. 1 No. 18A-10, Edificio Ip, CP 111711, Bogot\'a, Colombia}
\affiliation[21]{Observatorio Astron\'omico, Universidad de los Andes, Cra. 1 No. 18A-10, Edificio H, CP 111711 Bogot\'a, Colombia}
\affiliation[22]{Institut d'Estudis Espacials de Catalunya (IEEC), 08034 Barcelona, Spain}
\affiliation[23]{Institute of Cosmology and Gravitation, University of Portsmouth, Dennis Sciama Building, Portsmouth, PO1 3FX, UK}
\affiliation[24]{Institute of Space Sciences, ICE-CSIC, Campus UAB, Carrer de Can Magrans s/n, 08913 Bellaterra, Barcelona, Spain}
\affiliation[25]{Department of Physics and Astronomy, University of Southern California, Los Angeles, CA 90089, USA}
\affiliation[26]{Center for Cosmology and AstroParticle Physics, The Ohio State University, 191 West Woodruff Avenue, Columbus, OH 43210, USA}
\affiliation[27]{Department of Physics, The Ohio State University, 191 West Woodruff Avenue, Columbus, OH 43210, USA}
\affiliation[28]{The Ohio State University, Columbus, 43210 OH, USA}
\affiliation[29]{School of Mathematics and Physics, University of Queensland, 4072, Australia}
\affiliation[30]{Department of Physics and Astronomy, University of California, Irvine, 92697, USA}
\affiliation[31]{Sorbonne Universit\'{e}, CNRS/IN2P3, Laboratoire de Physique Nucl\'{e}aire et de Hautes Energies (LPNHE), FR-75005 Paris, France}
\affiliation[32]{Departament de F\'{i}sica, Serra H\'{u}nter, Universitat Aut\`{o}noma de Barcelona, 08193 Bellaterra (Barcelona), Spain}
\affiliation[33]{Instituci\'{o} Catalana de Recerca i Estudis Avan\c{c}ats, Passeig de Llu\'{\i}s Companys, 23, 08010 Barcelona, Spain}
\affiliation[34]{Astrophysics Research Centre, University of KwaZulu-Natal, Westville Campus, Durban 4041, South Africa}
\affiliation[35]{School of Mathematics, Statistics \& Computer Science, University of KwaZulu-Natal, Westville Campus, Durban 4041, South Africa}
\affiliation[36]{Department of Physics and Astronomy, Siena College, 515 Loudon Road, Loudonville, NY 12211, USA}
\affiliation[37]{Department of Physics, Cornell University, Ithaca, NY 14853, USA}
\affiliation[38]{Department of Astronomy, Cornell University, Ithaca, NY 14853, USA}
\affiliation[39]{Departamento de F\'{i}sica, Universidad de Guanajuato - DCI, C.P. 37150, Leon, Guanajuato, M\'{e}xico}
\affiliation[40]{Instituto Avanzado de Cosmolog\'{\i}a A.~C., San Marcos 11 - Atenas 202. Magdalena Contreras, 10720. Ciudad de M\'{e}xico, M\'{e}xico}
\affiliation[41]{Department of Physics and Astronomy, University of Waterloo, 200 University Ave W, Waterloo, ON N2L 3G1, Canada}
\affiliation[42]{Perimeter Institute for Theoretical Physics, 31 Caroline St. North, Waterloo, ON N2L 2Y5, Canada}
\affiliation[43]{Waterloo Centre for Astrophysics, University of Waterloo, 200 University Ave W, Waterloo, ON N2L 3G1, Canada}
\affiliation[44]{Instituto de Astrof\'{i}sica de Andaluc\'{i}a (CSIC), Glorieta de la Astronom\'{i}a, s/n, E-18008 Granada, Spain}
\affiliation[45]{Kavli Institute for Particle Astrophysics and Cosmology, 382 Via Pueblo Mall Stanford, CA 94305-4060, USA}
\affiliation[46]{Department of Physics and Astronomy, Sejong University, Seoul, 143-747, Korea}
\affiliation[47]{CIEMAT, Avenida Complutense 40, E-28040 Madrid, Spain}
\affiliation[48]{SLAC National Accelerator Laboratory, Menlo Park, CA 94025, USA}
\affiliation[49]{Kavli Institute for Particle Astrophysics and Cosmology and Department of Physics, Stanford University, Stanford, CA 94305, USA}
\affiliation[50]{Space Telescope Science Institute, 3700 San Martin Drive, Baltimore, MD 21218, USA}
\affiliation[51]{Department of Physics, University of Michigan, Ann Arbor, MI 48109, USA}
\affiliation[52]{University of Michigan, Ann Arbor, MI 48109, USA}
\affiliation[53]{Physics and Astronomy Department, Stony Brook University, Stony Brook, NY 11794}
\affiliation[54]{Department of Physics \& Astronomy, Ohio University, Athens, OH 45701, USA}
\affiliation[55]{Instituto de F\'isica, Pontificia Universidad Cat\'olica de Valpara\'iso, Casilla 4059, Valpara\'iso, Chile}
\affiliation[56]{National Astronomical Observatories, Chinese Academy of Sciences, A20 Datun Rd., Chaoyang District, Beijing, 100012, P.R. China}

\emailAdd{nsailer@berkeley.edu}

\abstract{
We infer the growth of large scale structure over the redshift range $0.4\lesssim z \lesssim 1$ from the cross-correlation of spectroscopically calibrated Luminous Red Galaxies (LRGs) selected from the Dark Energy Spectroscopic Instrument (DESI) legacy imaging survey with CMB lensing maps reconstructed from the latest \textit{Planck} and ACT data.
We adopt a hybrid effective field theory (HEFT) model that robustly regulates the cosmological information obtainable from smaller scales, such that our cosmological constraints are reliably derived from the (predominantly) linear regime.
We perform an extensive set of bandpower- and parameter-level systematics checks to ensure the robustness of our results and to characterize the uniformity of the LRG sample.
We demonstrate that our results are stable to a wide range of modeling assumptions, finding excellent agreement with a linear theory analysis performed on a restricted range of scales. 
From a tomographic analysis of the four LRG photometric redshift bins we find that the rate of structure growth is consistent with $\Lambda$CDM with an overall amplitude that is $\simeq5-7\%$ lower than predicted by primary CMB measurements with modest $(\sim2\sigma)$ statistical significance.
From the combined analysis of all four bins and their cross-correlations with \textit{Planck} we obtain $S_8 = 0.765\pm0.023$, which is less discrepant with primary CMB measurements than previous DESI LRG cross \textit{Planck} CMB lensing results. 
From the cross-correlation with ACT we obtain $S_8 = 0.790^{+0.024}_{-0.027}$, while when jointly analyzing \textit{Planck} and ACT we find $S_8 = 0.775^{+0.019}_{-0.022}$ from our data alone and $\sigma_8 = 0.772^{+0.020}_{-0.023}$ with the addition of BAO data.
These constraints are consistent with the latest \textit{Planck} primary CMB analyses at the $\simeq 1.6-2.2\sigma$ level, and are in excellent agreement with galaxy lensing surveys.
}

\begin{document}
\maketitle
\flushbottom

\section{Introduction}
\label{sec:introduction}

The evolution of large scale structure (LSS) fluctuations offers a unique window into fundamental physics, the formation of galaxies and their associated clusters \cite{1999coph.book.....P,2003moco.book.....D}. $\Lambda$CDM accurately predicts the evolution of matter perturbations from their primordial seeds to the present day on sufficiently large scales. Consequentially, large scale late time structure growth measurements can be used as a powerful consistency check of $\Lambda$CDM conditioned on primary CMB observations. More generally, structure growth is sensitive to extensions of the standard cosmological model including but not limited to dark matter interactions (e.g. \cite{Buen-Abad:2015ova,Buen-Abad:2017gxg,Archidiacono:2019wdp,Becker:2020hzj,Archidiacono:2022iuu,Lague:2024sox}), deviations from general relativity (e.g. \cite{Wenzl:2024sug}) and modifications to the expansion history tracing back to deep within the radiation dominated epoch (e.g. \cite{Sailer:2021yzm,Goldstein:2023gnw}).
 
There is a wide range of observational handles on the amplitude of low redshift density fluctuations, conventionally parameterized by $\sigma_8$ (or $S_8\equiv \sigma_8\sqrt{\Omega_m/0.3}$) which by definition sets the amplitude of \textit{linear} matter density fluctuations at the present day. 
Analyses of cluster counts \cite{Planck:2015lwi,DES:2020ahh,2023arXiv231003944S,2024arXiv240208458G,2024arXiv240102075B}, peculiar velocity surveys \cite{Howlett:2022len,Saulder:2023oqm,Stahl:2021mat}, Sunyaev-Zel'dovich (SZ) effects \cite{1972CoASP...4..173S,Horowitz:2016dwk}, redshift space distortions (RSD) \cite{Ivanov:2019pdj,DAmico:2019fhj,Colas:2019ret,Troster:2019ean,Chen:2021wdi,Ivanov:2021fbu,Zhang:2021yna,Kobayashi:2021oud,eBOSS:2020yzd,eBOSS:2020gbb} and gravitational lensing measurements \cite{Heymans:2013fya,Heymans:2020gsg,KiDS:2021opn,DES:2021wwk,DES:2021bvc,DES:2021vln,More:2023knf,Miyatake:2023njf,Sugiyama:2023fzm,Dalal:2023olq,Li:2023tui,Kilo-DegreeSurvey:2023gfr,Planck:2018lbu,Carron:2022eyg,ACT:2023dou} cover a broad range of scales with different systematic uncertainties and assumptions required to infer $\sigma_8$ (or $S_8$) from each observable.
Several recent analyses of low redshift tracers report $S_8$ constraints that are slightly lower than predicted from $\Lambda$CDM conditioned on primary CMB data from \textit{Planck}, albeit at modest statistical significance. 
Examples include an analysis of cluster counts from DES Y1 observations \cite{DES:2020ahh}; the peculiar velocity field derived from the Democratic Samples of Supernovae \cite{Stahl:2021mat}; 
reanalyses of BOSS full shape and post reconstruction data \cite{Ivanov:2019pdj,DAmico:2019fhj,Colas:2019ret,Chen:2021wdi,Ivanov:2021fbu} using Effective Field Theory (EFT) based models; galaxy shear and its cross-correlations with galaxy positions from KiDS \cite{Heymans:2020gsg,KiDS:2021opn}, DES Y3
\cite{DES:2021wwk,DES:2021bvc,DES:2021vln}, and HSC \cite{Miyatake:2023njf,Sugiyama:2023fzm,Dalal:2023olq,Li:2023tui} data; correlations between \textit{Planck} CMB lensing and DESI Legacy Survey galaxies \cite{Kitanidis:2020xno,Hang:2020gwn,White:2021yvw}, \textit{Planck} lensing and unWISE galaxies \cite{Krolewski:2021yqy}, \textit{Planck}+SPT lensing and DES galaxies (including shear) \cite{DES:2022xxr,DES:2022urg}, \textit{Planck}+ACT lensing and KiDS shear \cite{Robertson:2020xom}, ACT lensing and DES Y3 galaxies \cite{ACT:2023ipp}; and combinations of the above \cite{Garcia-Garcia:2021unp,Chen:2022jzq,Chen:2024vuf}, all of which find $S_8$ constraints that are $\sim5-10\%$ lower than \textit{Planck} at modest $\simeq2-3\sigma$ significance (with the exception of DES Y1 cluster counts, which claim a $\sim5\sigma$ tension). 

The collection of these low $S_8$ constraints (among others, see e.g. \cite{Abdalla:2022yfr} for a more comprehensive review) has become known as the $S_8$ ``tension." However, not all low redshift probes prefer low $S_8$ values. For example, an analysis of tSZ clusters identified from \textit{Planck} and ACT data \cite{Horowitz:2016dwk}, SPT clusters with DES and HST weak lensing \cite{2024arXiv240102075B}, ACT CMB lensing correlated with BOSS galaxies \cite{Darwish:2020fwf}, \textit{Planck} CMB lensing correlated with DES Y1 galaxy positions and shear \cite{Xu:2023qmp}, and \textit{Planck} CMB lensing correlated with Quaia quasars \cite{Piccirilli:2024xgo} are all consistent with \textit{Planck} to well within $1\sigma$, while the latest eROSITA cluster analysis \cite{2024arXiv240208458G} finds $S_8=0.86\pm0.01$.
Moreover, recent reanalyses of datasets claiming low $S_8$ values have reported less discrepant results than found previously. Examples include a reanalysis of DES Y1 clusters \cite{2023arXiv231003944S} that properly forward models cluster selection effects, a joint analysis of DES and KiDS data \cite{Kilo-DegreeSurvey:2023gfr} (which finds that their $S_8$ constraints vary at the $\simeq1\sigma$ level depending on the linear alignment model assumed), and a reanalysis of CMB lensing (\textit{Planck} PR4 + ACT DR6) correlated with unWISE galaxies \cite{ACT:2023oei}.
In addition, it has been noted \cite{Zhang:2021yna} that recent EFT-based analyses of BOSS data, which favor lower $S_8$ values than more traditional RSD analyses \cite{eBOSS:2020yzd,eBOSS:2020gbb} or halo model based methods \cite{Kobayashi:2021oud}, are susceptible to ``volume effects" that complicate the interpretation of marginal posteriors (exaggerating the ``standard" parameter based tension metric).

Notably, analyses of the CMB lensing power spectrum $C^{\kappa\kappa}_\ell$ from the latest \textit{Planck} \cite{Carron:2022eyg} and ACT \cite{ACT:2023kun} data are also in excellent agreement with the \textit{Planck} primary CMB.
CMB lensing directly probes the (Weyl) potential on predominantly linear scales and at late times ($z\simeq 0.5-5$) with a well-characterized source distribution, making it straightforward to model.
Additionally, the CMB lensing convergence is measured by utilizing very well understood statistical properties of the primary CMB, such that its calibration is known analytically as a function of cosmological parameters. 
Together these properties make CMB lensing arguably the most pristine probe of low redshift matter fluctuations. However, given that CMB lensing is a projected probe over a wide range of redshifts, it is difficult to make precise statements about specific late-time epochs (e.g. $0.4 < z < 1$) from CMB lensing alone. This in contrast to the tomographic CMB lensing analyses mentioned above (with mixed $S_8$ results) and explored in this work, which isolate the CMB lensing contribution from a desired redshift range via cross-correlations with a second LSS tracer.

\begin{figure}[!h]
    \centering
    \includegraphics[width=\linewidth,valign=c]{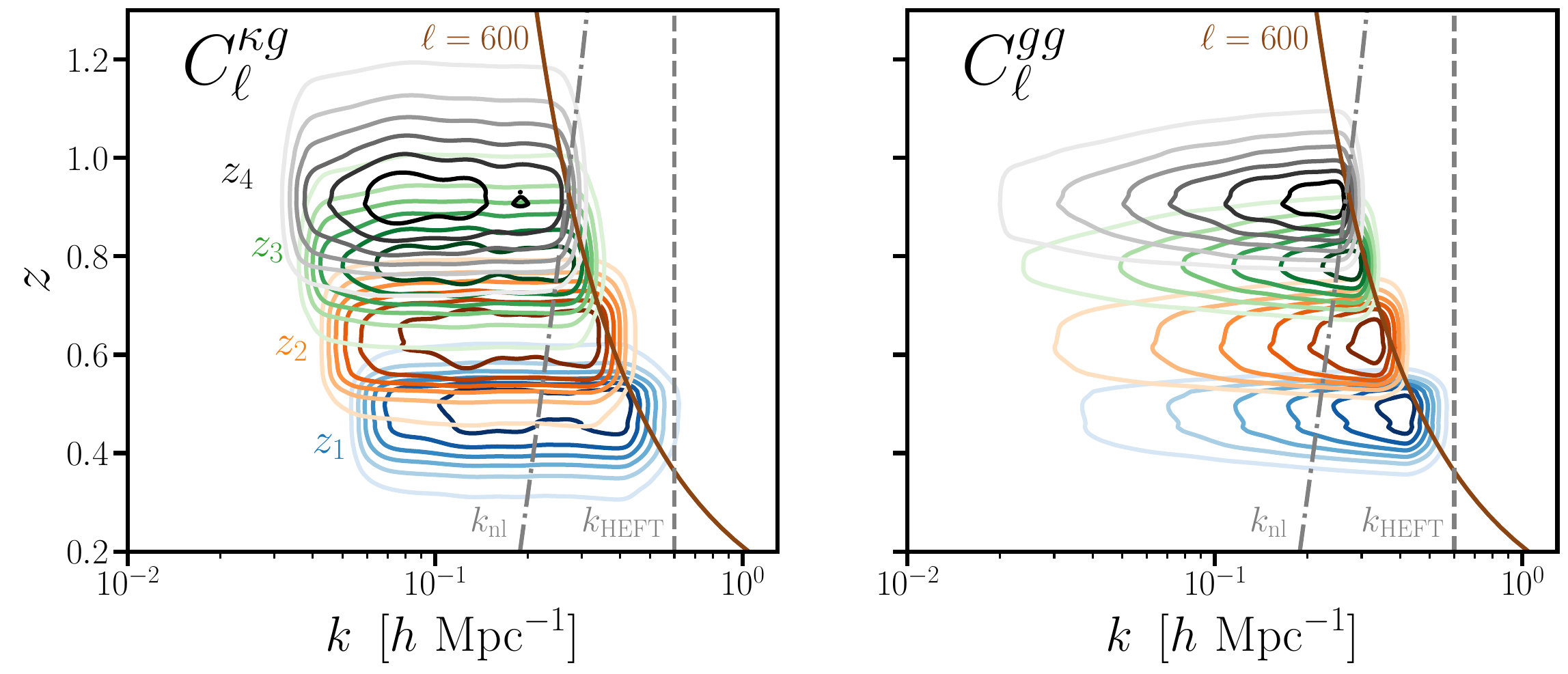}
    \caption{
    Contour plots of the peak normalized ${\rm SNR}(k,z)$ for the galaxy-CMB lensing cross-correlations ($C^{\kappa g}_\ell$, left) and the galaxy auto-correlations ($C^{gg}_\ell$, right) when taking $\ell_\text{max}=600$.
    ${\rm SNR}(k,z)$ is defined as the contribution to the signal-to-noise ratio per $dz\,d\ln k$, such that the total (signal-to-noise ratio)$^2$ is ${\rm SNR}^2 = \int dz\,d\ln k\,\,{\rm SNR}^2(k,z)$. See Appendix \ref{sec:snr} for a more detailed discussion. The colors of the contours correspond to the redshift bin (see Fig.~\ref{fig:dNdz_variations} for the redshift distributions and Fig.~\ref{fig:best_fit} for the measurements themselves) while the shading indicates the value of each contour (from lighter to darker these values are 0.15, 0.3, 0.45, 0.6 0.75, and 0.9). The brown solid line corresponds to $k=600/\chi(z)$, while $k_{\rm nl} = [\int dk P_\text{lin}(k)/6\pi^2]^{-1/2}$ (grey dot dashed line) and $k_\text{HEFT}\simeq 0.6\,\, h\,{\rm Mpc}^{-1}$ (grey dashed line) is the maximum scale for which we expect HEFT to be valid to the percent level \cite{Kokron21} for LRG-like galaxies. 
    } 
\label{fig:snr_per_dzdk}
\end{figure}

In particular, a previous cross-correlation analysis of DESI Luminous Red Galaxies (LRGs) with \textit{Planck} PR3 CMB lensing reported $S_8 = 0.725\pm0.030$ \cite{White:2021yvw}, roughly $3\sigma$ lower than favored by the primary (\textit{Planck} 2018) CMB. 
Motivated by the availability of lower noise CMB lensing maps from \textit{Planck} \cite{Carron:2022eyg} and ACT \cite{ACT:2023dou}, several improvements to the LRG sample \cite{Zhou:2023gji}, a more accurate treatment of mode couplings arising from e.g. masks in CMB lensing estimators, and the development of a Hybrid EFT emulator \cite{DeRose:2023dmk} we perform an improved analysis of the cross-correlation of DESI LRGs with CMB lensing.
The LRG sample has subpercent stellar contamination, is highly robust to the systematic weight treatment, and has a spectroscopically calibrated redshift distribution, while our fiducial HEFT model robustly marginalizes over the large-scale impact of small-scale astrophysical uncertainties, making our analysis tailored to mitigate systematic uncertainties.
In Fig.~\ref{fig:snr_per_dzdk} we show the contribution to the signal-to-noise ratio (SNR) for each of our measurements (see \S\ref{sec:cl_estimation} and the companion paper \cite{Kim2024}) per unit redshift and wavenumber for our fiducial analysis choices. Our data are primarily sensitive to redshifts $0.4 \lesssim z \lesssim 1$ and for all measurements 70\% of the SNR comes from $k<0.3\,h\,{\rm Mpc}^{-1}$. Given that we marginalize over higher-order astrophysical uncertainties, raw SNR (particularly at higher $\ell$) does not necessarily translate to a higher precision $S_8$ measurement, making our constraint predominately sensitive to linear scales ($k<0.1\,h\,{\rm Mpc}^{-1}$).

In a companion paper \cite{Kim2024} we present the fiducial power spectra measurements, the hybrid covariance used in our fiducial analysis, consistency tests exploring the robustness of the cross-correlation of DESI galaxies with the ACT DR6 lensing map and constraints on our best-constrained parameter combination $S^\times_8 \equiv \sigma_8(\Omega_m /0.3)^{0.4}$ using the likelihood presented here. In this work we cross-validate the power spectra measurements with an independent pipeline, investigate the robustness of the DESI LRG sample and its cross-correlation with \textit{Planck} PR4, present our theory model, likelihood implementation and validation, and our primary cosmological constraints on $S_8$ and $\sigma_8$, including the addition of BAO data.

The remainder of this paper is organized as follows. In \S\ref{sec:data} we summarize the data used in our analysis. In \S\ref{sec:cl_estimation} we outline the methodology for ancillary power spectra measurements and covariance estimation used in our \textit{Planck} PR3 reanalysis and systematics tests. We discuss the modeling (including alternatives to HEFT) in \S\ref{sec:modeling} and present our likelihood and associated tests in \S\ref{sec:verification_and_volume}. In \S\ref{sec:systematics} we perform additional systematics tests for the galaxy auto-spectra and the cross-correlation with \textit{Planck} PR4. Our main cosmological results are given in \S\ref{sec:results}. We discuss our results in the context of previous constraints in \S\ref{sec:comparison} and conclude with \S\ref{sec:discussion_and_conclusions}. 

\section{Data}
\label{sec:data}

Our analysis utilizes a photometric sample of Luminous Red Galaxies (LRGs) from the DESI Legacy Imaging Survey DR9 \cite{DESI:2022gle,2019AJ....157..168D,Zhou:2023gji} and CMB lensing convergence maps reconstructed from \textit{Planck} \cite{Planck:2018lbu,Carron:2022eyg} and Atacama Cosmology Telescope (ACT) \cite{ACT:2023dou,ACT:2023ubw,ACT:2023kun} data.
DESI is a highly multiplexed spectroscopic survey that is capable of measuring 5000 objects at once \cite{2016arXiv161100037D,2023AJ....165....9S,2023arXiv230606310M} and is currently operating on the Mayall 4-meter telescope at Kitt Peak National Observatory \cite{2022AJ....164..207D}. DESI is currently conducting a five-year survey and will obtain spectra for approximately 40 million galaxies and quasars \cite{2016arXiv161100036D,2023AJ....165..144G,2023AJ....166..259S}, enabling constraints on the nature of dark energy through its impact on the universe's expansion history \cite{2013arXiv1308.0847L}.

Some key properties of the LRG sample are summarized in Table~\ref{tab:sample_properties}.
The LRG footprint, which is illustrated in the left panel of Fig.~\ref{fig:dNdz_variations}, covers\footnote{These numbers differ from Table 1 of \cite{Zhou:2023gji}, which take into account the fractional coverage in each pixel, whereas we treat each pixel as observed (1) or unobserved (0). The sky coverage quoted in the text are the relevant numbers for the clustering analysis performed here. The fractional coverage has been accounted for in the systematics weights and is reflected e.g. in the (increased) shot noise of the galaxies.} 18200 deg$^2$ with a surface density of 500 deg$^{-2}$ with approximately 16600 and 7900 deg$^2$ of overlap with \textit{Planck} PR4 and ACT DR6 respectively. 
In \S\ref{sec:LRGs} we describe the LRG sample, including the photometric selection, systematic weights, redshift distribution and magnification bias estimation. We briefly discuss the \textit{Planck} and ACT CMB lensing maps in \S\ref{sec:planck_kappa} and \S\ref{sec:act_kappa} respectively, and refer the reader to the companion paper \cite{Kim2024} and refs.~\cite{Planck:2018lbu,Carron:2022eyg,ACT:2023dou} for a more detailed discussion.

\subsection{Luminous Red Galaxies}
\label{sec:LRGs}

The LRG samples are a subset of the Legacy Survey (LS) DR9 \cite{2019AJ....157..168D}, which is currently being used for targeting by the DESI spectroscopic redshift survey. The data are comprised of optical ($g,r,z$) imaging from the Beijing–Arizona Sky Survey (BASS \cite{2017PASP..129f4101Z}), Mayall $z$-band Legacy Survey (MzLS \cite{2019AJ....157..168D}), Dark Energy Camera Legacy Survey (DECaLS \cite{2019AJ....157..168D}) and the Dark Energy Survey (DES \cite{DES:2016jjg}), as well as
four mid-infrared ($W1-W4$) bands from the Wide-field Infrared Survey Explorer (WISE \cite{2010AJ....140.1868W}). 

\textbf{Photometric selection:} We use the ``Main LRG sample" whose photometric selection is described in detail in \cite{DESI:2022gle}. Briefly, the selection is composed of three color cuts on extinction-corrected $g,\,r,\,z$ and $W1$ magnitudes to mitigate stellar contamination, remove galaxies below $z\lesssim0.4$ and produce a roughly constant number density out to $z\sim 0.8$, in addition to a cut in $z$-band fiber magnitude to produce a tail in the redshift distribution extending just beyond $z=1$. Due to differences in the photometry of the different imaging surveys, the selection in the ``Northern" (BASS+MzLS) and ``Southern" (DECaLS+DES) regions differ slightly in their implementation (see Eqs. 1 \& 2 of \cite{DESI:2022gle}) in an effort to homogenize the LRG sample across the full footprint. 

\textbf{Photometric redshifts:} The LRG sample has been subdivided into four photometric redshift bins \cite{Zhou:2023gji} that we label $z_1$ through $z_4$. We note that the photometric redshifts presented in \cite{Zhou:2023gji} differ slightly from those released in \cite{Zhou:2020nwq} and used in previous analyses (e.g. \cite{White:2021yvw}). 
A detailed description of the photo-$z$ algorithm and its latest improvements is presented in Appendix B of \cite{Zhou:2023gji}. Most notably, the spectroscopic training data have been updated to include redshifts from DESI's Survey Validation \cite{DESI:2023dwi} and Early Data Release \cite{DESI:2023ytc}. We also note that the definition of the photo-$z$ bins in the Northern and Southern regions differ slightly in an effort to homogenize the sample (see Table 2 of \cite{Zhou:2023gji}).

\textbf{Image quality cuts:} We use the LRG masks publicly available at this \href{https://data.desi.lbl.gov/public/papers/c3/lrg_xcorr_2023/}{URL} \cite{Zhou:2023gji}, which have been constructed to mitigate contamination from unwanted imaging artifacts, stars, and other undesired astrophysical contaminants (e.g. large galaxies, star clusters, and planetary nebulae). These include the BRIGHT, GALAXY and CLUSTER \href{https://www.legacysurvey.org/dr9/bitmasks/\#maskbits}{bit masks} used in the LS DR9, in addition to a \href{https://data.desi.lbl.gov/public/ets/vac/lrg_veto_mask/v1/}{``veto" mask} constructed from unWISE \cite{2019PASP..131l4504M}, WISE \cite{2010AJ....140.1868W}, Gaia/Tycho-2 \cite{2018A&A...616A...1G,2021A&A...650C...3G}, and visually-inspected LS DR9 data to remove problematic regions not captured by the LS DR9 bit masks. For a more detailed description of the veto mask, see Section 2.4 and Appendix D of \cite{DESI:2022gle}. With this set of photometric selections and masks, only 0.3\% of the LRGs have been classified as stars by DESI's current spectroscopic data. 
In addition the following image quality cuts have been imposed: pixels with $E(B-V)\geq 0.15$ (determined with the extinction map from ref. \cite{SFD}, hereafter SFD) are masked; pixels where the stellar density is larger than 2500 deg$^{-2}$ (determined by the Gaia \cite{Myers:2022azg} stellar density map) are masked; and only pixels with at least two exposures in the $g$, $r$, and $z$ bands are included. 

\begin{figure}[!h]
    \centering
    \includegraphics[width=0.45\linewidth,valign=c]{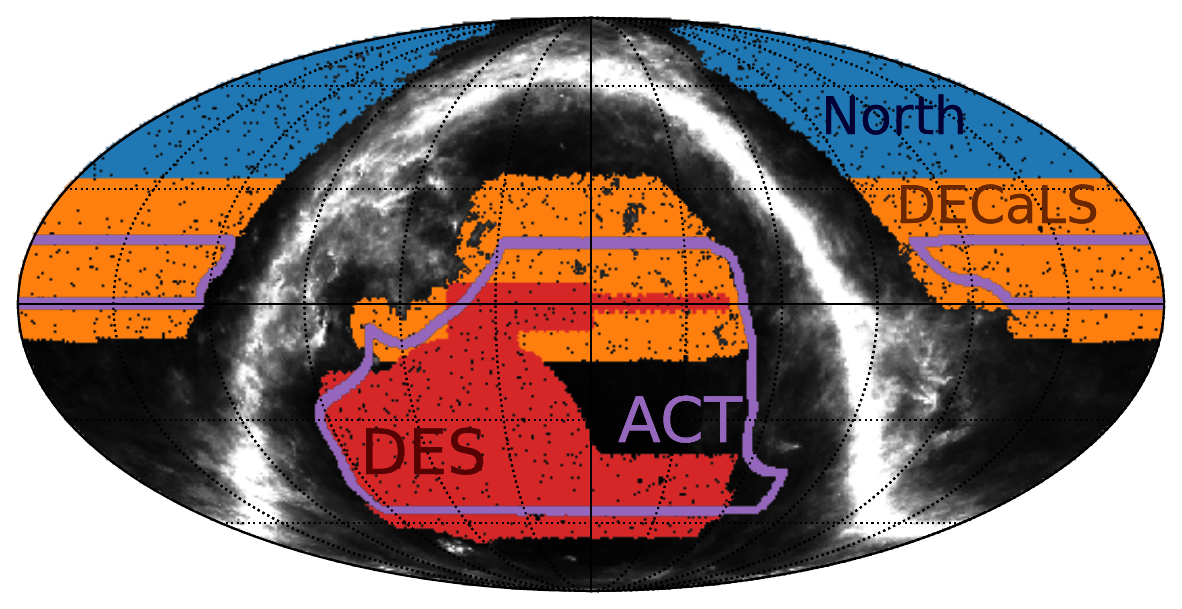}
    \includegraphics[width=0.45\linewidth,valign=c]{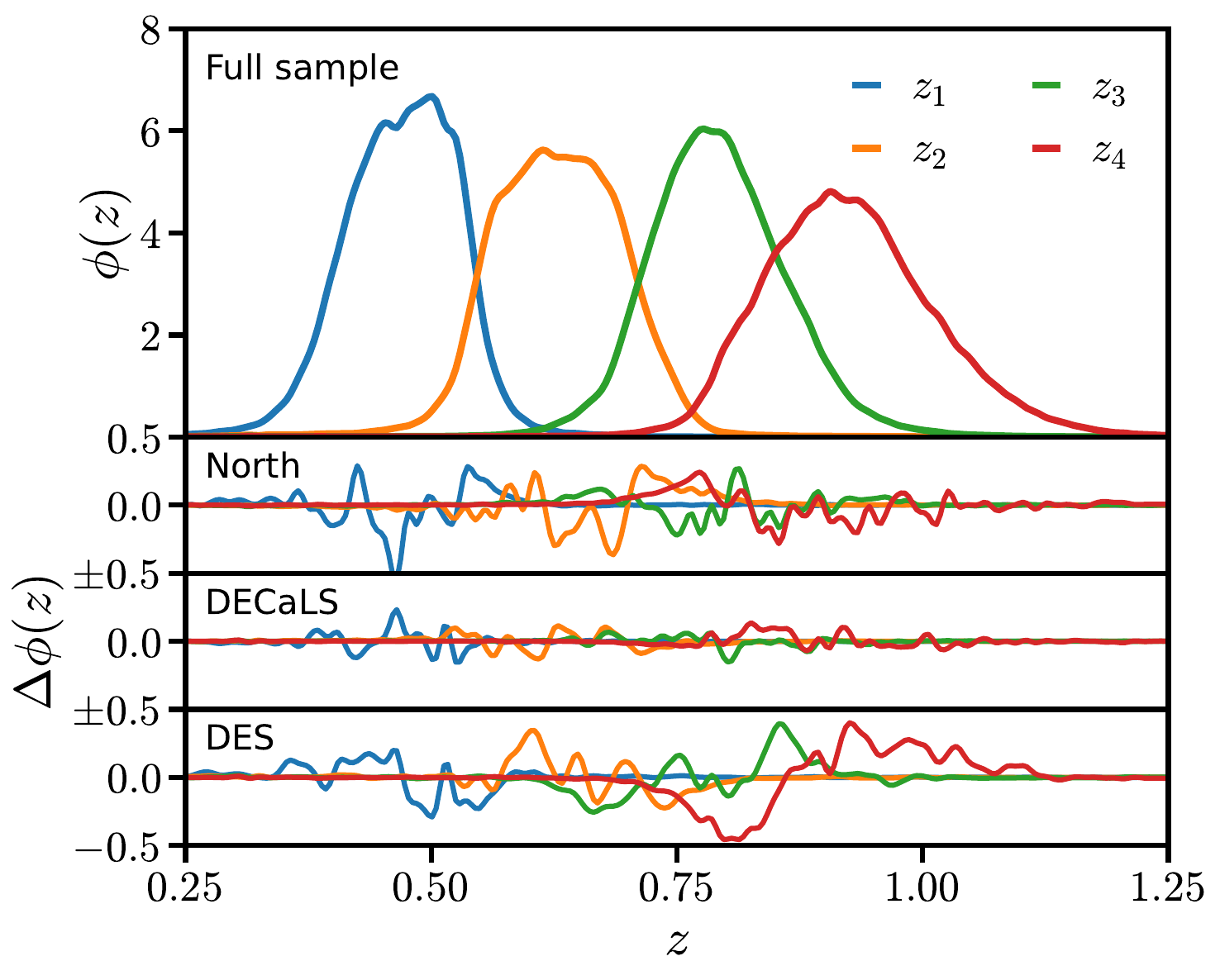}
    \caption{ 
    \textit{Left}: The imaging footprints of our LRG sample (blue, orange, red) and the ACT lensing footprint (purple). The SFD $E(B-V)$ map \cite{SFD} is shown in greyscale.
    \textit{Right}:
    In the top panel we show the (spectroscopically-calibrated) normalized redshift distributions of the four LRG samples (labeled by color) averaged over the full photometric footprint from \cite{Zhou:2023gji}. The bottom three panels show the variation of the redshift distribution across imaging regions: $\Delta \phi(z) = \phi^X(z) - \phi^\text{full}(z)$ where $X=$ North, DECaLS, or DES.
    } 
\label{fig:dNdz_variations}
\end{figure}

\textbf{Imaging weights:} Ref.~\cite{Zhou:2023gji} removes observational modulations to the LRG density maps following a standard procedure that we now describe.
Weights are constructed under the assumption that variations in the local mean LRG density due to instrumental and galactic effects can be modeled as a random downsampling of the ``true" LRG population. Under this assumption these density trends can be removed via division of the galaxy density map by a suitable weight map reflecting this downsampling factor. In practice, these weights are taken to be a linear combination of a set of templates whose coefficients are chosen via linear regression to remove trends in LRG density with the templates. The fiducial weights provided in \cite{Zhou:2023gji} use seven templates: depth and seeing in the three optical bands ($g,r,z$) and $E(B-V)$ \cite{SFD}. Ref. \cite{Zhou:2023gji} also provides a set of weights that do not use an $E(B-V)$ template, which we use in \S\ref{sec:systematics} to assess the importance of Galactic extinction and CIB contamination in the SFD map as a potential systematic. On the scales relevant for our analysis the systematic weights impact the galaxy auto-spectra by less than a percent (Fig.~\ref{fig:cgg_variations_compilation}). 

\textbf{Map making:} 
We use the LRG density contrast maps publicly available at this \href{https://data.desi.lbl.gov/public/papers/c3/lrg_xcorr_2023/}{URL}.
Here we briefly summarize the map making performed by ref.~\cite{Zhou:2023gji}. The LRGs were binned into HEALPix \cite{Gorski:2004by} pixels with \verb|nSide = 2048|. Likewise a pixelized map of weighted randoms was constructed using the previously described imaging weights. The masks and image quality cuts (discussed above) were then applied to both the LRG and weighted random maps. In an effort to mitigate the presence of many small holes in the LRG mask, pixels whose (weighted) random density was less than $1/5$ of the mean (weighted) random density over the footprint satisfying the image quality cuts were additionally masked. We follow \cite{White:2021yvw,Zhou:2023gji} and choose to not apodize the LRG masks. The LRG density contrast is defined as the LRGs divided by the weighted randoms, mean subtracted and normalized to the mean.

\begin{table}[!h]
    \centering
    \begin{tabular}{c|ccccccc}
         Sample & $z_{\rm eff}$ & $b^E_1$ & $\bar{z}$ & $\Delta z$ &  $s_\mu$ & $\bar{n}_\theta$ [deg$^{-2}$] & $10^6\,{\rm SN}^{\rm 2D}$\\
         \hline
         $z_1$ & 0.470 & 1.8 & 0.470 & 0.063 & 0.972 & 81.9 & 4.07 \\
         $z_2$ & 0.625 & 2.0 & 0.628 & 0.074 & 1.044 & 148.1 & 2.25 \\
         $z_3$ & 0.785 & 2.2 & 0.791 & 0.078 & 0.974 & 162.4 & 2.05 \\
         $z_4$ & 0.914 & 2.4 & 0.924 & 0.096 & 0.988 & 148.3 & 2.25 \\
    \end{tabular}
    \caption{
    A summary of the LRG sample properties. The effective redshift $z_{\rm eff}$ is calculated with Eq.~\eqref{eq:zeff}. We list the best-fit Eulerian linear bias $b_1^E = 1+b_1^L$ (where $b_1^L$ is the Lagrangian linear bias) values from our fiducial PR4+DR6 analysis. The mean redshift $\bar{z}$, standard deviation $\Delta z$, number count slope $s_\mu$, surface density $\bar{n}_\theta$ and shot noise ${\rm SN}^{\rm 2D}$ are taken from \cite{Zhou:2023gji}.
    }
    \label{tab:sample_properties}
\end{table}

\textbf{Redshift distribution:} The redshift distributions of the four photometric LRG samples have been directly calibrated in \cite{Zhou:2023gji} using 2.3 million LRG redshifts from DESI's Survey Validation \cite{DESI:2023dwi} and first year (Y1) of observations.
These redshift distributions account for the aforementioned imaging weights, masks and image quality cuts, as well as the spectroscopic systematics weights for DESI Y1 observations.
In addition, each spectroscopic galaxy is weighted by its inverse success rate (as defined in section 4.4 of \cite{DESI:2022gle}) to mitigate the impact of redshift failures\footnote{The spectroscopic sample used to calibrate these redshift distributions has a redshift failure rate of 1.3\% and catastrophic failure rate (defined as deviating by more than 1000 km/s from the true redshift) of 0.2\%.}.
The normalized redshift distributions of the four LRG samples, averaged over the full imaging footprint (determined by area weighting the redshift distributions calibrated on the North, DECaLS and DES footprints), is shown in the top right panel of Fig.~\ref{fig:dNdz_variations}. In the bottom three panels we show the variations of the redshift distribution in the three imaging regions. We propagate these variations to $\Delta C^{gg}_\ell$ in \S\ref{sec:systematics}, and examine their impact on cosmological constraints in \S\ref{sec:results}.

\textbf{Number count slope:} We make use of the number count slopes $s_\mu$ measured in \cite{Zhou:2023gji} for the ``combined" sample, which are listed in Table~\ref{tab:sample_properties}. 
The logarithmic change to the total number of LRGs for a small change in magnitude due to magnification is defined as $s_\mu = d\log_{10} N/dm$.
These are measured via finite difference by shifting the ($g,\,r,\,z,\,W1$) LRG magnitudes by $\delta m=\pm0.01$, computing the appropriate $z$-band fiber magnitude shift $\delta m_\text{fiber}$ for each galaxy (see Appendix C of \cite{Zhou:2023gji} for details regarding the $\delta m_\text{fiber}$ calculation), reapplying the photometric selection, recomputing the photometric redshifts with the shifted magnitudes and galaxy sizes, and finally rebinning the LRGs into photometric redshift bins.

\textbf{Changes to the LRG sample since White et al. \cite{White:2021yvw}:} These are listed in Appendix A of \cite{Zhou:2023gji} and briefly summarized here. The LRG photo-$z$'s have been improved using more training data, the redshift distributions are estimated with a larger number of spectra, the zero point offset for DEC $<-29.25\degree$ in the $r$ and $z$ bands has been corrected using the offset map from \cite{2023RNAAS...7..105Z}, and the imaging systematics no longer include a template for WISE $W1$ depth (due to a lack of an observed density trend with $W1$ depth and a discrepancy in the $W1$ depths in the data and randoms). With these changes, the LRG auto-correlation differs by at most 2\% from the measurements used in \cite{White:2021yvw}, which is primarily driven by the improved photo-$z$'s and hence narrower redshift distributions.

\subsection{\textit{Planck} CMB lensing}
\label{sec:planck_kappa}

We consider both \textit{Planck} 2018 \cite{Planck:2018lbu} (PR3) and PR4 \cite{Carron:2022eyg} CMB lensing measurements in this work. For both releases we low pass filter the CMB lensing convergence multipoles\footnote{As in \cite{White:2021yvw} we multiply the reconstructed CMB lensing multipoles $\hat{\kappa}_{\ell m}$ by $\text{Exp}[-(\ell/2500)^6]$. This filter impacts the cross-correlation measurements by less than $0.02\%$ for $\ell<600$, thus we neglect this filtering in our forward model. We do however include the filter in the fiducial CMB lensing noise curves used for our covariance.} $(\hat{\kappa}_{\ell m})$ and rotate them from Galactic to equatorial coordinates using \verb|healpy|'s \verb|rotate_alm| routine before constructing \verb|nSide = 2048| HEALPix convergence maps via \verb|alm2map|.
Following \cite{White:2021yvw} we apodize the (binary) CMB lensing masks with a 30 arcmin ``C2" filter using the \verb|NaMaster| \href{https://github.com/LSSTDESC/NaMaster}{\faGithub} \cite{Alonso:2018jzx} routine \verb|mask_apodization| before rotating the masks from galactic to equatorial coordinates with the \verb|healpy| routine \verb|rotate_map_alms|.

\textbf{Public Release 3 (PR3):} 
We reanalyze the cross-correlation with PR3 to make direct contact with a previous analysis \cite{White:2021yvw} (which found $S_8 = 0.725\pm0.030$) using a similar sample of LRGs. In \S\ref{sec:pr3} we quantify the impact of the improved LRG sample, new binning scheme, and updated modeling choices on these results.
The \textit{Planck} PR3 lensing measurement \cite{Planck:2018lbu} reconstructs the lensing convergence from a linear combination of minimum variance quadratic estimators \cite{Okamoto:2003zw} whose inputs are Wiener-filtered and inverse-variance-weighted \verb|SMICA| \cite{Planck:2018yye} temperature and polarization multipoles with $100\leq\ell\leq2048$. Specifically, we use the minimum variance lensing reconstruction\footnote{COM\_Lensing\_4096\_R3.00 from the \href{https://pla.esac.esa.int/\#home}{\textit{Planck} legacy archive}.} along with the provided CMB lensing mask and effective reconstruction noise curve $N^{\kappa\kappa}_\ell$.

\textbf{Public Release 4 (PR4):} 
The \textit{Planck} PR4 lensing map \cite{Carron:2022eyg} is reconstructed using a more optimal Global Minimum Variance estimator \cite{Maniyar:2021msb} from the latest NPIPE processing pipeline, which uses $\sim8\%$ additional measurement time and a more optimal (anisotropic) filtering scheme (using the same scales $100\leq\ell\leq2048$ as PR3). The resulting CMB lensing map is signal dominated (per mode) out to\footnote{We use context rather than different symbols (e.g. with $L$ instead of $\ell$, as in the companion paper \cite{Kim2024}) to differentiate CMB lensing and primary CMB multipoles, since we use $L$ in \S\ref{sec:cl_estimation} to denote bandpowers. From \S\ref{sec:cl_estimation} onward $\ell$ always refers to a CMB lensing multipole.} $\ell\sim70$, compared with $\ell\sim40$ with the previous PR3 map, resulting in an overall 20\% improvement to the signal-to-noise ratio and a $42\sigma$ detection significance of the CMB lensing auto-correlation. Specifically, we use the reconstructed convergence multipoles, mask and reconstruction noise curve provided on NERSC\footnote{See \href{https://github.com/carronj/planck\_PR4\_lensing}{github.com/carronj/planck\_PR4\_lensing} for more information. Specifically, we use PR4\_klm\_dat\_p.fits, mask.fits.gz and PR4\_nlkk\_p.dat.}. 

\subsection{ACT DR6 CMB lensing}
\label{sec:act_kappa}

Our headline analysis uses the latest CMB lensing measurement\footnote{We use the baseline reconstructed multipoles ($\hat{\kappa}_{\ell m}$, kappa\_alm\_data\_act\_dr6\_lensing\_v1\_baseline.fits) and corresponding mask (mask\_act\_dr6\_lensing\_v1\_healpix\_nside\_4096\_baseline.fits) that are available at this \href{https://portal.nersc.gov/project/act/dr6_lensing_v1/maps/baseline/}{URL}. We downgrade the mask from an nSide of 4096 to 2048 using healpy's ud\_grade routine.} (which detected the CMB lensing auto-correlation with a 43$\sigma$ significance) from the Atacama Cosmology Telescope (ACT) DR6 \cite{ACT:2023dou,ACT:2023ubw,ACT:2023kun}, which we low pass filter\footnote{We set $\hat{\kappa}_{\ell m}=0$ for $\ell>3000$, and leave modes with $\ell\leq3000$ untouched.} before constructing a $\kappa$ map using \verb|healpy|'s \verb|alm2map| routine with \verb|nSide = 2048|. The data used to construct this map consists of nighttime observations made through 2021 at 98 and 150 GHz. The CMB lensing convergence is reconstructed from a (nearly optimal) linear combination of temperature- and polarization-based quadratic estimators, which in turn have been separated into a linear combination of estimators measured with disjoint data splits \cite{Madhavacheril:2020ido} to mitigate noise biases.  All estimators use the scales $600<\ell<3000$ to mitigate contamination from Galactic emission and extragalactic foregrounds. To further mitigate extragalactic contamination in the temperature-based reconstruction, a profile-hardened estimator \cite{2013MNRAS.431..609N,Osborne:2013nna,Sailer:2020lal} has been adopted, which has been shown to efficiently suppress tSZ- and CIB-induced biases without requiring a finely-tuned model for the mean tSZ intensity profile \cite{Sailer:2022jwt}.

\section{Power spectra, window functions and covariances}
\label{sec:cl_estimation}

\begin{figure}[!h]
    \centering
    \includegraphics[width=\linewidth,valign=c]{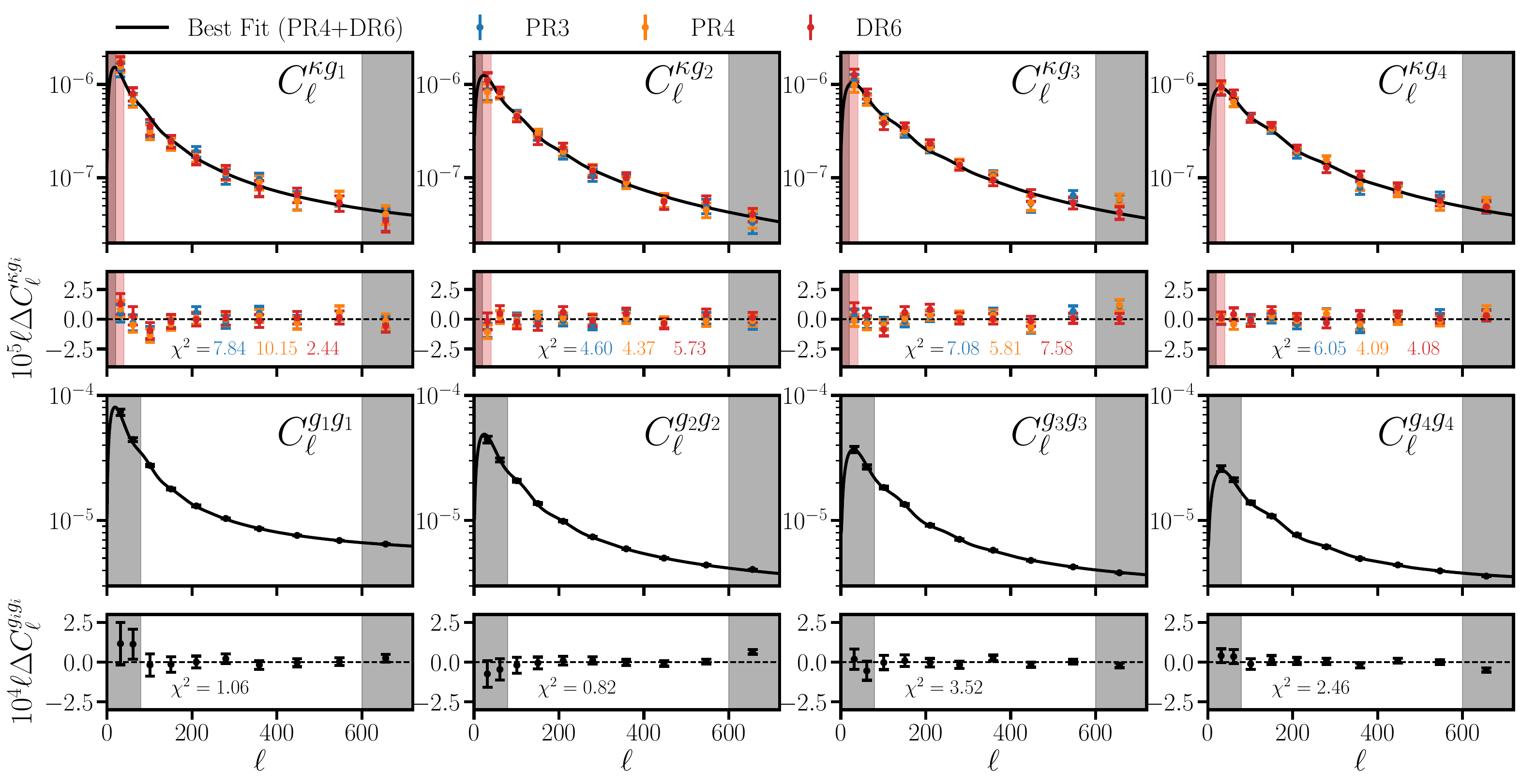}
    \caption{
    In the top row we plot the LRG cross-correlation ($C^{\kappa g}_\ell$) with \textit{Planck} PR3 (blue), PR4 (orange) and ACT DR6 (red), while in the third row we plot the LRG auto-spectra ($C^{gg}_\ell$). 
    With the exception of the PR3 cross-correlation (computed following \S\ref{sec:cl_estimation}) these spectra and their covariances are taken from the companion paper \cite{Kim2024} (and cross-validated in Appendix \ref{sec:joshuas_vs_noah_measurements}).
    The hybrid covariance computed in the companion paper \cite{Kim2024} finds that neighboring bandpowers are correlated by at most $4\%$.
    The black solid lines are the best-fit HEFT prediction when jointly fitting PR4 and DR6 using all four redshift bins.
    We show the residuals with the best-fit prediction in the second and fourth rows. For each measurement we quote its individual $\chi^2$ value (indicated by color for the cross-correlation measurements). 
    The grey shaded regions correspond to the scales omitted from our fiducial HEFT fits to the galaxy auto- and cross-correlation with \textit{Planck}, while the shaded red regions are additionally excluded in our fits to the ACT cross-correlation (see Table~\ref{tab:scalecuts}). There are 9 data points for each \textit{Planck} cross-correlation, 8 for each ACT cross-correlation and 7 for each galaxy auto-correlation. 
    We estimate the goodness of fit in \S\ref{sec:fit_good}.
    } 
\label{fig:best_fit}
\end{figure}

The bandpowers and accompanying covariance used in our fiducial \textit{Planck} PR4 and ACT DR6 analyses are discussed in detail in the companion paper \cite{Kim2024} and illustrated in Fig.~\ref{fig:best_fit}. 
In this work we independently measure these power spectra and analytically estimate their covariance (following a slightly different approach from the companion paper \cite{Kim2024}) as a cross-check. 
We compare cosmological constraints from the two approaches in Appendix \ref{sec:joshuas_vs_noah_measurements} and find negligible differences between them (in particular, the $S_8$ constraints are identical to three decimal places). In sections \ref{sec:verification_and_volume}, \ref{sec:systematics}  and \ref{sec:results} we perform systematics tests, estimate volume effects, fit to mocks, reanalyze the \textit{Planck} PR3 cross-correlation and perform a suite of parameter-based consistency checks $-$ all of which require additional measurements and covariance estimation that were done using the methods summarized below.

Masking and Fourier transforms are both linear operations. It follows that the measured power spectrum\footnote{We use tildes to differentiate a measured from an underlying power spectrum.} of a masked LSS tracer ($\tilde{C}_\ell$) is linearly related to the underlying power spectrum ($C_\ell$). Typically one bins the measured spectrum into bandpowers (which we label $\tilde{C}_L$, where capital $L$ stands for bandpower), such that $\langle \tilde{C}_L\rangle = \sum_{\ell} W_{L\ell} C_\ell$ where the ``window function" $W_{L\ell}$ is a functional of the mask(s), definition of the bandpowers and any other (optional) linear operations that the user applies. We measure ``mask-deconvolved" bandpowers\footnote{Defined such that $W_{L \ell}$ is a tophat in $\ell$ for each $L$ when the underlying power spectrum is piece-wise constant in each bandpower.} following the MASTER algorithm \cite{Hivon:2001jp} as implemented in \verb|NaMaster| \href{https://github.com/LSSTDESC/NaMaster}{\faGithub} \cite{Alonso:2018jzx}. Specifically we use the \verb|compute_full_master| routine with uniform weights and 33 non-overlapping bin edges that are linearly spaced in $\sqrt{\ell}$ across the fiducial ($20\leq\ell<600$) analysis range:
\begin{equation*}
\begin{aligned}
    &[10, 20, 44, 79, 124, 178, 243, 317, 401, 495, 600, 713, 837, 971, 1132, 1305, 1491, 1689,\\
    &\,1899, 2122, 2357, 2605, 2865, 3137, 3422, 3719, 4028, 4350, 4684, 5030, 5389, 5760, 6143].
\end{aligned}
\end{equation*}
We follow the default \verb|NaMaster| convention that each bandpower is inclusive (exclusive) for the minimum (maximum) bin edge, i.e. for two bin edges $L^{\rm edge}_{\rm min}$ and $L^{\rm edge}_{\rm max}$ the bandpower is given the average over all $C_\ell$'s satisfying $L^{\rm edge}_{\rm min}\leq \ell < L^{\rm edge}_{\rm max}$, such that our bin edges define 32 \textit{disjoint} bandpowers.
Despite the fact that we only use modes with $20\leq\ell<600$ in our analysis we compute bandpowers down to $L_{\rm edge}=10$ and up to $L_{\rm edge}=\,\,$\verb|3*nSide-1| to mitigate numerical artifacts in the \verb|NaMaster| implementation. Window functions are obtained from the \verb|get_bandpower_windows| routine, which are computed out to $\ell_{\rm max}=6143$. In \S\ref{sec:results}, when ``convolving" our theory prediction with the window function we truncate this sum at $\ell_{\rm max}=1000$. For the relevant bandpowers in our analysis, the window functions are $<10^{-6}$ for $\ell>1000$. 

After measuring a cross-correlation with a CMB lensing map we apply a ``normalization" correction. The implementation of this correction differs slightly from that used in the companion paper \cite{Kim2024}, which we discuss in Appendix~\ref{sec:joshuas_vs_noah_measurements}.

We estimate the bandpower covariance with \verb|NaMaster|'s \verb|gaussian_covariance| routine, which in our case requires fiducial spectra for the CMB lensing power spectrum $C^{\kappa\kappa}_\ell$, CMB lensing reconstruction noise $N^{\kappa\kappa}_\ell$, cross-correlation between CMB lensing estimators (relevant only when combining PR4 and DR6), the galaxy power spectra $C^{g_ig_j}_\ell$ (for $i,j=1,2,3,4$) and their cross-correlation with CMB lensing $C^{\kappa g_i}_\ell$. Fiducial spectra are provided up to $\ell_{\rm max}=3000$. We take $C^{\kappa\kappa}_\ell$ to be the fiducial CMB lensing power spectrum used in the ACT DR6 lensing simulations \cite{ACT:2023dou}, and $N^{\kappa\kappa}_\ell$ to be the effective (sky-averaged) lensing reconstruction noise for a given CMB experiment (see sections \ref{sec:planck_kappa} and \ref{sec:act_kappa}). The cross-correlation between the \textit{Planck} PR4 and ACT DR6 lensing reconstructions is calculated in the companion paper \cite{Kim2024} (see also \cite{ACT:2023dou,ACT:2023oei}) using the FFP10 simulations \cite{Planck:2018lkk}. Fiducial curves for $C^{g_ig_i}_\ell$ and $C^{\kappa g_i}_\ell$ are obtained by fitting to each measurement \textit{individually} as we now describe\footnote{We use the cross-correlation with PR4 when estimating covariances. We note that the exact procedure for obtaining fiducial theory spectra is not important provided that they are in good agreement with the data (once binned with the respective window functions). For example, \cite{White:2021yvw} simply fit a polynomial to the data.}. For each measured $\tilde{C}_L$ (either $\tilde{C}^{g_ig_i}_L$ or $\tilde{C}^{\kappa g_i}_L$ for a single $i$) and HEFT prediction (see \S\ref{sec:modeling}) $C_\ell$ we assign a loss function: $\sum_L (\tilde{C}_L - \sum_\ell W_{L\ell}C_\ell)^2/\tilde{C}_L^2$, where the sum over $\ell$ runs from $0$ to $3000$ and the sum over bandpowers runs up to $L_{\rm edge, max} = 2865$. In the HEFT prediction we hold the cosmology fixed to that used in the ACT DR6 lensing simulations\footnote{$A_s = 2.15086\times10^{-9}$, $\omega_c = 0.1203058$, $n_s = 0.9625356$, $h = 0.6702393$, $\omega_b = 0.02219218$, $M_\nu = 60$ meV} and use the measured redshift distributions of the LRGs (Fig.~\ref{fig:dNdz_variations}). A fiducial theory prediction is obtained by varying the nuisance terms to minimize the loss function\footnote{As a technical detail, we actually minimize the sum of the loss function and the $\chi^2$ associated with the nuisance parameter priors (listed in Table~\ref{tab:priors}). Doing so stabilizes the minimization.}. Fiducial curves for the galaxy cross-spectra ($C^{g_ig_j}_\ell$ for $i\neq j$) are obtained via Eq.~\eqref{eq:cgigj} by linearly interpolating the best-fit nuisance terms from the galaxy auto fits with redshift. 

We have checked that replacing the loss function with a Gaussian likelihood (using an analytic covariance with fiducial curves estimated via the preceding paragraph) and repeating the procedure above results in subpercent changes to the best-fit $C^{g_ig_j}_\ell$'s and $C^{\kappa g_i}_\ell$'s.

\section{Modeling}
\label{sec:modeling}

\subsection{Aemulus $\nu$}
\label{sec:aeumulus}

The models discussed in the following subsections are built on the \verb|Aemulus| $\nu$ simulations \cite{DeRose:2023dmk}, which are a set of 150 $N$-body simulations spanning a broad range of cosmologies (in particular, $1.1 < 10^9 A_s < 3.1$ and $0.08<\omega_c<0.16$) including both massive neutrinos and variations to the dark energy equation of state. 
The simulations treat the CDM+baryon ($cb$) and neutrino fields as two sets of $1400^3$ collisionless particles that are evolved in a $1.05h^{-1}\text{Gpc}$ box, yielding a $cb$ mass resolution of $3.51\times10^{10} (\Omega_{cb}/0.3)h^{-1}M_\odot$. The initial displacements ($z_\text{ini}=12$) of the $cb$ field are calculated using third-order Lagrangian Perturbation Theory (as implemented in \verb|monofonic| \cite{Michaux:2020yis,Hahn:2020lvr}) from an initial Gaussian density field whose power spectrum is taken to be $P_{cb}(k,z_\text{ini}) = D_{cb}(k,z_{\rm ini})^2 P_{cb,\text{lin}}(k,z=0)$, where $P_{cb,\text{lin}}(k,z=0)$ is computed with \verb|CLASS| \cite{CLASS} and the scale-dependent growth factor is computed using first-order Newtonian fluid approximation as implemented in \verb|zwindstroom| \cite{Elbers:2020lbn}, while the initial conditions for the neutrinos are set with \verb|fastDF| \cite{Elbers:2022tvb}. The $cb$ and neutrino particles are evolved from $z_\text{ini}=12$ to $z=0$ using a modified version of \verb|Gadget-3| \cite{yu_feng_2018_1451799} that enables modifications and includes relativistic corrections to the background evolution $H(z)$.

\subsection{Hybrid Effective Field Theory}
\label{sec:heft}

We adopt a Hybrid Effective Field Theory (HEFT) model \cite{Modi:2019qbt,Kokron21,Hadzhiyska21,Zennaro:2021pbe,Pellejero-Ibanez:2022efv,Nicola:2023hsd} in our fiducial analysis. Within this framework the ``initial" (Lagrangian) galaxy density is modeled as a biased tracer of the Lagrangian CDM+baryon density. Proto-galaxies at Lagrangian position $\bm{q}$ are advected to their later (real-space) positions $\bm{x}(t) = \bm{q}+\bm{\Psi}_{cb}(\bm{q},t)$ following the (non-linear) displacement vector $\bm{\Psi}_{cb}(\bm{q},t)$ of the CDM+baryon field \cite{Villaescusa-Navarro:2013pva,Castorina:2013wga,Castorina:2015bma,Villaescusa-Navarro:2017mfx,LoVerde:2014pxa,Munoz:2018ajr,Fidler:2018dcy,Chen:2022jzq} that is measured from simulations. From number conservation, the (real-space) late-time galaxy density contrast is given by
\begin{equation}
    1+\delta_g(\bm{x})
    =
    \int d^3\bm{q}
    \,
    F(\bm{q})
    \,
    \delta^D(\bm{x}-\bm{q}-\bm{\Psi}_{cb})
\end{equation}
where the weight function $F(\bm{q})$ characterizes the initial galaxy density fluctuations and $\delta^D$ is the Dirac delta function. Here we take $F(\bm{q})$ to be a perturbative expansion in scalar combinations of the tidal $(\partial_i\partial_j \Phi)$ and velocity $(\partial_i v_j)$ tensors to next-to-leading order in derivatives and perturbations \cite{2009JCAP...08..020M}
\begin{equation}
\label{eq:bias_expansion}
    F(\bm{q}) = 
    1 + b^L_1 \delta_{cb}(\bm{q})
    + \frac{b^L_2}{2} \big(\delta^2_{cb}(\bm{q}) - \langle\delta^2_{cb}\rangle\big)
    + b^L_s \big(s^2_{cb}(\bm{q}) - \langle s^2_{cb}\rangle\big) 
    +
    \frac{b^L_{\nabla^2}}{4} \big(\nabla^2\delta_{cb}(\bm{q}) - \langle\nabla^2\delta_{cb}\rangle\big)
    +\mathcal{E}(\bm{q})
\end{equation}
where $s^2_{cb} = \sum_{ij} s^{ij}_{cb} s^{ij}_{cb}$ is the square of the shear field $s^{ij}_{cb} = (\partial_i\partial_j/\partial^2-\delta^K_{ij}/3)\delta_{cb}$ and $\mathcal{E}(\bm{q})$ is a small-scale stochastic component. Within this approximation the late-time galaxy density contrast takes the form
\begin{equation}
    \delta_g(\bm{k}) = \left(1-\frac{b^L_{\nabla^2} k^2}{4}\right)\delta_{cb}(\bm{k}) + \sum_{X} b^L_X F_X(\bm{k}) +
    \mathcal{E}(\bm{k})
\end{equation}
where $X\in\{1,2,s\}$, $F_X(\bm{k})$ corresponds to the respective term (post-advection and Fourier transformed) in the weight function, and we have approximated the post-advection $\nabla^2\delta_{cb}$ term as $-k^2\delta_{cb}(\bm{k})$. The galaxy power spectrum is then
\begin{equation}
\label{eq:Pgg}
    P_{gg}(k) = \left(1-\frac{\alpha_a k^2}{2}\right)P_{cb}(k) + 2\sum_{X} b^L_X P_{cb,X}(k)+ \sum_{XY} b^L_X b^L_Y P_{XY}(k) + \text{SN}^\text{3D}
\end{equation}
where $P_{cb}$ is the power spectrum of the CDM+baryon field, $P_{cb,X}$ is the cross-correlation of the CDM+baryon field with $F_X$, $P_{XY}$ is the cross-spectrum of $F_X$ and $F_Y$, SN$^\text{3D}$ is a shot noise contribution, we introduced the ``counterterm" $\alpha_a$ for the auto-correlation, and we have neglected $\mathcal{O}(k^2\delta^3_{cb})$ contributions. Likewise the cross-correlation of the galaxies with the matter is approximately
\begin{equation}
\label{eq:Pgm}
    P_{gm}(k) = \left(1-\frac{\alpha_x k^2}{2}\right)P_{cb,m}(k) + \sum_{X} b^L_X P_{m,X}(k)
\end{equation}
where $P_{cb,m}$ is the cross spectrum of the CDM+baryon field with the total matter field, $P_{m,X}$ is the cross correlation of the total matter field with $F_X$, and we have defined the counterterm $\alpha_x$ for the cross-correlation.

The power spectra ($P_{cb}, P_{cb,X}$, $P_{XY}$, $P_{cb,m}$, and $P_{m,X}$) appearing in Eqs.~\eqref{eq:Pgg} and \eqref{eq:Pgm} are calculated in the \verb|Aemulus| $\nu$ simulations described in \S\ref{sec:aeumulus}. The sample variance of these measurements on perturbative scales has been suppressed using the Zel'dovich density field (implemented using the \verb|ZeNBu| \href{https://github.com/sfschen/ZeNBu}{\faGithub} and \verb|velocileptors| \href{https://github.com/sfschen/velocileptors}{\faGithub} \cite{Chen:2020fxs} codes) as a control variate\footnote{Specifically, \cite{DeRose:2023dmk} estimates a power spectrum (e.g. $P_{X,Y}$) using $\hat{P}^\text{CV}_{X,Y}\equiv \hat{P}^\text{Aemulus}_{X,Y} - \beta (\hat{P}^\text{LPT}_{X,Y} - P^\text{LPT}_{X,Y})$, where $\hat{P}^\text{Aemulus}_{X,Y}$ is the measured power spectrum from the Aemulus $\nu$ simulation, $\hat{P}^\text{LPT}_{X,Y}$ is the measured power spectrum of the (Zel'dovich) LPT evolution of the same initial conditions used in the simulation and $P^\text{LPT}_{X,Y}$ is the analytic prediction for the ensemble average (which is known to machine precision). The variance of $\hat{P}^\text{CV}_{X,Y}$ is (optimally) suppressed by a factor of $1-\rho^2_{\text{Aemulus},\text{LPT}}$ relative to the variance of $\hat{P}^\text{Aemulus}_{X,Y}$ if one chooses $\beta = \text{Cov}[\hat{P}^\text{Aemulus}_{X,Y},\hat{P}^\text{LPT}_{X,Y}]/\text{Var}[\hat{P}^\text{LPT}_{X,Y}]$, where $\rho_{\text{Aemulus},\text{LPT}}$ is the correlation coefficient of the Aemulus and LPT measurements.}.
We use the \verb|Aemulus| $\nu$ emulator \href{https://github.com/AemulusProject/aemulus_heft}{\faGithub} in our calculations, which has been shown to be accurate to within $0.25\%$ for the scales $(k\lesssim0.6\,h\,{\rm Mpc}^{-1})$ and redshifts $(z\lesssim1)$ relevant for our analysis \cite{DeRose:2023dmk}. 

While we have introduced two counterterms ($\alpha_x$ and $\alpha_a$) in Eqs.~\eqref{eq:Pgg} and \eqref{eq:Pgm}, these parameters are not independent in the limit that the bias expansion (Eq.~\ref{eq:bias_expansion}) accurately describes the \textit{field-level} galaxy distribution and the simulated $\delta_m(\bm{x})$ accurately describes the true matter distribution\footnote{This is analogous to multi-tracer studies where the coefficients of counterterm-like contributions to the cross-correlation of two biased tracers can be expressed in terms of counterterm coefficients appearing in the two individual auto-spectra (shown explicitly in Appendix C of \cite{Ebina:2024ojt}).}. The former approximation is valid up to baryonic feedback and the finite mass resolution of the simulations. Both of these effects are small on the scales of interest and have the qualitative impact of damping small-scale power that we approximate by introducing an effective ``derivative bias" in the matter distribution: $\delta_m(\bm{k})\to \big(1-\tilde{\epsilon} k^2/4\big)\delta_m(\bm{k})$. Baryonic feedback scenarios capable of resolving the $S_8$-tension in galaxy-lensing measurements (Fig 6 of ref.~\cite{Amon:2022azi}, C-OWLS AGN with $\log_{10}(\Delta T_\text{heat}/{\rm K})=8.7$ at $z=0$) require a $\sim10\%$ suppression to the matter power spectrum at $k\simeq0.4$ $h\,$Mpc$^{-1}$ corresponding to $\tilde{\epsilon}\simeq1.3$ $h^{-2}$Mpc$^2$, while the effective grid scale in the \verb|Aemulus| $\nu$ simulations is $\approx 1\,h^{-1}$Mpc corresponding to $\tilde{\epsilon}\simeq1$ $h^{-2}$Mpc$^2$. With the inclusion of the effective derivative bias in the matter distribution the $\mathcal{O}(k^2)$ components of the galaxy-auto and cross-correlation with matter are
\begin{equation}
\begin{aligned}
\label{eq:ksq_contributions}
    P_{gg}(k) 
    &\supset
    2\times
    \bigg\langle 
    \bigg[-\frac{b^L_{\nabla^2}k^2}{4}\delta_{cb}(\bm{k})\bigg]
    \times
    \bigg[
    \delta_{cb}(\bm{k})
    +
    b^L_1 F_1(\bm{k})
    \bigg]^*
    \bigg\rangle'
    \\
    &=
    -\frac{b^E_1 b^L_{\nabla^2} k^2}{2} P_{cb}(k)
    \\
    P_{gm}(k) 
    &\supset
    \bigg\langle
    \bigg[
    \bigg(
    1-
    \frac{b^L_{\nabla^2}k^2}{4}
    \bigg)
    \delta_{cb}(\bm{k}) 
    + b_1^L F_1(\bm{k})
    \bigg]
    \bigg(
    1
    -
    \frac{\tilde{\epsilon}k^2}{4}
    \bigg)
    \delta^*_m(\bm{k})
    \bigg\rangle'
    \\
    &\supset
    -\frac{(b^E_1\tilde{\epsilon}+b^L_{\nabla^2})k^2}{4} P_{cb,m}(k)
\end{aligned}
\end{equation}
where we neglect $\mathcal{O}(k^2 \delta^3)$ contributions (in particular, $F_1(\bm{k}) = \delta_{cb}(\bm{k})$ under this approximation), we introduced the Eulerian bias $b_1^E \equiv 1+b_1^L$, the symbol $\supset$ denotes a subset of the terms appearing in the power spectra predictions (in this case, the $k^2$ terms), and primed brackets correspond to an ensemble average with the momentum-conserving delta function removed, e.g.\ $\langle \delta(\bm{k}_1) \delta(\bm{k}_2)\rangle \equiv (2\pi)^3 \delta^D(\bm{k}_1+\bm{k}_2) \langle \delta(\bm{k}_1) \delta(\bm{k}_2)\rangle'$. 
Comparing Eq.~\eqref{eq:ksq_contributions} to Eqs.~\eqref{eq:Pgg} and \eqref{eq:Pgm} we see that $\alpha_a = b^E_1 b^L_{\nabla^2}$ and $2\alpha_x = b_1^E \tilde{\epsilon} + b^L_{\nabla^2}$, which in turn implies 
\begin{equation}
\label{eq:counterterm_relationship}
    \alpha_x = \frac{\alpha_a}{2 b^E_1} + \epsilon,
\end{equation}
where $\epsilon\equiv b_1^E \tilde{\epsilon}/2$ ($b_1^E/2\simeq 1$ for the LRGs considered here) has a ``typical" value of $\simeq 1\,h^{-2}$Mpc$^2$, which we marginalize over with an informative Gaussian prior centered at zero with width $\simeq 2\,h^{-2}$Mpc$^2$ in our fiducial analysis (see \S\ref{sec:priors}).
We note that this prior is sufficiently broad to permit an enhancement of small-scale power (within one prior $\sigma$) rather than a suppression, even when taking into account the suppression already present in the model prediction arising from the effective grid scale in the \verb|Aemulus| $\nu$ simulations.

\subsection{Linear theory}
\label{sec:linthy}

Non-linear corrections to the matter power spectrum are $\leq 1.5\%$ for $k\leq 0.1\,h\,{\rm Mpc}^{-1}$ and $z>0.3$. Provided that the length scale associated with the bias expansion (e.g. the Lagrangian radius of LRG host halos) is smaller than that associated with non-linear (gravitational) dynamics\footnote{We note that this approximation is implicitly made when fitting to higher $k$ with HEFT than with ``pure" perturbation theory. A caveat to this claim is that some higher-order contributions, such as $\langle \delta^2 \delta^2\rangle$, are non-zero in the $k\to 0$ limit. However, these contributions are scale-independent in this limit and thus degenerate with shot noise.}, one can accurately model the galaxy distribution using linear theory on these scales, which we consider as an alternative to our fiducial HEFT model in \S\ref{sec:results}. 

Rather than using pure linear theory for the galaxy cross- and auto-correlations, we choose to use the HEFT prediction with $b_2^L=b_s^L=0$, and adjust our scale cuts such that $\ell_{\rm max}(z) \simeq  \chi(z) \times (0.1\,h\,{\rm Mpc}^{-1})$ (see \S\ref{sec:scale_cuts} and Table~\ref{tab:scalecuts}). This is done purely for convenience (e.g. the emulator is faster than a \verb|CLASS| evaluation), since swapping the HEFT predictions for linear theory on these scales would have a negligible impact on our constraints. To approximately account for the residual systematic error in $P_{gg}$ and $P_{gm}$ from neglecting higher-order contributions on these scales we choose to marginalize over $\alpha_a$ and $\alpha_x$ independently with an informative prior chosen such that a 1.5\% correction is allowed at $k= 0.1\,h\,{\rm Mpc}^{-1}$ at $1\sigma$ (see \S\ref{sec:priors} and Table~\ref{tab:priors}). As for our fiducial model, we use the non-linear $P_{mm}$ prediction when calculating magnification contributions (\S\ref{sec:3Dto2D}).

\subsection{Model independent constraints}
\label{sec:modelind}

As yet another alternative to HEFT and linear theory, we consider a ``model independent" (or ``fixed shape") approach to constrain $\sigma_8(z)$. Under this approach we use the same scale cuts as for our linear theory analysis, fix the cosmology, and allow the linear bias for the galaxy auto ($b^{L,a}_1$) and cross ($b^{L,x}_1$) to differ by a factor
\begin{equation}
\label{eq:alpha8}
    \alpha_8 \equiv
    \frac{1+b^{L,x}_1}{1+b^{L,a}_1}.
\end{equation}
Within linear theory one constrains $\sigma_8$ schematically through the ratio $C^{\kappa g}/\sqrt{C^{gg}} \sim \alpha_8\sigma_{8,\text{ fid}}$, where $\sigma_{8,\text{ fid}}$ is the value of $\sigma_8$ for the fiducial cosmology. In \S\ref{sec:alternative_models} we consider fitting to each redshift bin independently, and interpret our constraint on $\alpha_8\,\sigma_{8,\text{ fid}}(z_i)$ as a model independent measurement of $\sigma_8(z_i)$.

\subsection{From 3D to 2D}
\label{sec:3Dto2D}

We use the Kaiser-Limber approximation \cite{1953ApJ...117..134L,1992ApJ...388..272K} (hereafter Limber approximation) to model the angular galaxy auto- and cross-correlation:
\begin{equation}
\begin{aligned}
\label{eq:limber}
    C^{gg}_\ell 
    &= 
    \int_0^\infty \frac{d\chi}{\chi^2} 
    \bigg[
    \left[W^g(\chi)\right]^2 P_{gg}\left(k,z\right)
    +
    2
    W^g(\chi) W^\mu(\chi) P_{gm}(k,z)
    +
    \left[W^\mu(\chi)\right]^2 P_{mm}(k,z)
    \bigg]
    \\
    C^{\kappa g}_\ell
    &=
    \int_0^\infty \frac{d\chi}{\chi^2}
    W^\kappa(\chi)
    \bigg[
    W^g(\chi) P_{gm}(k,z)
    + 
    W^\mu(\chi) P_{mm}(k,z)
    \bigg]
\end{aligned}
\end{equation}
where $k=(\ell+1/2)/\chi$ \cite{LoVerde:2008re} and $z$ is implicitly a function of $\chi$. The projection kernels for the galaxies, magnification contribution, and CMB lensing are \cite{Villumsen:1995ar,BARTELMANN2001291,Moessner:1997qs}
\begin{equation}
\begin{aligned}
\label{eq:kernels}
    W^g(\chi) &= H(z)\phi(z)\\
    W^\mu(\chi) 
    &= (5s_\mu-2) \frac{3}{2}\Omega_m H_0^2 (1+z) \int_z^{\infty} dz' 
    \frac{\chi(z)(\chi(z')-\chi(z))}{\chi(z')} \phi(z')
    \\
    W^\kappa(\chi) &= \frac{3}{2}\Omega_m H_0^2 (1+z) \frac{\chi(\chi_*-\chi)}{\chi_*}
\end{aligned}
\end{equation}
where $\phi(z)\propto dN/dz$ is the normalized ($\int dz\phi(z)=1$) redshift distribution, $\chi_*$ ($z_*$) is the comoving distance (redshift) to the surface of last scattering, and $s_\mu$ is the number count slope. 
Due to the narrow redshift distributions of the LRG bins, the magnification contributions to the galaxy auto-spectra amount to less than a percent of the total signal, while for the cross-correlation with CMB lensing the magnification contribution is at most $\simeq10\%$ for the highest redshift bin (and smaller for the remaining bins).
We compute background quantities (i.e $\Omega_m$, $\chi_*$, $H(z)$, and $\chi(z)$) relevant for Limber integration from our sampled cosmological parameters (\S\ref{sec:priors} and Table~\ref{tab:priors}) using \verb|CLASS| \cite{CLASS}.

\begin{figure}[!h]
    \centering
    \includegraphics[width=\linewidth,valign=c]{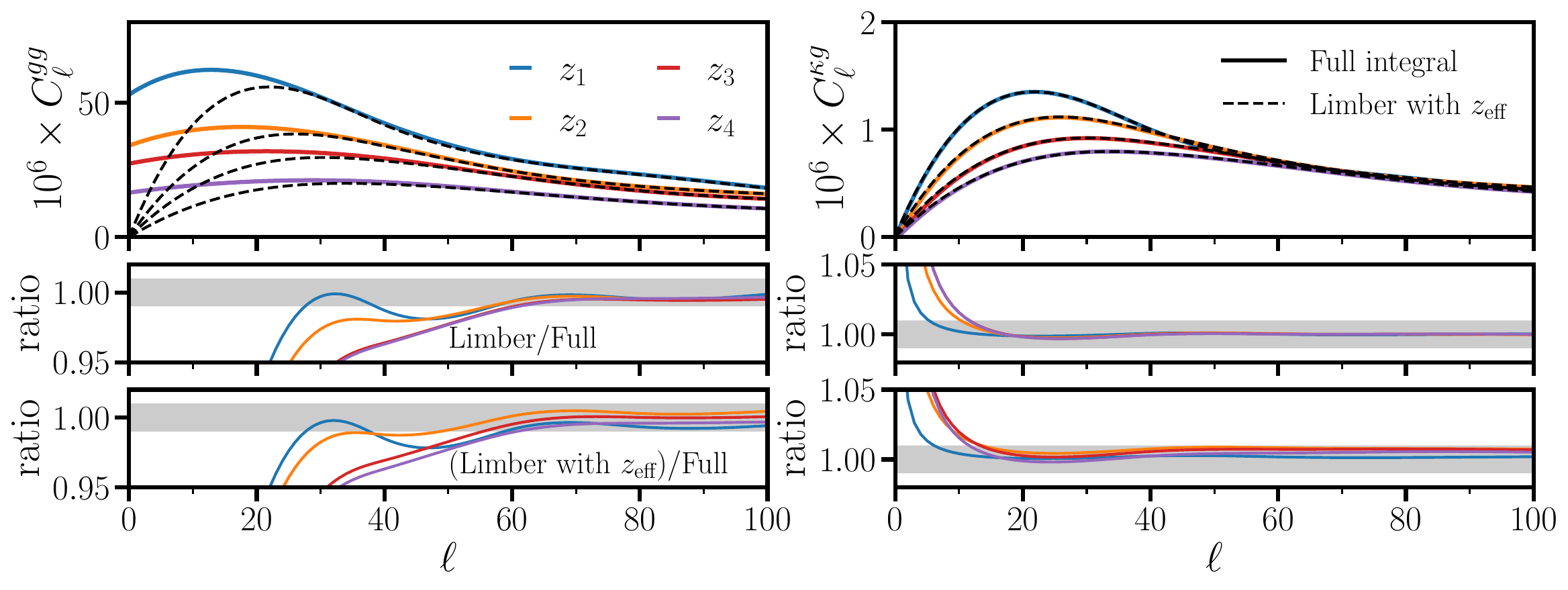}
    \caption{ 
    Estimation of ``beyond Limber" effects within linear theory for both the galaxy auto- and cross-correlation (left and right). The middle panels show the ratio of the Limber approximation to the full integral, while the bottom panels show the ratio when evaluating the power spectra at a fixed redshift $z_\text{eff}$, defined in Eq.~\eqref{eq:zeff}. The shaded bands correspond to $1\pm0.01$.
    } 
\label{fig:beyondLimber}
\end{figure}

In practice, we neglect the evolution of $P_{gg}$ and $P_{gm}$ in the Limber integrals, and evaluate them at the effective redshift
\begin{equation}
\label{eq:zeff}
    z_\text{eff} 
    = \frac{\int d\chi\, z(\chi)\, [W^g(z)]^2/\chi^2}{\int d\chi [W^g(z)]^2/\chi^2}.
\end{equation}
This approximation, which we expect to be accurate to subpercent precision for our LRG sample (see Fig.~\ref{fig:beyondLimber}) due to its narrow redshift distributions, has the advantage of making our analysis largely agnostic to the assumed redshift evolution of the galaxy nuisance parameters within each redshift bin. Instead, we treat e.g. $b_1^L(z_i)$ as a single number evaluated at the effective redshift for the $i$'th redshift bin.  

Deviations from the Limber approximation are primarily relevant on large scales where linear theory is a good approximation. Within this approximation and neglecting magnification bias, the spherical harmonic coefficients $g_{\ell m}$ of the projected galaxy density contrast can be split in two pieces $g^\text{real}_{\ell m} + g^\text{RSD}_{\ell m}$ where \cite{Fisher:1993pz,2007MNRAS.378..852P} 
\begin{equation}
\begin{aligned}
\label{eq:rsd_limber}
    g^X_{\ell m} 
    &\equiv 
    \int \frac{d^3 k}{(2\pi)^3}
    Y^*_{\ell m}(\hat{\bm{k}})
    \int d\chi \mathcal{W}^X_\ell(k,\chi) \delta_g(\bm{k})
    \\
    \mathcal{W}^\text{real}_\ell(k,\chi)
    &=
    4\pi i^\ell W^g(\chi) j_\ell(k\chi)
    \\
    \mathcal{W}^\text{RSD}_\ell(k,\chi)
    &=
    4\pi i^\ell \frac{1}{k}
    \frac{dW^{g}}{d\chi} j'_\ell(k\chi) \beta(z),
\end{aligned}
\end{equation}
$\delta_g(\bm{k})$ is the real-space 3D galaxy density contrast, $\beta(z) = f(z)/b_1^E(z)$ where $f(z)$ is the linear growth rate and $j'_\ell(x) = \partial_x j_\ell(x)$ is the derivative of the spherical Bessel function of the first kind. The cross-correlation $C^{XY}_\ell \equiv \langle g^X_{\ell m} g^{Y*}_{\ell m}\rangle$ is given by the ``full integral"
\begin{equation}
    C_\ell^{XY}
    =
    \int \frac{k^2 dk}{(2\pi)^3} 
    \int d\chi d\chi'
    \mathcal{W}^X_\ell(k,\chi)
    \mathcal{W}^Y_\ell(k,\chi')
    P_{gg}(k;\chi,\chi')
\end{equation}
where $\langle\delta_g(\bm{k},\chi)\delta_g(\bm{k}',\chi') \rangle \equiv (2\pi)^3 \delta^D(\bm{k}+\bm{k}')P_{gg}(k;\chi,\chi')$.

In Fig.~\ref{fig:beyondLimber} we show the full integral for both the auto- and cross-correlations assuming linear theory: $P_{gg}(k;\chi,\chi') = D(z)b(z)D(z')b(z')P_0(k)$ where $D(z)$ is the linear growth factor normalized to $D(0)=1$ and $P_0(k)$ is the $cb$ linear power spectrum at $z=0$ (we are ignoring shot noise). We take $b(z)$ to be a linear interpolation of the best-fit values listed in Table 1 of \cite{White:2021yvw}. We also show the Limber approximations, including the effective redshift approximation. For $\ell \lesssim 75$, the largest beyond Limber effect is due to RSD, while above $\ell\gtrsim75$ the largest effect is due to the effective redshift approximation, which is still $<1\%$ and neglected in our analysis. 

\subsection{Pixelization}
\label{sec:pixelization}

The LRGs considered here have been placed on a HEALPix grid\footnote{See \cite{BaleatoLizancos:2023jbr} for an alternative approach that bypasses placing galaxies on a discrete grid, and hence the pixel window function and aliasing.}. This pixelization has the approximate effect of convolving the galaxies with the pixel window function, but in addition gives rise to aliasing of power \cite{Das:2008qm,Sefusatti:2015aex,BaleatoLizancos:2023jbr,Sailer2024}. The combination of these effects can be well approximated by taking $C^{\kappa g}_\ell \to w_\ell C^{\kappa g}_\ell$ and $C^{gg}_\ell\to w_\ell^2(C^{gg}_\ell - \text{SN}^{\rm 2D}) + \text{SN}^{\rm 2D}$, where $w_\ell$ is the pixel window function (as determined by \verb|healpy|'s \verb|pixwin| for the appropriate \verb|nSide|). Previous analyses (e.g. \cite{White:2021yvw,Krolewski:2021yqy}) have approximately corrected for the pixel window function at the data level, which requires assuming a fiducial value for the projected shot noise. This is somewhat undesirable (but still reasonable) given that one fits for the shot noise in practice.
Here we opt to forward model the impact of pixelization using the aforementioned approximation. 
For \verb|nSide = 2048| the window function differs from one by less than $0.4\%$ for $\ell<600$, and as a result this improved treatment of the pixel window function has a minuscule impact on our final constraints.

\section{Likelihood and pipeline checks}
\label{sec:verification_and_volume}

Parameter inference is performed with the \verb|cobaya| \cite{Torrado:2020dgo,2019ascl.soft10019T} sampling framework.
We make our likelihood publicly available\footnote{with the exception of the neural network weights for the HEFT emulator, which will be made public upon the publication of an upcoming DESI-DES Y3 cross-correlation analysis \cite{DeRose24b}.} (\verb|MaPar| \href{https://github.com/NoahSailer/MaPar/tree/main}{\faGithub}) and summarize its code structure in Appendix~\ref{sec:mapar}. Chains are sampled using \verb|cobaya|'s Markov Chain Monte Carlo Metropolis sampler \cite{Lewis:2002ah,Lewis:2013hha} and are considered converged when the Gelman-Rubin \cite{gelmanrubin,brooksgelman} statistic satisfies $R-1\leq0.01$.
We obtain marginal distributions with \verb|GetDist| \cite{Lewis:2019xzd} with the first 30\% of the chains removed as burn-in. 
Best-fit points are obtained using \verb|cobaya|'s default minimizer (see Appendix~\ref{sec:anamarg} for a discussion of analytic minimization for linear model parameters).

We adopt a Gaussian likelihood throughout and analytically marginalize over all parameters that linearly appear in our theory prediction: $\alpha_a(z_i)$, $\alpha_x(z_i)$ and shot noise. In our fiducial analysis analytic marginalization reduces the number of directly sampled parameters from 30 (2 cosmological and $7\times 4$ nuisance parameters) to 18 which dramatically improves the convergence time of our chains. 
We discuss our implementation of analytic marginalization in Appendix~\ref{sec:anamarg}. In the same Appendix we explicitly verify that our implementation of analytic marginalization is in agreement with the brute-force approach.

In \S\ref{sec:results} we place tight constraints on $\sigma_8$ after including a BAO prior, which efficiently breaks the $\Omega_m-\sigma_8$ degeneracy. 
To mimic this in both our ``volume effects" tests (\S\ref{sec:volume}) and fits to mock data (\S\ref{sec:buzzard_fits}) we include a mock BAO prior when quoting $\sigma_8$ constraints. We construct this prior using the covariances from \cite{2011MNRAS.416.3017B,Ross:2014qpa,BOSS:2016wmc} and adjust the central value of each distance measurement to the predicted $\Lambda$CDM value assuming a Buzzard cosmology (see \verb|MaPar/mocks/mockBAO/|).

\subsection{Scale cuts}
\label{sec:scale_cuts}

Throughout \S\ref{sec:results} we use (a subset of) the bandpowers discussed in \S\ref{sec:cl_estimation} spanning the range $20\leq \ell<600$.
The fiducial scale cuts used for each our modeling choices (sections \ref{sec:heft}, \ref{sec:linthy} and \ref{sec:modelind}) are summarized in Table~\ref{tab:scalecuts}. 
As discussed in \S\ref{sec:3Dto2D}, we adopt the Limber approximation and neglect the impact of redshift space distortions in our model, which thus acts as a systematic on large scales. For both the galaxy auto- and cross-correlation with \textit{Planck}, we adopt an $\ell_{\rm min}$ ($20$ for $\kappa g$ and $79$ for $gg$) where the ``beyond Limber" corrections are less than a percent (see Fig.~\ref{fig:beyondLimber}). 
On large scales the ACT DR6 lensing reconstruction is contaminated by a mean field contribution that greatly exceeds the CMB lensing signal for $\ell \lesssim 30$ (e.g. Fig. 9 of \cite{ACT:2023dou}), making it difficult accurately correct for in this regime. Following suit with the DR6 auto-correlation measurement \cite{ACT:2023dou} (which adopted $\ell_{\rm min}=40$) we choose to adopt a larger $\ell_{\rm min} = 44$ for the ACT cross-correlation to mitigate spurious correlations with the mean field.

\begin{table}[!h]
    \centering
    \resizebox{\textwidth}{!}{\begin{tabular}{cc||cccc|cccc|cccc}
    & & \multicolumn{4}{c|}{\textit{Planck} $C^{\kappa g}_\ell$}  & \multicolumn{4}{c|}{ACT $C^{\kappa g}_\ell$} & \multicolumn{4}{c}{$C^{g g}_\ell$}\\
    && $z_1$ & $z_2$ & $z_3$ & $z_4$ & $z_1$ & $z_2$ & $z_3$ & $z_4$ & $z_1$ & $z_2$ & $z_3$ & $z_4$ \\
    \hline
    \hline
    \rowcolor{gray!20}
    & $\ell_{\rm min}$ & 20 & 20 & 20 & 20 & 44 & 44 & 44 & 44 & 79 & 79 & 79 & 79\\
    \rowcolor{gray!20}
    \multirow{-2}{*}{HEFT (fiducial)} & $\ell_{\rm max}$ & 600 & 600 & 600 & 600 & 600 & 600 & 600 & 600 & 600 & 600 & 600 & 600\\
    \hline
    \multirow{2}{*}{Linear theory} & $\ell_{\rm min}$ & $-$ & 20 & 20 & 20 & $-$ & 44 & 44 & 44 & $-$ & 79 & 79 & 79\\
    & $\ell_{\rm max}$ & $-$ & 178 & 243 & 243  & $-$ & 178 & 243 & 243 & $-$ & 178 & 243 & 243\\
    \hline
    \multirow{2}{*}{Model independent} & $\ell_{\rm min}$ & 20 & 20 & 20 & 20 & 44 & 44 & 44 & 44 & 79 & 79 & 79 & 79\\
    & $\ell_{\rm max}$ & 178 & 178 & 243 & 243 & 178 & 178 & 243 & 243 & 178 & 178 & 243 & 243\\
    \end{tabular}}
    \caption{The fiducial scale cuts used for each of our modeling choices. We drop the lowest redshift bin when using linear theory, and note that $\ell_{\rm min}$ and $\ell_{\rm max}$ refer to bandpower bin edges as opposed to their ``centers" (as was done in \cite{White:2021yvw}).}
    \label{tab:scalecuts}
\end{table}

At high $\ell$ our scale cuts vary with the model being used. Previous work \cite{Kokron21} has found that our fiducial HEFT model (\S\ref{sec:heft}) is accurate to subpercent precision for $k\leq 0.6\,h\,{\rm Mpc}^{-1}$ when fitting to dark matter halos with the same characteristic masses $(10^{12.5} < \log_{10} M_h/M_\odot < 10^{13})$ and redshifts as expected from recent HOD fits \cite{Yuan:2023ezi} for the host halos of our LRG sample.
This motivates $\ell_{\rm max} \simeq \chi(z_{\rm min}) \times (0.6\,h\,{\rm Mpc}^{-1})$ where $z_{\rm min}$ is the lower edge of a given redshift bin. In particular, for the first redshift bin $z_{\rm min}\sim0.35$ and thus $\ell_{\rm max}\simeq 600$. In principle one could reliably extend to a higher $\ell_{\rm max}$ for the higher redshift samples, however in practice we expect limited gains from higher $\ell$ due to shot noise, lensing reconstruction noise, and degeneracies with galaxy bias. For simplicity we adopt a redshift-independent $\ell_{\rm max}=600$ in our fiducial HEFT analysis for both the galaxy auto- and cross-correlation.

As discussed in \S\ref{sec:linthy} we tune our linear theory scale cuts such that $\ell_{\max}\simeq \chi(z_{{\rm eff},i})\times(0.1\,h\,{\rm Mpc}^{-1})$, where the differences between the linear and non-linear matter power spectrum are $\leq 1.5\%$. For the lowest redshift bin applying this scale cut removes all but one bandpower for the galaxy auto-correlation ($79 < \ell < 124$), which we opt to discard when quoting linear theory constraints. For bins $z_2$, $z_3$ and $z_4$ we adopt $\ell_{\max} = 178,\,243,\,243$ corresponding to $k_{\rm max}\simeq 0.11$, $0.13$, $0.11\,h\,{\rm Mpc}^{-1}$. We use the same $\ell_{\rm max}$'s used in our linear theory fits  for our model independent constraints. Unlike for linear theory, we consider a model independent $\sigma_8(z_1)$ measurement in the lowest redshift bin, for which we take $\ell_{\max}=178$. In \S\ref{sec:buzzard_fits} we verify that all of our models recover unbiased results on simulations when adopting these scale cuts.

\subsection{Fiducial cosmology, sampled parameters and priors}
\label{sec:priors}

Our data (Fig.~\ref{fig:best_fit}) are particularly powerful at extracting the relative amplitude between $C^{\kappa g}_\ell$ and $\sqrt{C^{gg}_\ell}$ on large scales, roughly corresponding to $S_8$.
Since our goal is to assess the consistency of a low redshift structure growth measurement with that predicted within $\Lambda$CDM conditioned on primary CMB data, we fix the remaining relevant $\Lambda$CDM parameters to their \textit{Planck} 2018 \cite{Planck:2018vyg} mean values\footnote{We use the values from the TT,TE,EE+lowE column in Table 2 of \cite{Planck:2018vyg}, which contains no low redshift information modulo lensing and other secondary anisotropies.}. 
Specifically, we fix the baryon abundance $\Omega_b h^2 = 0.02236$, spectral index $n_s = 0.9649$, $\Omega_m h^3 = 0.09633$ as a proxy for the angular acoustic scale $\theta_*$ \cite{2dFGRSTeam:2002tzq,Planck:2018vyg} and additionally fix the sum of the neutrino masses to $\sum m_\nu = 0.06$ eV. We directly sample the (log) primordial power-spectrum amplitude $\ln(10^{10}A_s)$ and dark matter abundance $\Omega_c h^2$ with uniform $\mathcal{U}(2,4)$ and $\mathcal{U}(0.08,0.16)$ priors respectively. When quoting model-independent constraints these parameters are fixed to $3.045$ and $0.1202$ respectively \cite{Planck:2018vyg} and we vary $\alpha_8$ (Eq.~\ref{eq:alpha8}) with a $\mathcal{U}(0.5,1.5)$ prior.
We note that the $\ln(10^{10}A_s)$ prior extends beyond the range of the \verb|Aemulus| $\nu$ simulations (\S\ref{sec:aeumulus}), however, our data are sufficiently constraining such that for nearly all\footnote{The one exception is when analyzing $z_1$ independently without a BAO prior. Restricting the $\ln(10^{10}A_s)$ prior to $\mathcal{U}(2.4,3.4)$ (corresponding to the Aemulus $\nu$ range) truncates the $\sigma_8-\Omega_m$ contour shown in Fig.~\ref{fig:zi_OmM_sigma8}, however, the mean $S_8$ is largely unaffected ($\Delta S_8 = 0.003$, corresponding to a $<\sigma/10$ shift).} of the results presented in \S\ref{sec:results} the $2\sigma$ credible intervals lie within the \verb|Aemulus| $\nu$ range.

\begin{table}[!h]
    \centering
    \resizebox{\textwidth}{!}{\begin{tabular}{c||c||>{\columncolor[gray]{0.9}}ccc}
    Parameter & Description & \multicolumn{3}{c}{Prior or fixed value} \\
    \hline
    \hline
    $\cdots$ & $\cdots$ & HEFT (fiducial) & Linear theory & Model independent \\
    \hhline{~|~|-|-|-|}
    $\ln(10^{10}A_s)$ & ln(primordial amplitude) & $\mathcal{U}(2,4)$ & $\mathcal{U}(2,4)$ & 3.045\\
    $\Omega_c h^2$ & dark matter abundance & $\mathcal{U}(0.08,0.16)$ & $\mathcal{U}(0.08,0.16)$ & 0.1202\\
    $\alpha_8$ & Eq.~\eqref{eq:alpha8} & $-$ & $-$ & $\mathcal{U}(0.5,1.5)$ \\
    \hline
    \hline
    $b^L_1(z_i)$ & linear (Lagrangian) bias& $\mathcal{U}(0,3)$ & $\mathcal{U}(0,3)$ & $\mathcal{U}(0,3)$\\
    $b^L_2(z_i)$ & quadratic bias& $\mathcal{U}(-5,5)$ & $0$ & $0$\\
    $b^L_s(z_i)$ & shear bias & $\pmb{\mathcal{N}(0,1)}$ & $0$ & $0$\\
    $s_\mu(z_i)$ & number count slope & $\pmb{\mathcal{N}({\rm Tab.~\ref{tab:sample_properties}}, 0.1)}$ & $\pmb{\mathcal{N}({\rm Tab.~\ref{tab:sample_properties}}, 0.1)}$ & $\pmb{\mathcal{N}({\rm Tab.~\ref{tab:sample_properties}}, 0.1)}$ \\
    ${\rm SN}^{\rm 2D}(z_i)$ & projected shot noise & $\pmb{\mathcal{N}_r({\rm Tab.~\ref{tab:sample_properties}}, 0.3)}$ & $\pmb{\mathcal{N}_r({\rm Tab.~\ref{tab:sample_properties}}, 0.3)}$ & $\pmb{\mathcal{N}_r({\rm Tab.~\ref{tab:sample_properties}}, 0.3)}$\\
    $\alpha_a(z_i)\,\,[h^{-2}\,{\rm Mpc}^2]$ & counterterm (auto)& $\mathcal{N}(0,50)$ & $\pmb{\mathcal{N}(0,3)}$ & $\pmb{\mathcal{N}(0,3)}$\\
    $\alpha_x(z_i)\,\,[h^{-2}\,{\rm Mpc}^2]$ & counterterm (cross)& Eq.~\eqref{eq:counterterm_relationship} & $\pmb{\mathcal{N}(0,3)}$ & $\pmb{\mathcal{N}(0,3)}$\\
    $\epsilon(z_i)\,\,[h^{-2}\,{\rm Mpc}^2]$ & Eq.~\eqref{eq:counterterm_relationship} & $\pmb{\mathcal{N}(0,2)}$ & $-$ & $-$\\
    \end{tabular}}
    \caption{Sampled parameters and their associated priors for each of our modeling choices. Parameters with argument $z_i$ (for $i=1,2,3,4$) are redshift dependent, e.g. $b_s^L(z_1)$ and $b_s^L(z_2)$ are treated as separate parameters that are individually sampled. With the exception of shot noise and number count slope, the priors for all nuisance parameters are identical in each redshift bin. We center the prior means for $s_\mu(z_i)$ and ${\rm SN}^{\rm 2D}(z_i)$ around those listed in Table~\ref{tab:sample_properties}. $\mathcal{U}(X,Y)$ denotes a uniform prior between $X$ and $Y$, $\mathcal{N}(\mu,\sigma)$ is a Gaussian prior with mean $\mu$ and standard deviation $\sigma$, and $\mathcal{N}_r(\mu,\sigma)\equiv \mathcal{N}(\mu,\sigma\mu)$. Priors are in \textbf{bold font} when the posterior of a nuisance parameter is (approximately) prior-dominated.}
    \label{tab:priors}
\end{table}

Priors on galaxy-induced nuisance parameters vary with the model being used and are summarized in Table~\ref{tab:priors}. In all scenarios we place uniform priors on the linear Lagrangian bias (in each redshift bin) between $0$ and $3$, place a Gaussian prior on shot noise centered around its Poisson value (Table~\ref{tab:sample_properties}) with a (relative) $30\%$ width, and a Gaussian prior on the number count slopes centered around their measured values \cite{Zhou:2023gji} (Table~\ref{tab:sample_properties}) with width 0.1, which is $\simeq10\times$ larger than the Poisson errors estimated by ref. \cite{Zhou:2023gji}.

For our fiducial HEFT model we adopt a uniform prior on $(b_2^L)$ between $-5$ and $5$. 
This prior is largely ``uninformative" in the sense that the data are readily capable of distinguishing e.g. $b_2^L=-5$ from $b_2^L=+5$, resulting in marginal posteriors that are significantly narrower than the prior (see Fig.~\ref{fig:giant_triangle} in Appendix~\ref{sec:big_triangle}). 
We find that $b_s^L$ is highly-degenerate with $b_2^L$ (see Fig.~\ref{fig:anamarg} in Appendix~\ref{sec:anamarg}) and that the best-fit values of $b_s^L$ to mock LRGs (\S\ref{sec:verification_and_volume}) are typically of $\mathcal{O}(0.1)$. 
For these reasons we apply an informative $\mathcal{N}(0,1)$ Gaussian prior on each $b_s^L(z_i)$, and explore the impact of widening this prior in \S\ref{sec:param_based_consistency_tests}.
We adopt a wide Gaussian prior with mean zero and width $50\,\,h^{-2}\,{\rm Mpc}^2$ on the (auto) counterterm $\alpha_a(z_i)$, while the (cross) counterterm $\alpha_x(z_i)$ is determined via Eq.~\eqref{eq:counterterm_relationship} with a Gaussian prior on $\epsilon$ centered at $0$ with width $2\,\,h^{-2}\,{\rm Mpc}^2$. 
The prior on the former is largely uninformative $-$ $\alpha_a = 50 \,\,h^{-2}\,{\rm Mpc}^2$ corresponds to a $\mathcal{O}(1)$ change in $P_{gg}$ at $k=0.2\,h\,{\rm Mpc}^{-1}$ (see Fig.~\ref{fig:anamarg} in Appendix~\ref{sec:anamarg}) $-$ while the latter is an informative prior on the $\alpha_a-\alpha_x$ relationship, where the value of $2\,\,h^{-2}\,{\rm Mpc}^2$ has been chosen to reflect uncertainties arising from \verb|Aemulus| $\nu$'s finite grid resolution and potential baryonic feedback (see the discussion in \S\ref{sec:heft}).

For the linear theory and model independent constraints we set $b_2^L=b_s^L=0$. We vary $\alpha_a$ and $\alpha_x$ independently with informative priors on both parameters to roughly account for the residual systematic error arising from non-linear evolution. Motivated by the discussion in \S\ref{sec:linthy} we adjust our prior such that a $1.5\%$ variation in $P_{gm}$ (or $P_{gg}$) at $k=0.1\,h\,{\rm Mpc}^{-1}$ is allowed within one (prior) sigma, corresponding to a $3 \,\,h^{-2}\,{\rm Mpc}^2$ width.

The priors discussed above are physically plausible (for the LRG sample \cite{Zhou:2020nwq}) while at the same time sufficiently restrictive to mitigate ``volume effects" (see \S\ref{sec:volume}) arising from unconstrained degeneracy directions, which complicate the interpretation of marginal posteriors. In part due to the rise in EFT-like models being applied to cosmological datasets (for which there are dozens of poorly constrained nuisance parameters and hence sizable ``volume effects"), the discussion of priors has recently become an active topic in the literature \cite{Zhang:2021yna,Hadzhiyska:2023wae,Donald-McCann:2023kpx,Euclid:2023tog,Braganca:2023pcp,Maus2024}. Previous work has found that a Jeffreys prior \cite{1946RSPSA.186..453J}, or ``partial" Jeffreys prior \cite{Hadzhiyska:2023wae,Donald-McCann:2023kpx} have been effective in reducing volume effects. Others have suggested imposing a ``perturbativity prior" \cite{Braganca:2023pcp} on higher-order corrections to penalize regions of parameter space for which higher-order corrections are non-perturbative. We choose not to explore these approaches in this work. However we comment on the expected size of volume effects in \S\ref{sec:volume}.

\subsection{Code checks}

As a check of our Limber code we replaced the 3D power spectrum predictions in Eq.~\eqref{eq:limber} with a linear bias model where $P_{mm}(k,z)$ is calculated using \verb|CLASS|' \cite{CLASS} default version of \verb|HaloFit| \cite{Takahashi:2012em}. When integrating over $\chi$ we include the redshift evolution of the power spectrum and set $b(z) = 2$, $s_\mu(z) =1$ and $dN/dz$ to the measured redshift distribution of the third $(z_3)$ LRG photo-$z$ bin. We found that our predicted $C^{gg}_\ell$ and $C^{\kappa g}_\ell$ agreed with \verb|CCL|'s \cite{LSSTDarkEnergyScience:2018yem} \verb|angular_cl| method to within $\simeq0.3\%$, which is of the same order as difference between \verb|CLASS|' default version of \verb|HaloFit| and \verb|CCL|'s default non-linear matter power spectrum. We repeated this exercise with $s_\mu(z)=0.4$ to explicitly test our calculation of the magnification contribution and found similar agreement.

As an additional check, we compared our (fiducial) $C_\ell$'s predicted with the \verb|Aemulus| $\nu$ emulator to those predicted using \verb|velocileptors| \cite{Chen:2020fxs} and found excellent $\mathcal{O}(0.1\%)$ agreement (i.e. of order the emulator error) in the $\ell\to0$ limit (more precisely, $\ell < 100$) for a broad range of nuisance term values, ensuring that our definitions of e.g. $b_2^L$ are consistent with those used in previous works \cite{White:2021yvw}, and that our summation of power spectrum monomials (e.g. Eq.~\ref{eq:Pgg}) has been performed self-consistently.

\subsection{``Volume effect" estimation}
\label{sec:volume}

The high accuracy and flexibility of our fiducial HEFT model comes at the cost of introducing 16 additional nuisance parameters in our fiducial setup: $b^L_2$, $b^L_s$, $\alpha_a$, $\alpha_x$ in each of the four redshift bins. Given the large dimensionality of our posterior (30 parameters when analyzing all four bins) and the highly-compressed nature of our dataset (power spectra of projected LSS tracers), one may naturally worry that our analysis is susceptible to ``volume effects": shifts in the marginalized posteriors of cosmological parameters away from their values at the maximum \textit{a posteriori} (MAP) sourced by poorly-constrained degeneracy directions with other parameters that cover a large volume of parameter space \cite{Hadzhiyska:2023wae,Maus2024}. 

\begin{figure}[!h]
    \centering
    \includegraphics[width=\linewidth,valign=c]{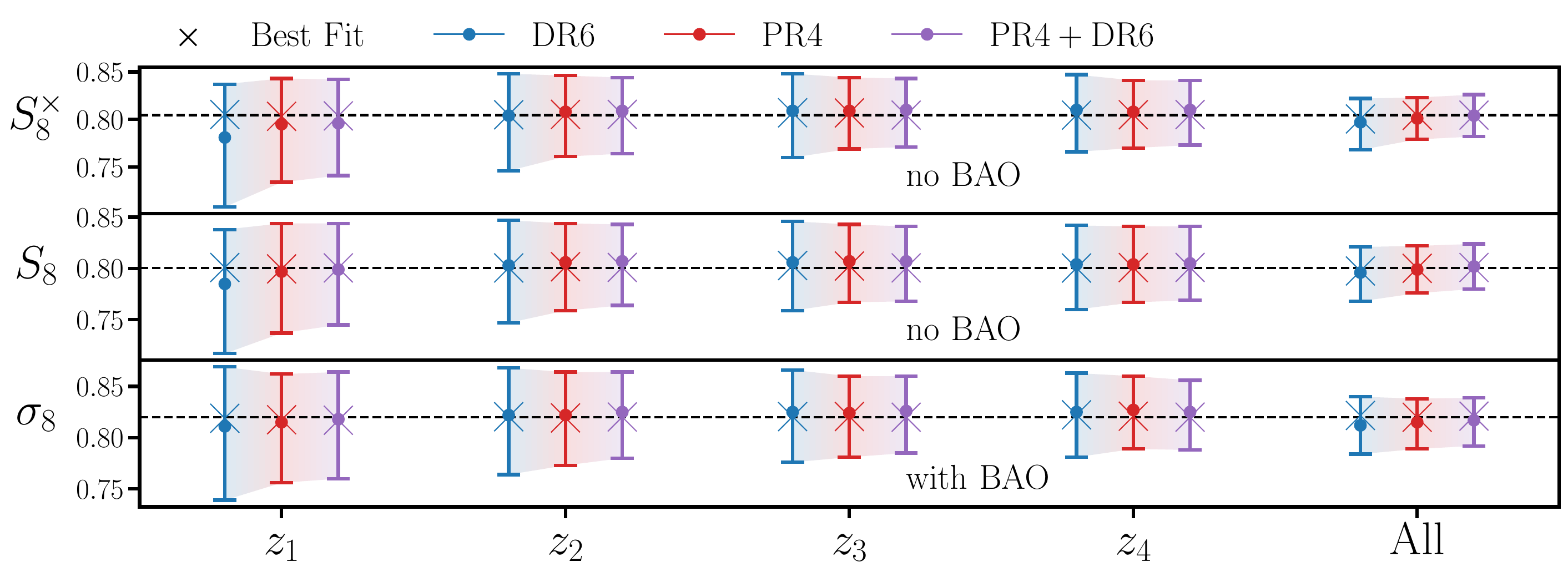}
    \caption{
    We estimate the size of ``volume effects" for our fiducial analysis choices by fitting to a noiseless model prediction assuming a PR4 (blue), DR6 (red) or joint PR4$+$DR6 (purple) covariance. We consider fitting to individual redshift bins ($z_i$ on the $x$-axis) or jointly fitting to all redshift bins (All).
    We show the marginal mean (circles), $\pm 1\sigma$ errors (caps) and best-fit value ($\times$'s) for $S_8^\times\equiv \sigma_8(\Omega_m/0.3)^{0.4}$ (top), $S_8$ (middle) and $\sigma_8$ (bottom). When quoting $\sigma_8$ constraints we include a mock BAO prior (see \S\ref{sec:verification_and_volume}). The horizontal dashed lines indicate the cosmological parameters ($\Omega_m=0.286$ and $\sigma_8=0.82$) used in the model prediction.
    }
\label{fig:model_fit}
\end{figure}

In particular, volume effects were present in the previous PR3 cross-correlation analysis \cite{White:2021yvw} for which the $S_8$ mean (0.725) was roughly $1\sigma$ lower than the best fit value. As discussed in \cite{White:2021yvw}, the primary culprit for these volume effects was a degeneracy with the counterterm $\alpha_x$, which was treated as a separate parameter from $\alpha_a$ and marginalized over with a broad and largely-uninformative prior. Here we adopt a prior relating $\alpha_a$ and $\alpha_x$ (Eq.~\ref{eq:counterterm_relationship}). This prior, along with the extended dynamic range that HEFT affords over ``pure" perturbation theory (increasing $\ell_{\rm max}$ to 600) efficiently mitigate the impact of volume effects on our analysis.

To estimate the size of residual volume effects we fit to a noiseless model prediction. We generate this prediction with our fiducial HEFT model (\S\ref{sec:heft}) using the measured redshift distributions and magnification biases of the LRGs. We adopt the Buzzard cosmology (\S\ref{sec:buzzard_fits}) in our prediction and adjust the remaining nuisance parameters to roughly match the data\footnote{Specifically, we take $b_1^L = (0.872,1.01,1.19,1.27)$, $10\times b_2^L=(-2.37,-1.63,0.553,2.96)$, $10^3\times b_s^L = (3.82,-2.03,-20.7,-22.7)$, $\alpha_a = (-2.79, -3.86, -5.71, -6.45)$, and $10^6\times \text{SN}^{\rm 2D} = (4.01,2.24,2.09,2.32)$ in each redshift bin respectively, and fix $\alpha_x(z) = \alpha_a(z)/2(1+b_1^L(z))$.}. Finally we ``convolve" the model predictions (see \S\ref{sec:pixelization}) with the pixel window function and bin them into bandpowers using the same binning as for the data (\S\ref{sec:scale_cuts}). Fig.~\ref{fig:model_fit} summarizes our results when fitting to the model prediction using our fiducial analysis choices (HEFT with scales cuts and priors listed in Tables \ref{tab:scalecuts} and \ref{tab:priors} respectively). We consider the cases of fitting to each redshift bin individually and jointly fitting to all bins ($x$-axis), and for each scenario consider a DR6, PR4 or joint PR4$+$DR6 covariance (blue, red and purple). In all scenarios, volume effects (difference between the mean and truth) are $<\sigma/3$, while for the ``baseline" scenario of PR4+DR6 with all four redshift bins the volume effects are all $\leq \sigma/10$. 

We additionally checked that our linear theory (\S\ref{sec:linthy}) and our model-independent (\S\ref{sec:modelind}) approaches yield negligible volume effects on cosmological parameters when fitting to a model prediction using a joint PR4+DR6 covariance. This is unsurprising given that these models have far fewer nuisance parameters than our fiducial HEFT model. 

While the tests considered here provide an estimate for the expected ``volume effect" size, they are not exhaustive in the sense that the size (and direction) of these shifts can in principle depend on the noise (or systematic) realization in the data (e.g. \cite{Chintalapati:2024lvi}). To quickly diagnose the presence of these shifts, we always report the location of the best-fit in addition to the mean when quoting marginal posteriors. 

\subsection{Fits to Buzzard mocks}
\label{sec:buzzard_fits}

We assess the accuracy of each of our modeling choices by fitting to mock data constructed from the \verb|Buzzard| simulations \cite{DES:2019jmj,DeRose:2021avs,Wechsler:2021esl}.
We briefly summarize these simulations and the mock LRG sample below, and refer the reader to ref.~\cite{DeRose24} and the references above for a more detailed discussion.

The \verb|Buzzard| DM particle lightcone is constructed from three separate N-body simulations spanning $0<z<0.32$, $0.32<z<0.84$ and $0.84<z<2.35$ respectively.
Particles ($1400^3$ to $2048^3$ particles depending on the redshift bin) are evolved with \verb|Gadget-2| \cite{Springel:2005mi} in a box size ranging from $1.05-4\,h^{-1}\,{\rm Gpc}$, yielding an effective halo mass resolution of $0.5\times10^{13}\,h^{-1}M_\odot$ for $z\leq 0.32$ and $10^{13}\,h^{-1} M_\odot$ for $z\lesssim 2$.
Galaxy positions, velocities, rest-frame magnitudes and SEDs are assigned with the \verb|Addgals| algorithm \cite{Wechsler:2021esl}.
The magnitudes and SEDs are then passed through DECam-like bandpasses that closely match those applied to the data, with slight modifications to better match the observed projected number density.
Photometric redshifts are assigned to each galaxy by adding a Gaussian photometric error $\sigma_z(z)$ to the true redshift $z$, where $\sigma_z(z)$ has been calibrated with the photo-$z$ errors measured by ref.~\cite{Zhou:2023gji}.
Galaxies are then assigned to photometric redshift bins defined similarly to those presented in ref.~\cite{Zhou:2023gji}.
The CMB lensing convergence map and magnification contributions are constructed within the Born approximation from the matter distribution in the simulation.

The galaxy auto- and cross-correlations with CMB lensing are binned into the same bandpowers discussed in \S\ref{sec:scale_cuts} using an independent pipeline rather than the one detailed detailed in the companion paper \cite{Kim2024} and \S\ref{sec:cl_estimation}. We adopt the same scale cuts as in Table~\ref{tab:scalecuts} and adjust the central values of the priors on the number count slope and shot noise to the measured values in the simulations\footnote{$s_\mu(z_i) = 1.04, 0.97, 0.81, 0.80$ and $10^6\,{\rm SN}^{\rm 2D} = 3.7786, 1.7858, 1.9084, 2.5209$}. Otherwise the priors are identical to those listed in Table~\ref{tab:priors} for each modeling choice. We adjust the fiducial values of $n_s=0.96$, $\Omega_bh^2=0.02254$, $\Omega_mh^3=0.098098$ and $\sum m_\nu=0$ to those used in the Buzzard simulations. We use the measured redshift distributions of the simulated LRGs when fitting to the mock measurements.

\begin{figure}[!h]
    \centering
    \includegraphics[width=0.74\linewidth,valign=b]{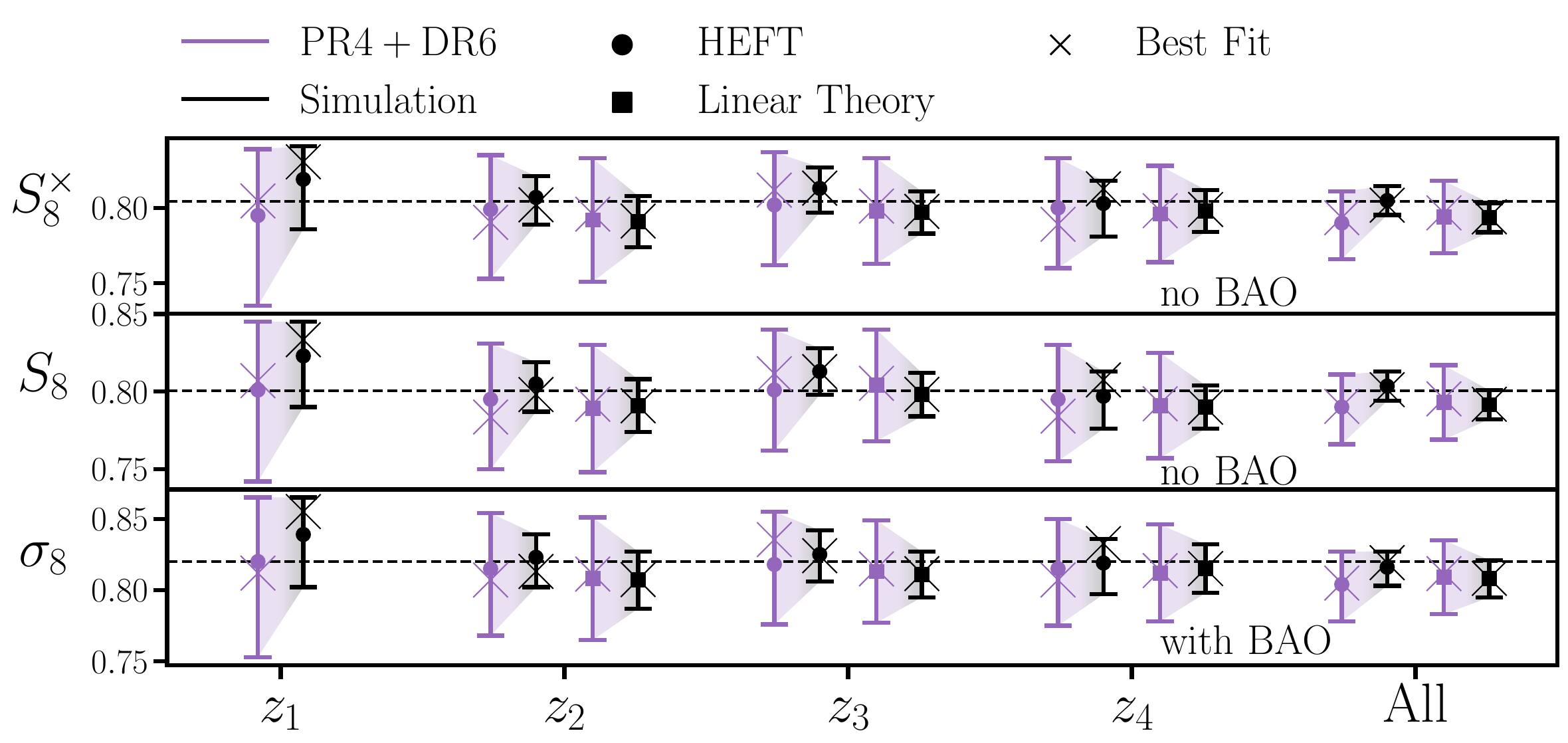}
    \includegraphics[width=0.24\linewidth,valign=b]{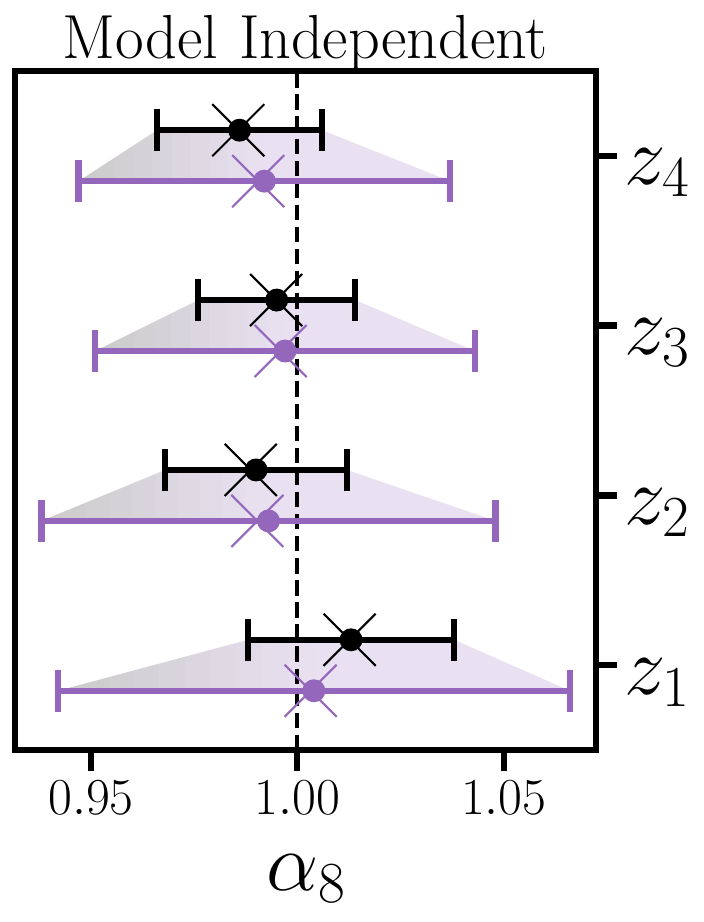}
    \caption{
    Cosmological parameter recovery when fitting to mock galaxy and CMB lensing measurements from the Buzzard simulations. 
    In the left plot we show $S^\times_8\equiv\sigma_8(\Omega_m/0.3)^{0.4}$ (top), $S_8$ (middle) and $\sigma_8$ (bottom) constraints for HEFT (circles) and linear theory (squares). 
    In purple we adopt a PR4+DR6 covariance, while in black we use a covariance representative of the Buzzard simulation.
    We include a mock BAO prior when quoting $\sigma_8$ constraints (see \S\ref{sec:verification_and_volume}). In the right panel we show the model-independent measurements of $\alpha_8$ (see Eq.~\ref{eq:alpha8}). 
    The location of the best fit is indicated with a $\times$.
    Dashed lines indicate the values ($\Omega_m = 0.286$, $\sigma_8=0.82$, $\alpha_8=1$) of the Buzzard cosmology. 
    }
\label{fig:buzzard}
\end{figure}

The mock power spectra are measured from seven quarter-sky cutouts, corresponding to $\simeq4\times$ the sky coverage of the cross-correlation with $\textit{Planck}$ PR4. The mock CMB lensing maps contain no lensing reconstruction noise. We analytically estimate the covariance of these mock measurement using the methods discussed in \S\ref{sec:cl_estimation}, where the power spectrum of the mock CMB lensing map is taken to be $C^{\kappa\kappa}_\ell$ without reconstruction noise and the ``mask" is taken to be the full sky. We multiply the resulting covariance matrix by $4/7$ to account for the Buzzard sky-coverage.

Our results are summarized in Fig.~\ref{fig:buzzard}, where we show the posteriors of $S^\times_8 \equiv \sigma_8 (\Omega_m/0.3)^{0.4}$ (top), $S_8$ (middle) and $\sigma_8$ (bottom) for our fiducial HEFT (errorbars with circles) and linear theory (errorbars with squares) models, while for the model-independent constraints (right panel) we show $\alpha_8$ constraints (Eq.~\ref{eq:alpha8}). For both HEFT and linear theory we consider fitting to each redshift bin individually or jointly fitting to all bins ($x$-axis), while for the model independent approach we only consider the former ($y$-axis). Since we do not consider fitting to $z_1$ with linear theory in \S\ref{sec:results} we correspondingly choose not to plot $z_1$ linear theory constraints in Fig.~\ref{fig:buzzard}. The true values of cosmological parameters are indicated by the thin black dashed lines.

In black we show the constraints when adopting a simulation-like covariance for which $\sim 1\sigma$ level fluctuations in the posteriors are expected, which is consistent with the observed scatter. We show the constraints for a joint PR4+DR6 covariance (still using the same mock measurements) in purple, for which we naively expect $\sim0.5\sigma$ fluctuations, in addition to (small, see \S\ref{sec:volume}) volume shifts. This is again consistent with the observed scatter in the mock fits. In particular, when jointly fitting to all four redshift bins using our fiducial HEFT model with a joint PR4+DR6 covariance our recovered cosmological parameters are within $0.7\sigma$ of their true values. We conclude that each of the considered models (with their fiducial scale cuts and priors) show no evidence of bias, and no signs of significant volume effects. 

\section{LRG systematics tests}
\label{sec:systematics}

In addition to the null tests presented in the companion paper \cite{Kim2024}, which primarily (but not exclusively) test for systematics in the ACT DR6 CMB lensing map, here we present an additional set of systematics tests for the LRG auto- and cross-correlation with the baseline \textit{Planck} PR4 $\kappa$ map. 

A significant challenge for current and future large-scale structure surveys is to ensure the uniformity of a galaxy sample's physical properties on different regions of the sky. The selection criterion and systematics weights used for our LRG samples \cite{Zhou:2020nwq} were primarily tuned to correct for correlations in LRG density $\bar{n}(\hat{\bm{n}})$ with a set of systematic templates. 
While enforcing a uniform $\bar{n}(\hat{\bm{n}})$ is a necessary condition, this does not necessarily ensure the uniformity of other physical properties, e.g. redshift distributions \cite{BaleatoLizancos:2023zpl} or linear bias, which have the potential to bias cosmological inference if these effects are large and not properly taken into account. 
In sections \ref{sec:imaging_footprint_variations}, \ref{sec:north_vs_south}, \ref{sec:dec_pm15},
and \ref{sec:strict_ebv_star} we quantify these effects by examining variations in the LRG auto- and cross-correlation with $\textit{Planck}$ PR4 on different footprints\footnote{Some of the footprints considered here (e.g. the Northern imaging region) have either minimal or no overlap with the ACT lensing footprint, which is why we only choose to examine variations in the PR4 cross-correlation. The robustness of the DR6 cross-correlation to additional footprint variations is explored in the companion paper \cite{Kim2024}.}. In \S\ref{sec:sysweights} we quantify the impact of systematic weights on our measurements.
Finally in \S\ref{sec:gxspec} we examine the consistency of the measured galaxy cross-spectra $C^{g_i g_j}_\ell$ with those predicted from the galaxy auto- and cross-correlation with CMB lensing alone.
We find significant evidence for variations in the LRGs' physical properties on different footprints (\S\ref{sec:imaging_footprint_variations}), which we expect to be our leading source of systematic error. In \S\ref{sec:param_based_consistency_tests} we estimate the impact of these variations on cosmological parameters, finding at most $\simeq0.2$ shifts to our fiducial $S_8$ constraints (some of which is statistical).

\subsection{Imaging footprints}
\label{sec:imaging_footprint_variations}

We first consider variations with the different imaging footprints (North, DECaLS and DES), which are shown in Fig.~\ref{fig:dNdz_variations}. The North mask is defined by the intersection of ${\rm DEC}>32.375\degree$ and the NGC. The (binary) DES mask is defined to be non-zero where there is a positive detection fraction in any of the DES DR2 $grizY$ bands. The DECaLS mask is defined to be non-zero everywhere except North and DES, with an additional cut on declination ${\rm DEC}>-15\degree$. Finally we multiply each imaging region mask by the fiducial LRG mask, producing binary masks for each region that are subsets of the full LRG footprint.

We (re)mask the LRG samples with each of the imaging region masks and measure their auto- and cross-correlation with PR4 using the methods discussed in \S\ref{sec:cl_estimation}. We approximate the covariance of these measurements using the analytic methods described in \S\ref{sec:cl_estimation}, from which we compute the variance of the difference: ${\rm Var}[X-Y] = {\rm Var}[X] + {\rm Var}[Y] - 2{\rm Cov}[X,Y]$. 

\begin{figure}[!h]
    \centering
    \includegraphics[width=\linewidth,valign=c]{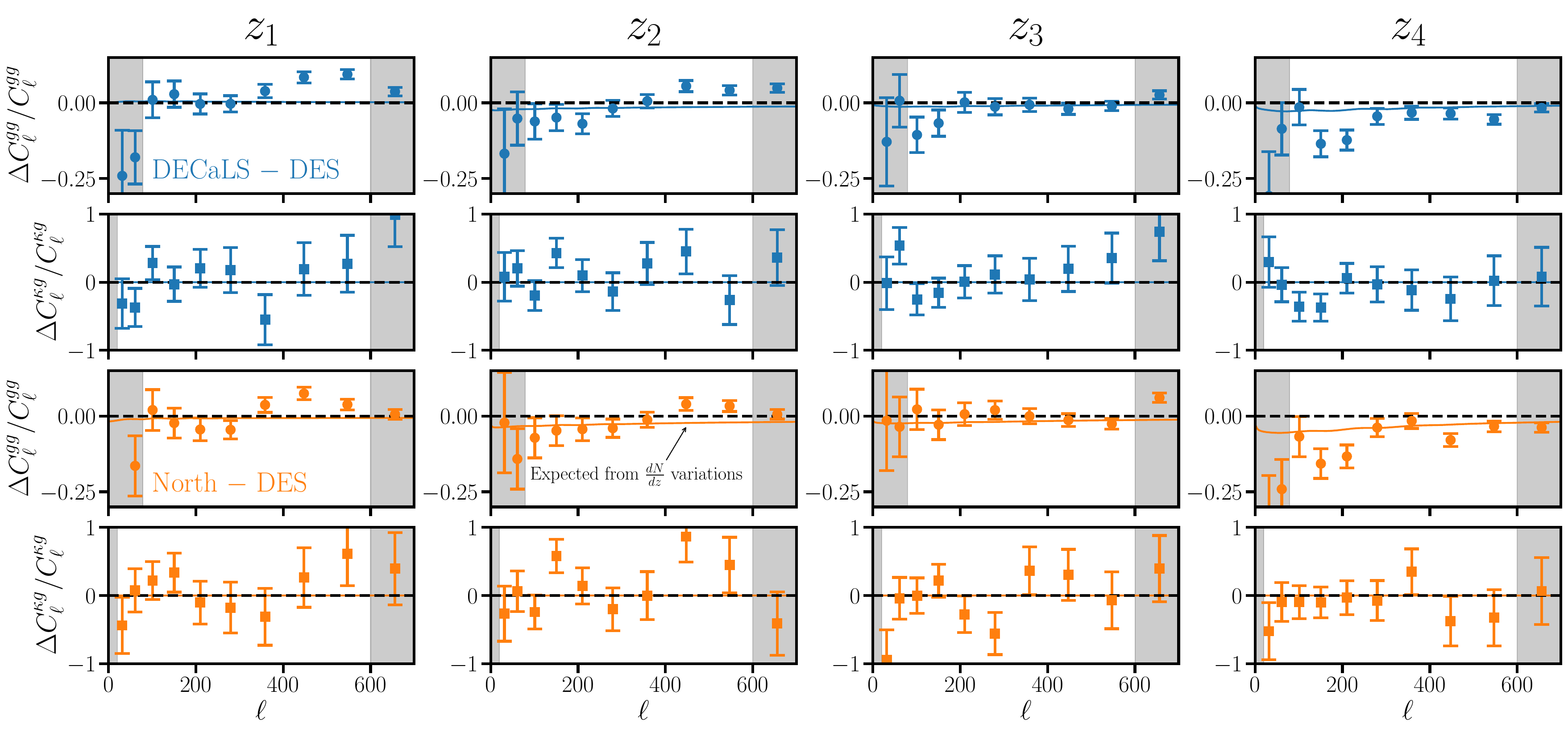}
    \caption{ 
    The relative variation in LRG power spectra (first and third rows) and their cross-correlation with PR4 (second and fourth rows) across different imaging footprints. 
    We divide by the (binned) $C_\ell$'s used for covariance estimation (see \S\ref{sec:cl_estimation}).
    Columns indicate the redshift bin. 
    In blue we show the difference between DECaLS and DES, while in orange we show the difference between North and DES. 
    The gray regions are excluded from the fitting range. 
    The PTEs for the cross-correlation measurements are sensible: 0.58, 0.44, 0.62, 0.55 for DECaLS$-$DES and 0.73, 0.11, 0.25, 0.84 for North$-$DES.
    The PTEs for the auto-correlation measurements are: $5.1\times10^{-11}$, $1.1\times10^{-3}$, 0.33, $5.1\times10^{-8}$ for DECaLS$-$DES and $1.1\times10^{-3}$, 0.075, 0.85 and $9.7\times10^{-8}$ for North$-$DES.
    The small PTEs indicate significant variations in the LRG sample's statistical properties on different imaging footprints.
    The solid (blue and orange) lines show the expected variation if the galaxies were drawn from the same 3D statistical distribution (taken to be the fiducial best fit to PR4+DR6) but projected to 2D using the different redshift distributions shown in Fig.~\ref{fig:dNdz_variations}.
    } 
\label{fig:imaging_footprint_variations}
\end{figure}

Our results are summarized in Fig.~\ref{fig:imaging_footprint_variations}, where we show the relative variation in $C^{gg}_\ell$ (first and third row) and $C^{\kappa g}_\ell$ (second and fourth row). In blue we show the relative difference between DECaLS and DES, while in orange we plot the difference between North and DES. 
We find that $C^{gg}_\ell$ is consistent across the DECaLS and North regions over our fiducial analysis range. However, we find significant variations between the DES footprint and elsewhere for all redshift bins except $z_3$ (i.e. the blue and orange points differ from zero in the same way). For $z_1$ this difference is primarily relevant on small scales and is thus likely due to mismatched shot noise. For $z_2$ and $z_4$ this difference is also relevant on large scales $(\ell\simeq150)$. For comparison we show the expected variation from changes in the redshift distribution alone (solid lines), which are not large enough to explain the observed discrepancies (especially for $z_4$). 
We hypothesize that the LRGs in the DES region have a slightly larger linear bias. 
This could potentially be the byproduct of the DES region's deeper $z-$band depth, which $z_4$ is most sensitive to. 
This is reflected in the systematics weight maps, where the DES footprint is clearly visible (see Fig.~8 of \cite{Zhou:2023gji}).
Lower photometric noise in the DES region would make it less likely for fainter objects to scatter into the sample, resulting in a net increase in the DES region's linear bias. 
Alternatively, it may be that the redshift distribution in the southern half of the DES footprint $({\rm DEC} < -15\degree)$ differs from that on the footprint where we have direct spectroscopic calibration of the redshift distribution (see \S\ref{sec:dec_pm15}).
We have repeated this exercise when additionally applying a stricter $E(B-V)\leq 0.05$ and stellar-density ($<1500$ per square degree) cuts and found qualitatively similar results, suggesting that this behavior is not sourced by Galactic contamination. 

We note, however, that the variations in the cross-correlation with \textit{Planck} PR4 are consistent with fluctuations (see the caption of Fig.~\ref{fig:imaging_footprint_variations} for the PTEs) but note that the SNR of the cross-correlation is low relative to the galaxy auto. 
Moreover, since our constraint on $S_8$ is not proportional to $C^{gg}_\ell$ but instead depends essentially on the ratio $C^{\kappa g}_\ell/\sqrt{C^{gg}_\ell}$,  biases to $S_8$ (or $\sigma_8$) may be significantly smaller than those impacting $C^{gg}_\ell$.
For example, if a LSS tracer on two disjoint regions of the sky (with sky coverage $f_{\rm sky, 1}$ and $f_{\rm sky, 2}$) has linear bias $b_1$ and $b_2$ respectively on each region, then the galaxy-auto spectrum is sensitive to the effective bias $b^2_{\rm eff, a} = \sum_i f_{\rm sky, i} b_i^2/\sum_i f_{\rm sky, i}$ while the cross-correlation with matter (CMB lensing) probes $b_{\rm eff, x} = \sum_i f_{\rm sky, i} b_i/\sum_i f_{\rm sky, i}$. In the limit that the bias variation is small, say $b_2 = b_1(1+\epsilon)$, the ratio $b_{\rm eff, x}/\sqrt{b_{\rm eff, a}^2} = 1 + \mathcal{O}(\epsilon^2)$. Thus the expected bias to $S_8$ is of order $\epsilon^2$. In Fig.~\ref{fig:imaging_footprint_variations} we see $\sim 10\%$ changes to $C^{gg}$ between North and DES for $z_4$, suggesting a $\sim 5\%$ change in the linear bias, for which the corresponding bias to $S_8$ is a quarter of a percent which is significantly smaller than our statistical errorbars (roughly $2.5\%$, see \S\ref{sec:results}). 

We explore the impact of these variations at the parameter-level in \S\ref{sec:param_based_consistency_tests} where we treat the North, DECaLS and DES regions as individual samples with their own corresponding nuisance parameters, and using the redshift distributions calibrated for each individual footprint. This prescription mitigates bias to $S_8$ from variations in linear bias or redshift distributions across footprints\footnote{This isn't necessarily true for variations (in linear bias or redshift distribution) within each footprint \cite{BaleatoLizancos:2023zpl}. We expect these to be negligibly small for our analysis.}. We find consistent results when treating the entire LRG footprint as a single sample as we do when treating each imaging region as its own sample. For simplicity we adopt the former as the fiducial scenario in \S\ref{sec:results}. 

\subsection{Northern and Southern galactic caps}
\label{sec:north_vs_south}

\begin{figure}[!h]
    \centering
    \includegraphics[width=\linewidth]{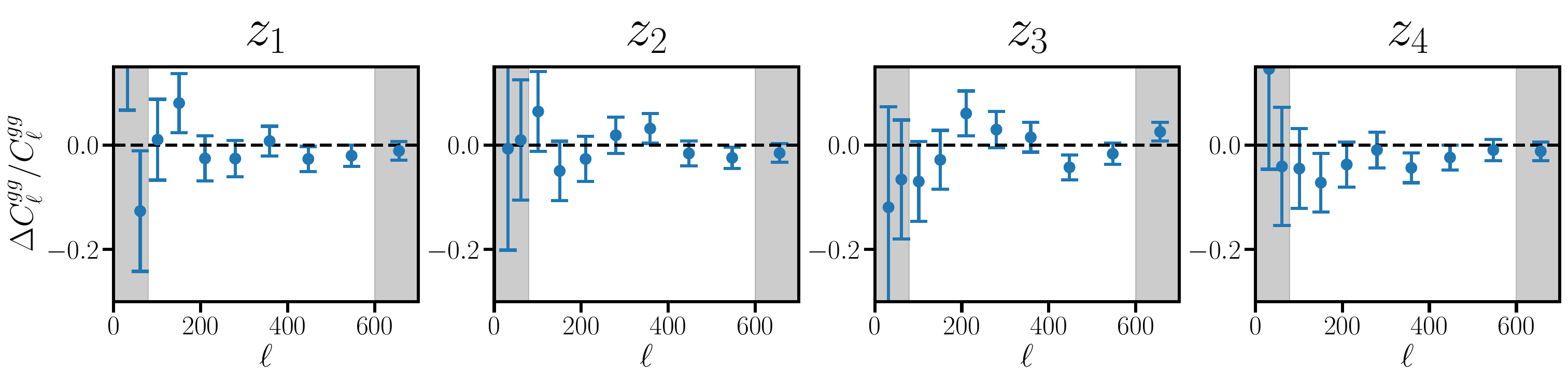}
    \caption{
    The relative difference in LRG clustering between the NGC and SGC regions of the DECaLS footprint. 
    We divide by the (binned) $C_\ell$'s used for covariance estimation (see \S\ref{sec:cl_estimation}).
    The gray regions indicate our fitting range.
    The PTEs are 0.62, 0.63, 0.32 and 0.43 respectively.
    } 
\label{fig:ngc_vs_sgc}
\end{figure}

Here we test for variations in the LRG clustering in the Northern and Southern galactic hemispheres. Since we have already established (\S\ref{sec:imaging_footprint_variations}) that there are significant deviations in $C^{gg}_\ell$ across different imaging regions (i.e. DES vs elsewhere), a strict comparison of the intersection of our full LRG mask with the NGC and SGC would yield qualitatively similar results to \S\ref{sec:imaging_footprint_variations}. To isolate the impact of NGC vs SGC, we instead choose to compare the LRG clustering on the northern and southern regions of the DECaLS footprint alone, which we show in Fig.~\ref{fig:ngc_vs_sgc}. The PTEs (listed in the caption) of these differences are all reasonable, suggesting that variations across different galactic caps are consistent with fluctuations.
The same is true for the cross-correlation with PR4 (and DR6, see the companion paper \cite{Kim2024}), which we have opted not to plot in Fig.~\ref{fig:ngc_vs_sgc}. 

\subsection{Above and below ${\rm DEC}=-15\degree$}
\label{sec:dec_pm15}

\begin{figure}[!h]
    \centering
    \includegraphics[width=\linewidth]{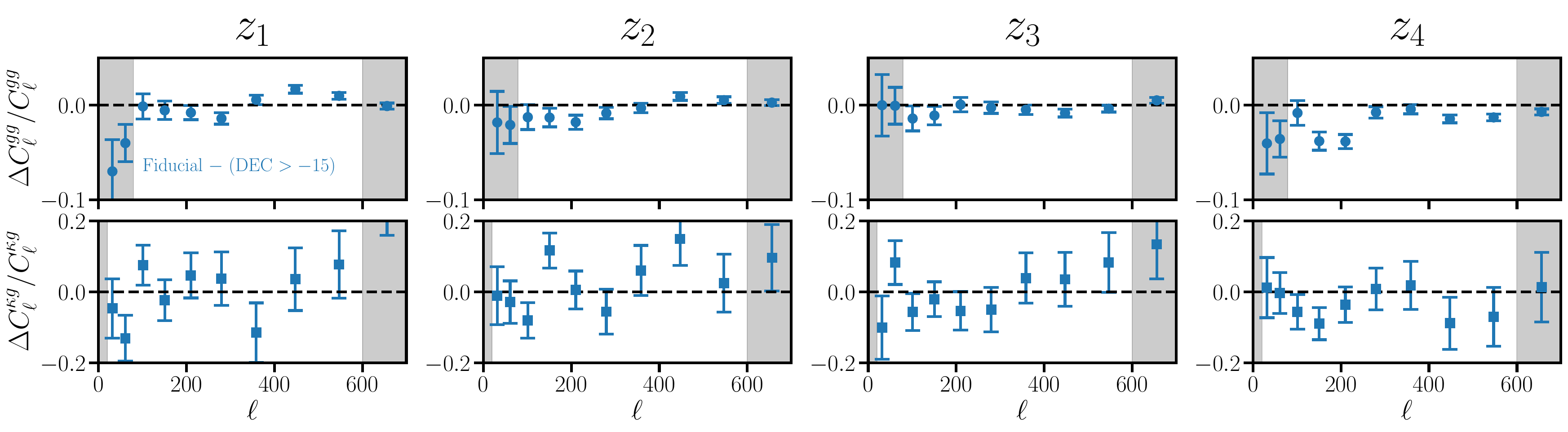}
    \caption{
    The relative difference in our fiducial LRG power spectra (first row) and their cross-correlation with PR4 (second row) with those measured with an additional ${\rm DEC}>-15\degree$ mask applied. 
    We divide by the (binned) $C_\ell$'s used for covariance estimation (see \S\ref{sec:cl_estimation}).
    The gray regions indicate our fitting range.
    The PTEs for the cross-correlations are 0.36, 0.13, 0.59 and 0.53 respectively. 
    With the exception of $z_3$, the PTEs for the auto-correlations are clearly inconsistent with fluctuations: $2.7\times10^{-5}$, $9.2\times10^{-5}$, 0.24 and $2.5\times10^{-13}$ respectively.
    } 
\label{fig:DECpm15}
\end{figure}

The LRG redshift distribution has been directly calibrated with DESI spectroscopy on a subset of the full (imaging) footprint that approximately corresponds to ${\rm DEC}>-15\degree$ (see e.g. Fig.~10 of \cite{Zhou:2023gji}). 
When treating the full LRG footprint as a single sample we implicitly assume that the redshift distribution below ${\rm DEC}<-15\degree$ (or other physical properties of the sample) is consistent with that measured on the calibrated subset. Here we empirically test for variations above and below ${\rm DEC} = -15\degree$ by comparing our fiducial auto- and PR4 cross-correlation measurements with those obtained when additionally masking pixels with ${\rm DEC} < -15\degree$.
These results are summarized in Fig.~\ref{fig:DECpm15}. As in \S\ref{sec:imaging_footprint_variations} we find that the variations in the PR4 cross-correlation are consistent with statistical fluctuations (see the caption of Fig.~\ref{fig:DECpm15} for PTEs). The variations to the galaxy auto-correlation are statistically significant, however the (relative) magnitude of these variations are milder than found in \S\ref{sec:imaging_footprint_variations} (at most $5\%$). We explore the impact of these variations at the parameter-level in \S\ref{sec:param_based_consistency_tests} and find statistically consistent results with and without masking the ${\rm DEC} < -15^\degree$ region.

\subsection{Stricter extinction and stellar density cuts}
\label{sec:strict_ebv_star}

\begin{figure}[!h]
    \centering
    \includegraphics[width=\linewidth]{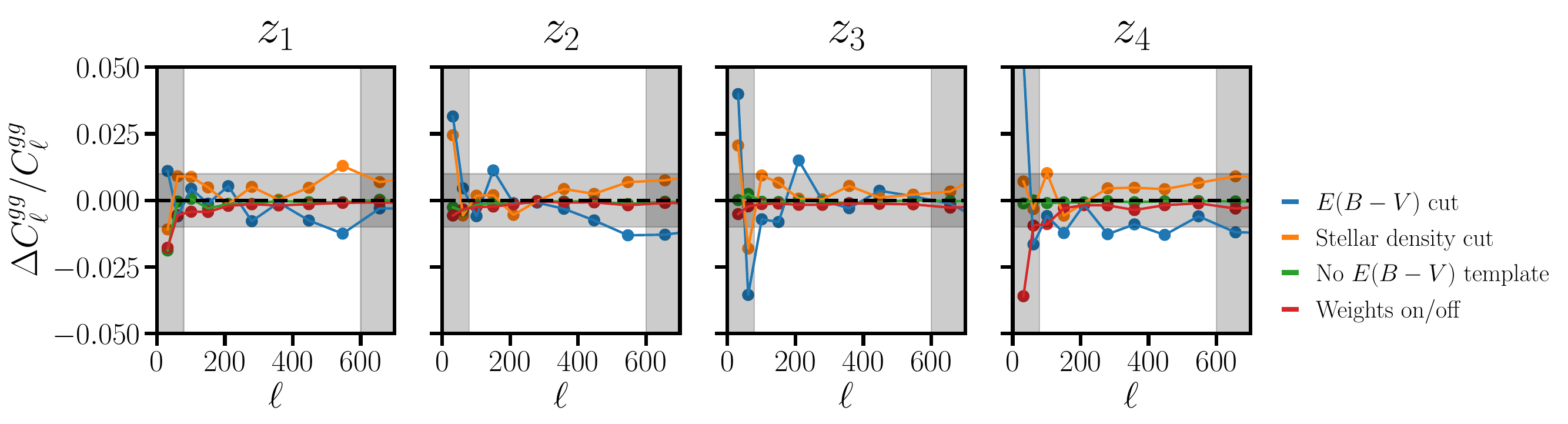}
    \caption{
    The relative change in the LRG auto spectra when adopting a stricter $E(B-V)$ cut (from 0.15 to 0.05; blue), adopting a stricter stellar density cut (2500 to 1500 deg$^{-2}$; orange), removing the SFD $E(B-V)$ map \cite{SFD} as a template when constructing systematics weights (green), and neglecting the systematics weights altogether (red). 
    Different panels indicate the redshift bin.
    The gray vertical regions indicate our fitting range, while the gray horizontal band indicates the interval $[-0.01,0.01]$.
    } 
\label{fig:cgg_variations_compilation}
\end{figure}

To check for potential signs of galactic contamination we consider making stricter cuts on extinction and stellar density. These results are shown in Fig.~\ref{fig:cgg_variations_compilation}. In blue we show the relative variation in $C^{gg}_\ell$ when we additionally mask regions with $E(B-V)\geq0.05$ using the $E(B-V)$ map from \cite{SFD}. In orange we show the variations when masking regions where the stellar density exceeds 1500 deg$^{-2}$ using the map from ref.~\cite{Myers:2022azg}. In both scenarios we find that the galaxy power spectra vary by $\simeq1\%$ on the scales of interest. We do not consider estimating the covariance of the measurements here as they are highly correlated, making our analytic estimation to the covariance matrix considerably less accurate. Regardless of the PTEs associated with these variations, their impact on our cosmological results would be negligible. 

\subsection{Impact of systematic weights}
\label{sec:sysweights}

The systematics weights published by ref.~\cite{Zhou:2023gji} are designed to remove spatial trends in the mean galaxy density with a set of seven templates: the SFD $E(B-V)$ map \cite{SFD} in addition to seeing and depth in each of the three optical $(g,r,z)$ bands. 
Here we quantify the impact of these weights on the galaxy auto-spectra, and in addition quantify the expected systematic error arising from residual observational trends. 

Ref. \cite{Zhou:2023gji} provides weights that only use seeing and depth as templates. We show the variations in the galaxy power spectra with and without including $E(B-V)$ as a systematics template in Fig.~\ref{fig:cgg_variations_compilation} (green points), while the red points in Fig.~\ref{fig:cgg_variations_compilation} show the impact of neglecting the systematics weights completely. These variations are at most a percent on the scales of interest, thus even modest (e.g. $\sim10\%$) errors in the systematics weight map would be expected to only yield $\sim0.1\%$ changes to our fiducial power spectra.
As the power spectra and masks of these sets of maps are nearly identical, it is unclear how one would estimate the covariance between them (even numerically). For this reason we do not attempt to assign errorbars to this plot. Regardless of the statistical significance of these deviations, they are small enough that they would have no significant impact on our results. 

\begin{figure}[!h]
    \centering
    \includegraphics[width=\linewidth]{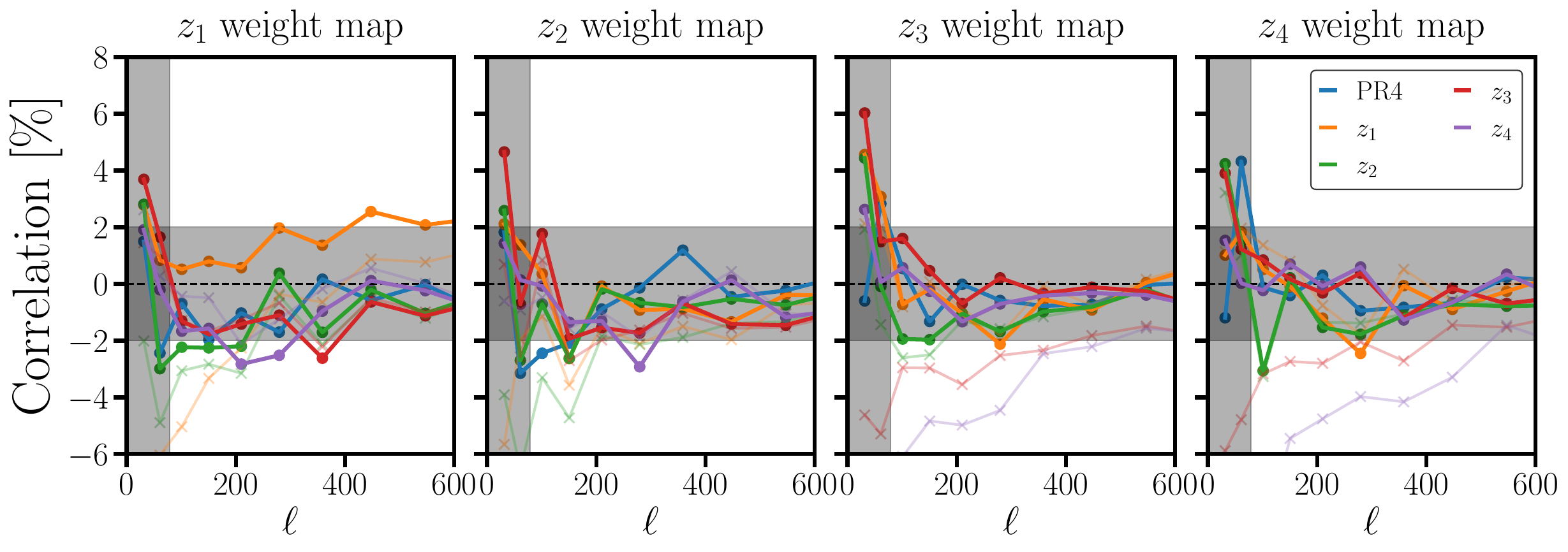}
    \caption{ 
    The correlation $(\equiv C^{ab}_\ell / \sqrt{C^{aa}_\ell C^{bb}_\ell})$ between the fiducial systematics weight maps (panels) and galaxy density contrast maps (colored lines and filled circles) after the systematics weights have been applied.
    We additionally show the correlations between the weight maps and \textit{Planck} PR4 convergence map (blue).
    The correlations between the weight maps and galaxy maps without any systematics weights applied are shown by the pale lines with $\times$ markers. 
    The gray vertical region indicates our fitting range for the galaxy auto-spectra. 
    The horizontal gray band corresponds to the interval $[-0.02,0.02]$, which is of the same order as the observed point-to-point scatter in the measured correlations.
    } 
\label{fig:corr_sysweights}
\end{figure}

In the ideal limit that systematics weight maps perfectly remove observational trends in the mean galaxy density, correlations between the weight maps and the (systematics-corrected) galaxy density contrast maps should vanish. 
We plot correlations between the systematic weight maps (panels) and the galaxy samples (colors) in Fig.~\ref{fig:corr_sysweights}.
On the scales relevant for our analysis we find $\simeq 1-2\%$ residual correlations between the weight maps and galaxy samples. 
In Fig.~\ref{fig:corr_sysweights} we also show the correlations with the galaxy samples before the systematics weight were applied (pale lines with $\times$'s).
On large scales ($100 \lesssim \ell \lesssim 200$) we find that applying systematics weights reduces the correlation between the galaxy samples and their respective weights maps by $\sim3-6\%$. 
As discussed previously (Fig.~\ref{fig:cgg_variations_compilation}), this $\sim3-6\%$ reduction in correlation results in subpercent changes to the galaxy auto-spectra.
Extrapolating this trend, we expect the residual $\simeq 1-2\%$ correlations with the weight maps to bias our galaxy auto-spectra by less than a third of a percent on the scales of interest.

\subsection{Galaxy cross-spectra}
\label{sec:gxspec}

As a consistency check we compare the measured galaxy cross-spectra to predictions from our fiducial model, fitted from the galaxy auto-spectra and cross-correlations with CMB lensing.
The galaxy cross-spectra could in principle be used to better constrain number count slopes and improve the degeneracy-breaking between cosmological and bias parameters.
However, given the heightened sensitivity of these cross-spectra to potential mischaracterization of the tails in the redshift distributions, and that the magnitude of the systematic contamination is expected to be similar to that in the galaxy auto-spectra while the cosmological signal is $\simeq10\times$ smaller, we choose not to fit to the galaxy cross-spectra in \S\ref{sec:results}. 

\begin{figure}[!h]
    \centering
    \includegraphics[width=\linewidth]{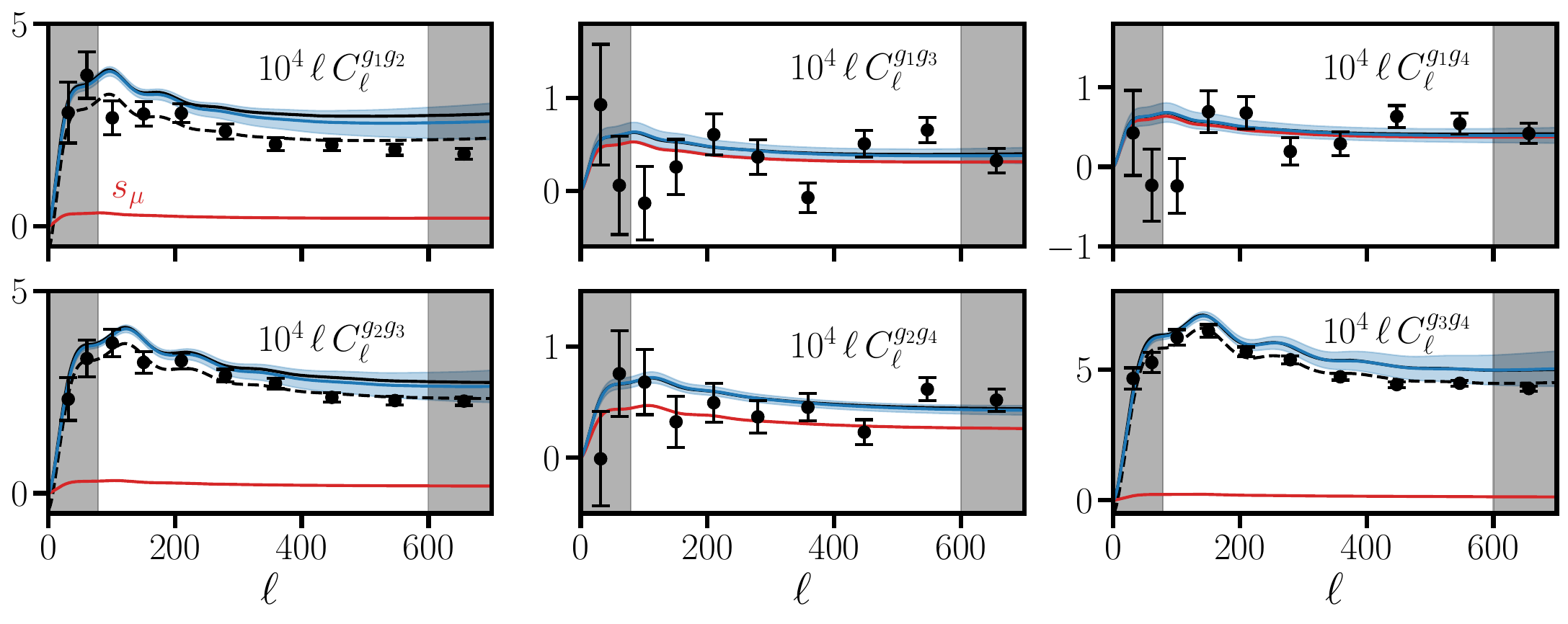}
    \caption{ 
    The measured galaxy cross-spectra (circles with errorbars). Black lines indicate the predicted (Eq.~\ref{eq:cgigj}) cross-spectra from the best-fit to our fiducial PR4+DR6 analysis (\S\ref{sec:pr4_dr6_results}), neglecting correlated shot noise. 
    We also show the mean (blue line) and $\pm1\sigma$ interval (shaded blue) associated with the PPD of that same fit (again neglecting shot noise).
    Note that in some panels (e.g. $C^{g_3g_4}_\ell$) the mean and best-fit curves overlap.
    By contrast, the cross-correlations between neighboring bins are all low compared to the prediction. This effect is not yet fully understood, but we subtract a $1/\ell$ contribution from the best fit prediction to obtain the dashed black curves and then check the impact of marginalizing over a bias of that form in \S\ref{sec:param_based_consistency_tests}.
    In red we show the magnification contribution to the best fit, i.e. the last two terms in Eq.~\eqref{eq:cgigj}.
    } 
\label{fig:cgigj}
\end{figure}

We show the measured galaxy cross-spectra in Fig.~\ref{fig:cgigj}. The errorbars of these measurements are estimated following \S\ref{sec:cl_estimation}.
We note that there is negligible redshift overlap between the lowest and highest redshift bins, however $C^{g_1 g_4}_\ell$ is non-zero (primarily) due to correlations of $z_1$ galaxies with the matter magnifying the $z_4$ sample. 
We detect a non-zero value of $C^{g_1 g_4}_\ell$, and hence a detection of magnification, with a statistical significance of $8\sigma$ on the fiducial scale range ($79 \leq \ell < 600$) adopted for our fits to the galaxy auto-spectra.

We predict the cross-spectra from fits to the galaxy auto- and cross-correlation with CMB lensing alone as follows.
Assuming that the galaxies in the four photometric redshift bins can be thought of as subsamples of a single smoothly evolving galaxy sample, the cross-correlation between two photometric redshift bins is given by
\begin{equation}
\label{eq:cgigj}
    C^{g_ig_j}_\ell 
    = 
    \int \frac{d\chi}{\chi^2} 
    \bigg[
    W^{g_i}W^{g_j}P_{gg}
    +
    \big(W^{g_i} W^{\mu_j}+W^{g_j}W^{\mu_i}\big) P_{gm}
    +
    W^{\mu_i} W^{\mu_j} P_{mm}
    \bigg]
\end{equation}
within the Limber approximation\footnote{Using Eq.~\eqref{eq:rsd_limber} we estimate that beyond-Limber corrections to the galaxy cross-spectra are $\leq 2\%$ for $\ell > 79$.}, where we have suppressed the explicit dependence of projection kernels on comoving distance and power spectra on wavenumber and redshift for simplicity. 
The projection kernels $W^{g_i}$ and $W^{\mu_i}$ are given by Eq.~\eqref{eq:kernels} where $\phi$ is replaced with the redshift distribution of the appropriate sample. 
Given a set of nuisance parameters (for each effective redshift) we can construct a plausible evolution for $b_1^L(z),\,b_2^L(z),\,b_s^L(z),\,\alpha_a(z),\,s_\mu(z),$ and $\epsilon(z)$ by interpolating their values at each effective redshift\footnote{We use scipy's interp1d with kind=`linear' and its default extrapolation for $z<0.47$ and $z>0.91$.}. Along with a set of cosmological parameters, this interpolation then determines the non-stochastic contribution to $C^{g_i g_j}_\ell$. 

Using the best-fit parameters to our baseline PR4+DR6 analysis (\S\ref{sec:pr4_dr6_results}) we predict $C^{g_i g_j}_\ell$ following the recipe above, which are shown as solid black lines in Fig.~\ref{fig:cgigj}. We also show the mean (blue line) and $\pm1\sigma$ (shaded blue) interval associated with the posterior predictive distribution (PPD) of the same baseline fit. This is obtained from 16384 draws of the posterior. We select 256 random samples from the chain, and generate 64 Monte-Carlo realizations of the linear parameters for each sample following the methods discussed in Appendix \ref{sec:anamarg}. 
We find that the measured $C^{g_1 g_2}_\ell$ is $\sim 20\%$ smaller than predicted. We find smaller deficits for the remaining two cross-spectra of neighboring redshift bins: approximately $12\%$ and $8\%$ for $C^{g_2 g_3}_\ell$ and $C^{g_3 g_4}_\ell$ respectively.
The remaining cross-spectra, whose signals all receive large (or are dominated by) magnification contributions, are all within reasonable agreement with the data\footnote{There is a single point in $C^{g_1 g_3}$ at $\ell\simeq350$ that fluctuated $\simeq 2.8\sigma$ lower than the prediction.}. 
As we discuss below and in \S\ref{sec:param_based_consistency_tests}, even if these deficits are the result of systematic contamination in the data we expect these contaminants to have a negligible impact on our cosmological constraints.

There are several important subtleties to bear in mind when comparing these predictions to the measured cross-spectra.
First, in Eq.~\eqref{eq:cgigj} we assume that the physical properties of the galaxy samples depend only on redshift, which may not necessarily be the case.
For example, the evolution of the linear bias in the high-$z$ tail of $z_1$ may disagree with the linear bias evolution in the low $z$ tail of $z_2$, which would not be captured by Eq.~\eqref{eq:cgigj}. 
Additionally, we assume a simple linear interpolation of nuisance parameters, while the true redshift evolution may be more complex.
Second, we should in principle allow for a free shot noise component for each cross-spectrum $-$ especially for neighboring bins with overlapping redshift distributions $-$ although Fig.~\ref{fig:cgigj} suggests that a negative shot noise contribution would be required to alleviate the observed deficits.
With these subtleties in mind, we caution the reader that the predictions made with Eq.~\eqref{eq:cgigj} are somewhat crude and should be interpreted as such.

On the other hand, it may be the case that these deficits are the result of some form of systematic modulation in the data.
First, these deficits could suggest miscalibrated magnification contributions, although we find this unlikely given that the magnification contribution for neighboring bins is small, and that the data for the remaining cross-spectra with significant magnification contributions are in reasonable agreement with our predictions.
Second, the tails of the redshift distributions may be slightly miscalibrated.
To get a sense for the error in the tails we swapped the fiducial redshift distributions to those calibrated on DES footprint\footnote{We repeated this using the redshift distributions calibrated on the North and DECaLS footprints but found smaller variations than for DES.} \cite{Zhou:2023gji} and found $\simeq 2$, 15 and 8\% variations in $C^{g_1 g_2}_\ell$, $C^{g_2 g_3}_\ell$ and $C^{g_3 g_4}_\ell$ respectively.
Third, the probability that a galaxy is assigned to e.g. bin 1 or bin 2 at a fixed redshift may be modulated by an observational effect (e.g. extinction) that could lead to negative correlations between neighboring redshift bins. 
Fourth, the deficits could indicate the presence of a systematic contaminant (not captured by the systematic weights) that is negatively correlated between neighboring redshift bins, or in the maximally pessimistic scenario, a contaminant that has added power to the galaxy auto-spectra, resulting in an overestimate of the cross-spectra.

Out of an abundance of caution, we entertain the maximally pessimistic scenario in more detail. 
We find that the residuals between the best-fit predictions and measured galaxy cross-spectra are well fit by a $1/\ell$ contribution 
(black dashed lines in Fig.~\ref{fig:cgigj}) with the magnitude of the coefficients ranging from $-4\times 10^{-5} $ to $-6\times 10^{-5}$.
A $\Delta C^{g_i g_i}_\ell = 5\times10^{-5}/\ell$ absolute bias corresponds to a relative bias of at most $3\%$ to any of the galaxy auto-spectra. 
In the simplistic picture where $S_8$ is derived from the ratio $ C^{\kappa g}/\sqrt{C^{gg}}$, a relative $3\%$ scale-invariant increase (neglecting shot noise) in $C^{gg}$ would bias $S_8$ low by $1.5\%$, corresponding to a $\simeq 0.6\sigma $ shift for our fiducial analysis.
However, this crude calculation overestimates the resulting bias to $S_8$ since the systematic contamination has a different scale-dependence than a rescaling of $S_8$, such that not all of the $\sim3\%$ bias would be projected to cosmological constraints. 
We confirm this suspicion in \S\ref{sec:param_based_consistency_tests} by adding $1/\ell$ terms to the galaxy power spectra and marginalizing over their coefficients with $\mathcal{N}(0,10^{-4})$ priors. Doing so yields negligible ($0.05\sigma$) shifts in our fiducial linear theory $S_8$ constraints. 

We conclude that the observed deficits are unlikely to source significant biases to our cosmological constraints. These tests highlight the powerful ability of spectroscopically-calibrated galaxy samples to self-consistently probe systematic contamination, and test for immunity to it at the cosmological parameter level (\S\ref{sec:param_based_consistency_tests}). This is in contrast to photometric samples, for which systematic contaminants in the galaxy cross-spectra are more difficult to identify in the presence of noisy redshift distributions.

\section{Results}
\label{sec:results}

\subsection{Blinding}
\label{sec:blinding}

We blinded both the cross-correlation with ACT DR6 and all cosmological parameters derived from ACT data until all of the scale cuts (\S\ref{sec:scale_cuts}) and priors (\S\ref{sec:priors}) were finalized, null and systematics tests were satisfactorily passed (see the companion paper \cite{Kim2024} as well as sections \ref{sec:systematics} and \ref{sec:param_based_consistency_tests}), and our likelihood had been tested with the \verb|Buzzard| mocks (\S\ref{sec:buzzard_fits}).
We emphasize that none of our fiducial analysis choices have changed post unblinding.
We defer a detailed discussion of the full blinding policy to the companion paper \cite{Kim2024}, and further note here that scale cuts and priors were chosen based on rough guidelines from previous works (e.g. \cite{Kokron21}) and were finalized before running tests on the latest \verb|Buzzard| mock measurements, rather than using the \verb|Buzzard| mocks to calibrate these choices. 
This improves the mock tests as a cross-check for two reasons: (1) the fits to the latest mock measurements were done semi-blindly\footnote{in the sense that our analysis choices were fixed before running on the finalized measurements and that none of these choices changed after running the mock tests} and (2) calibrating e.g. priors with the \textit{same set} of mocks used to validate the likelihood would in general reduce the scatter of the recovered cosmological parameters around their true values.

\subsection{\textit{Planck} PR3 reanalysis}
\label{sec:pr3}

\begin{figure}[!h]
    \centering
    \includegraphics[width=\linewidth]{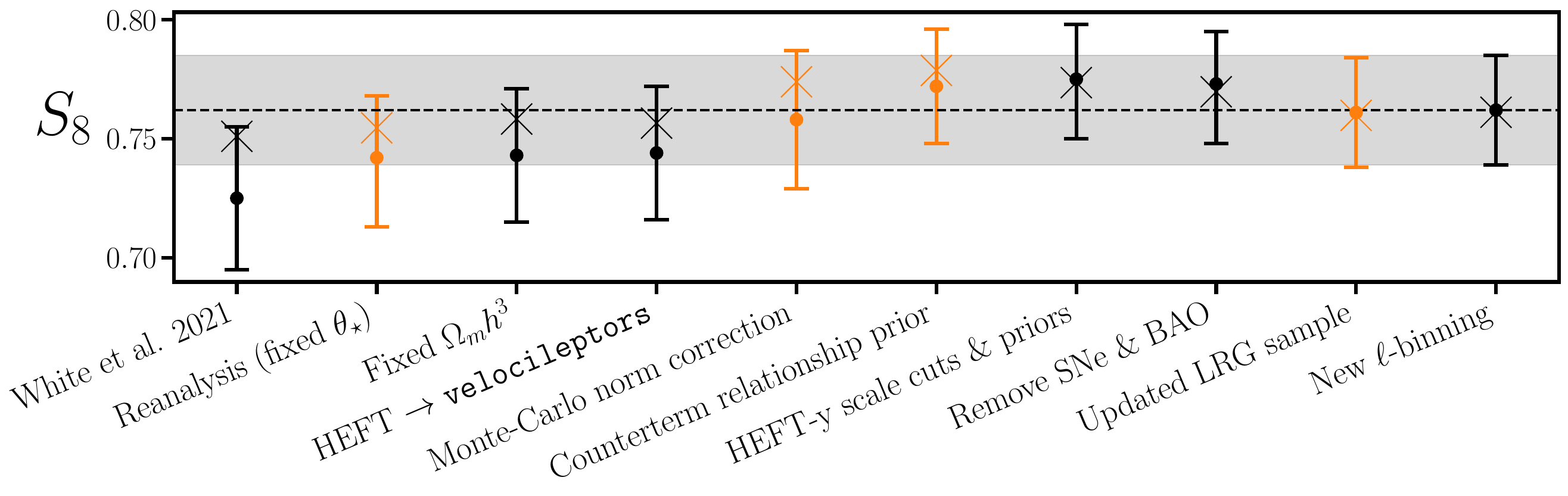}
    \caption{ 
    Connecting the dots between the White et al. \cite{White:2021yvw} $S_8$ constraints (far left) and those obtained from our fiducial PR3 reanalysis (far right).
    Highlighted in orange are the four significant changes in the analysis that are predominantly responsible for the shift in $S_8$.
    The shaded bands correspond to the $\pm1\sigma$ region of our fiducial PR3 constraints.
    Best-fit values are indicated by $\times$'s. 
    See the text in \S\ref{sec:pr3} for a detailed discussion of the analysis changes.
    } 
\label{fig:white21_reanalysis}
\end{figure}

White et al. \cite{White:2021yvw} found $S_8 = 0.725\pm0.030\,\, [0.751]$ (including BAO) from the cross correlation of \textit{Planck} PR3 CMB lensing with a previous version of the LRG sample considered here, where here and below the best-fit value is given in square brackets. 
With the latest LRG sample and the same CMB lensing map, our fiducial analysis choices yield
\begin{equation}
\label{eq:pr3_s8}
    \textit{Planck } \text{PR3: } S_8 = 0.762\pm 0.023\,\, [0.761]
\end{equation}
from our data alone, and 
\begin{equation}
\label{eq:pr3_sigma8}
    \textit{Planck } \text{PR3: } \sigma_8 = 0.758^{+0.023}_{-0.026}\,\, [0.765]
\end{equation}
once BAO data from 6dF \cite{2011MNRAS.416.3017B}, BOSS DR7 \cite{Ross:2014qpa} and DR12 \cite{BOSS:2016wmc} are included. 
The decrease in statistical error (from 0.030 to 0.023) is the result of the informative prior on the $\alpha_a-\alpha_x$ relationship (Eq.~\ref{eq:counterterm_relationship}). 
There are four significant changes (orange points in Fig.~\ref{fig:white21_reanalysis}) in our analysis that are responsible for the $\simeq 1\sigma$ upward shift in the \textit{mean} value of $S_8$:
\begin{itemize}
    \item consistent use of neutrino masses (see below): $\Delta S_8 = +0.017$ 
    \item including a ``normalization" correction to the PR3 cross-correlation: $\Delta S_8 = +0.015$
    \item adding an informative prior on the $\alpha_a-\alpha_x$ relationship (Eq.~\ref{eq:counterterm_relationship}), which efficiently mitigates ``volume effects" without shifting the best-fit significantly: $\Delta S_8 = +0.014$
    \item updating the LRG sample: $\Delta S_8 = -0.012$
\end{itemize}
We provide a more thorough description of additional changes below. Their impacts on $S_8$ constraints are summarized in Fig.~\ref{fig:white21_reanalysis}.

To smoothly connect our results with those found by ref.~\cite{White:2021yvw} we begin by reanalyzing the same data with nearly identical analysis choices.
We use the power spectra, window functions, covariance and redshift distributions provided at this \href{https://zenodo.org/record/5834378\#.ZRRqVtLMKEJ}{link}.
We adjust our scale cuts (for both the auto- and cross-correlation) to $\ell_{\rm min}(z_i) = 25$ and $\ell_{\rm max}(z_i) = 275,\,325,\,375,\,425$ for each redshift bin respectively\footnote{We note that ref.~\cite{White:2021yvw} quotes $\ell_{\rm max}$'s in terms of $\ell$ bin centers, whereas here we refer to bin edges.}.
Following \cite{White:2021yvw} we fix $n_s=0.97$, $\omega_b=0.022$, $\sum m_\nu = 0.06$ eV, $100\,\theta_\star = 1.04109$ and directly sample $\ln(10^{10}A_s)$ and $\Omega_c h^2$ with the priors listed in Table~\ref{tab:priors}. We fix $b_s(z_i)=0$, vary $\alpha_x$ directly (rather than $\epsilon$) with a $\mathcal{N}(0,50)$ prior and adjust the central value of our priors for shot noise and magnification bias to those listed in Table 1 of \cite{White:2021yvw}. 
Priors on the remaining nuisance parameters are the same as those listed in Table~\ref{tab:priors}.
Following \cite{White:2021yvw} we include supernovae \cite{Pan-STARRS1:2017jku} and BAO \cite{2011MNRAS.416.3017B,Ross:2014qpa,BOSS:2016wmc} data by default as an effective prior on $\Omega_m$. 
With these analysis choices we find $S_8 = 0.742^{+0.026}_{-0.029}\,\, [0.755]$ (second point in Fig.~\ref{fig:white21_reanalysis}). 
We have repeated this exercise with $\Omega_m h^3 = 0.09633$ fixed instead of $\theta_\star$, and using \verb|velocileptors| instead of Aemulus $\nu$ for the power spectrum predictions and found negligible changes: $S_8 = 0.743\pm 0.028\,\, [0.758]$ and $S_8 = 0.744\pm 0.028\,\, [0.757]$ respectively (third and fourth points in Fig.~\ref{fig:white21_reanalysis}). In what follows we always fix $\Omega_m h^3$ instead of $\theta_\star$, and continue to use the HEFT emulator rather than \verb|velocileptors|.

We note that the $S_8$ mean of our reanalysis is $\Delta S_8 = +0.017$ larger ($\sim 0.5\sigma$) than that found by ref.~\cite{White:2021yvw}. 
This difference is attributed to an inconsistent use of neutrino masses.
The 3D power spectrum emulators used by \cite{White:2021yvw} did not include massive neutrinos, while elsewhere in the likelihood (i.e. for computing background quantities and $\sigma_8$) the neutrino mass was set to 0.06 eV.
The lack of massive neutrinos in the emulators results in ``extra power" that is compensated for by lowering $S_8$.
The size of the resulting bias can be estimated as $\Delta S_8 = 0.5 \Delta P(k) / P(k) \simeq -4 f_\nu$ \cite{Hu:1997mj,Lesgourgues:2013sjj,Lesgourgues:2006nd}, which for $\Omega_m = 0.3$, $h = 0.67$ and $\sum m_\nu = 0.06$ eV gives $\Delta S_8 = - 0.019$, in good agreement with the observed offset. 

The PR3 cross-correlation used in \cite{White:2021yvw} did not include a Monte Carlo ``normalization" correction \cite{ACT:2023oei}. We compute this correction following Appendix~\ref{sec:joshuas_vs_noah_measurements}. Applying this to the White et al. data increases $S_8$ by approximately $0.5\sigma$ to $S_8 = 0.758\pm 0.029\,\, [0.774]$ (fifth point in Fig.~\ref{fig:white21_reanalysis}). 
Next we consider imposing the same prior on the $\alpha_a-\alpha_x$ relationship (Eq.~\ref{eq:counterterm_relationship}) as used in our fiducial analysis, which gives $S_8 = 0.772\pm 0.024\,\, [0.779]$ (sixth point in Fig.~\ref{fig:white21_reanalysis}). 
We attribute the $\Delta S_8 = +0.014$ shift in $S_8$ mean to the mitigation of ``volume effects", and note that we observe a significantly smaller (0.005) shift in the best-fit value.
Next we consider updating the scale cuts and priors on higher order biases to approximately match those used in our fiducial analysis. 
We adjust the values of fixed cosmological parameters to those listed in \S\ref{sec:priors} and adopt the fiducial priors listed in Table~\ref{tab:priors}, with the exception of shot noise and magnification whose central values are still centered on those listed in Table 1 of \cite{White:2021yvw}. 
We set $\ell_{\rm min}^{gg}(z_i) = 75$, $\ell_{\rm  min}^{\kappa g}(z_i) = 25$ and $\ell_{\rm max}(z_i) = 575$ for both the galaxy auto- and cross-correlation. 
Applying these changes gives $S_8 = 0.775^{+0.023}_{-0.025}\,\, [0.774]$ (seventh point in Fig.~\ref{fig:white21_reanalysis}).
This is in very good agreement with the previous result and suggests that variations in scale cuts and priors on higher order bias parameters have an insignificant impact on our $S_8$ constraints. We explore this in more detail in \S\ref{sec:param_based_consistency_tests}.
We next consider removing the supernovae and BAO priors (eighth point in Fig.~\ref{fig:white21_reanalysis}), which has a negligible impact on our results: $S_8 = 0.773^{+0.022}_{-0.025}\,\, [0.770]$. 

We remeasured the cross-correlation of PR3 with the latest version of our LRG sample \cite{Zhou:2023gji} using the same $\ell$-binning scheme as in \cite{White:2021yvw}.
We use the redshift distributions provided by \cite{Zhou:2023gji} and recompute the window functions and covariance matrix following \S\ref{sec:cl_estimation}.
The new data prefer a slightly lower value of $S_8$ when adopting the same scale cuts as in the previous paragraph: $S_8 = 0.761\pm 0.023\,\, [0.760]$.
Finally, we switch from linear $\Delta \ell = 50$ bins to the $\sqrt{\ell}$ bins discussed in \S\ref{tab:scalecuts} and adjust our scale cuts to match those in Table~\ref{tab:scalecuts}, such that the analysis choices now exactly match those used in our fiducial PR4 analysis below.
Updating the $\ell$ binning and scale cuts has a negligible impact on our constraints: $S_8 = 0.762\pm 0.023\,\, [0.761]$.

\subsection{\textit{Planck} PR4, ACT DR6 and their combination}
\label{sec:pr4_dr6_results}

\begin{figure}[!h]
    \centering
    \includegraphics[width=\linewidth]{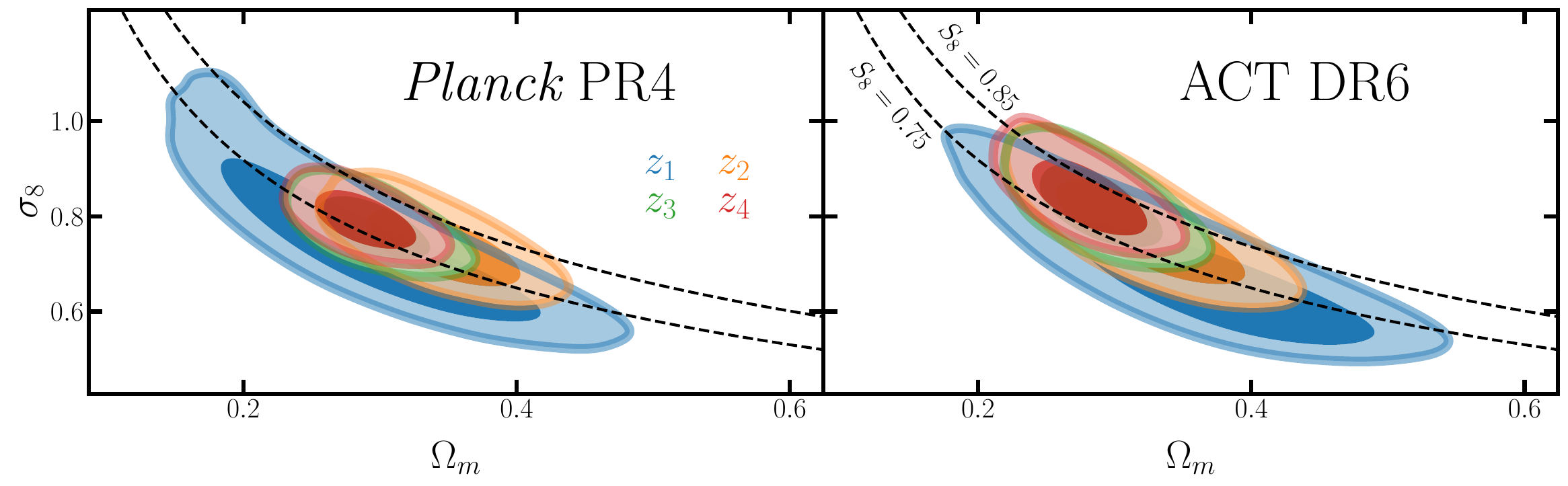}
    \caption{ 
    Fiducial cosmological constraints (without BAO) when analyzing each redshift bin separately (colors), both for \textit{Planck} PR4 (left) and ACT DR6 (right). 
    The dashed lines correspond to $S_8 = 0.75$ and 0.85.
    } 
\label{fig:zi_OmM_sigma8}
\end{figure}

In this section we present results using our fiducial HEFT model (\S\ref{sec:heft}), scale cuts (\S\ref{sec:scale_cuts}, Table~\ref{tab:scalecuts}) and priors (\S\ref{sec:priors}, Table~\ref{tab:priors}).
In Fig.~\ref{fig:zi_OmM_sigma8} we show the cosmological constraints obtained when fitting to each redshift bin individually (without BAO), considering separately the cases of cross-correlating with \textit{Planck} PR4 (left) and ACT DR6 (right).
The $\Omega_m-\sigma_8$ contours from each redshift bin are all in agreement to well within $1\sigma$, as apparent in Fig.~\ref{fig:zi_OmM_sigma8}. 

When jointly analyzing all four galaxy auto-spectra and their cross-correlations with \textit{Planck} PR4 we find
\begin{equation}
\label{eq:pr4_s8}
    \textit{Planck } \text{PR4: } S_8 = 0.765\pm0.023\,\, [0.764],
\end{equation}
from our measurements alone (i.e. no BAO), where the best-fit value is quoted in square brackets.
We note that this is in very good agreement with the $\textit{Planck}$ PR3 result (Eq.~\ref{eq:pr3_s8}), and that while the PR4 lensing reconstruction has improved significantly over PR3, we interestingly do not find a corresponding improvement in our $S_8$ constraint.
In combination with BAO data from 6dF \cite{2011MNRAS.416.3017B} SDSS DR7 \cite{Ross:2014qpa} and DR12 \cite{BOSS:2016wmc}, we are able to break the $\Omega_m-\sigma_8$ degeneracy, yielding 
\begin{equation}
\label{eq:pr4_sigma8}
    \textit{Planck } \text{PR4: } \sigma_8 = 0.762\pm0.023\,\, [0.765].
\end{equation}
We find a slight improvement in the PR4 $\sigma_8$ constraint over PR3 (cf. Eq.~\ref{eq:pr3_sigma8}).

When jointly analyzing the four galaxy auto-spectra and their cross-correlations with ACT DR6 (see Fig.~\ref{fig:best_fit}) we find
\begin{equation}
    \text{ACT DR6: }  S_8 = 0.790^{+0.024}_{-0.027}\,\, [0.789]
\end{equation}
while with the addition of BAO data we find
\begin{equation}
    \text{ACT DR6: } \sigma_8 = 0.787^{+0.025}_{-0.029}\,\, [0.794]
\end{equation}
We note that the DR6 constraints are weaker than PR4 despite the lower reconstruction noise of the DR6 map due to the larger overlap between the PR4 footprint and LRGs.

\begin{figure}[!h]
    \centering
    \includegraphics[width=0.49\linewidth]{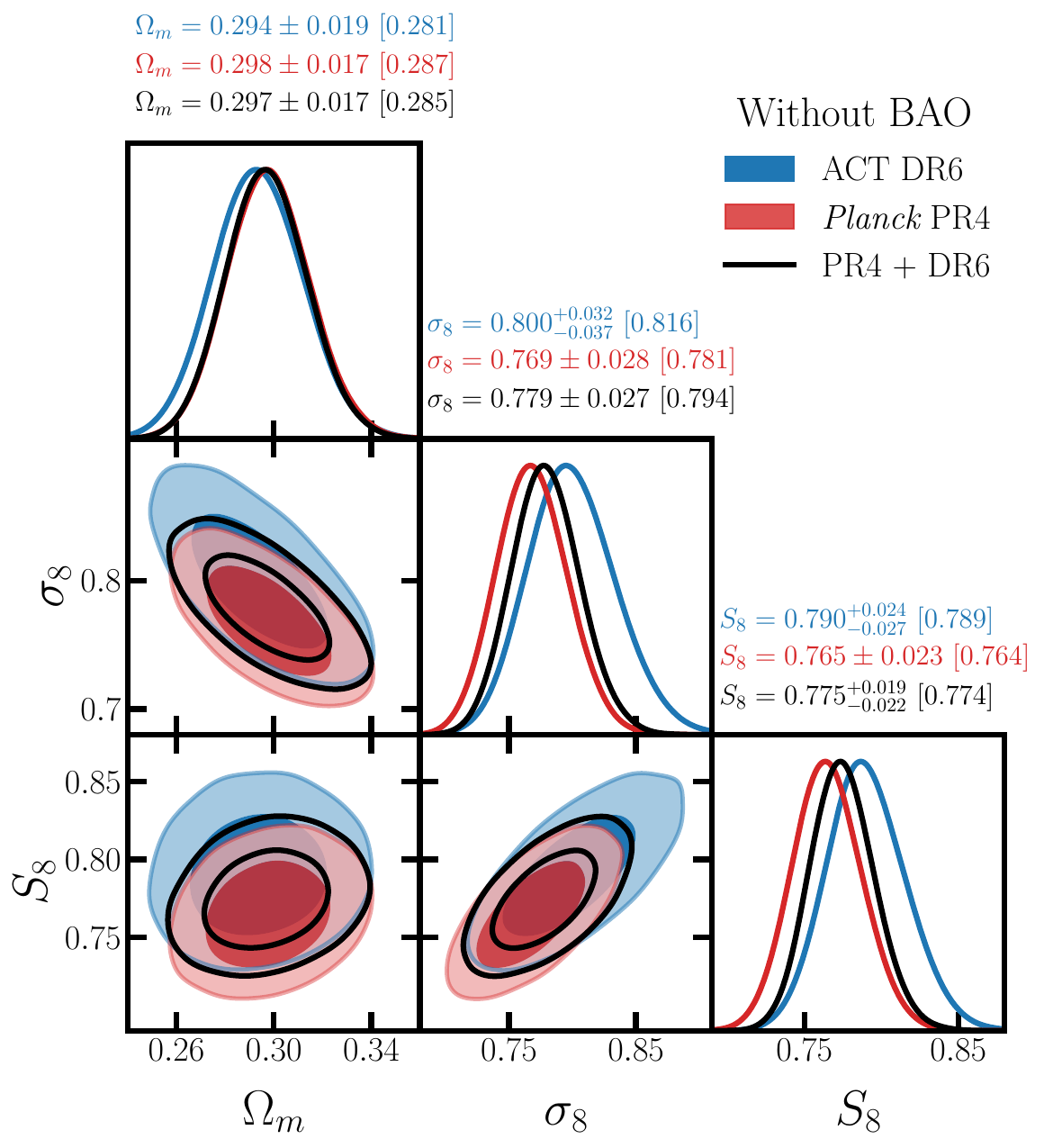}
    \includegraphics[width=0.49\linewidth]{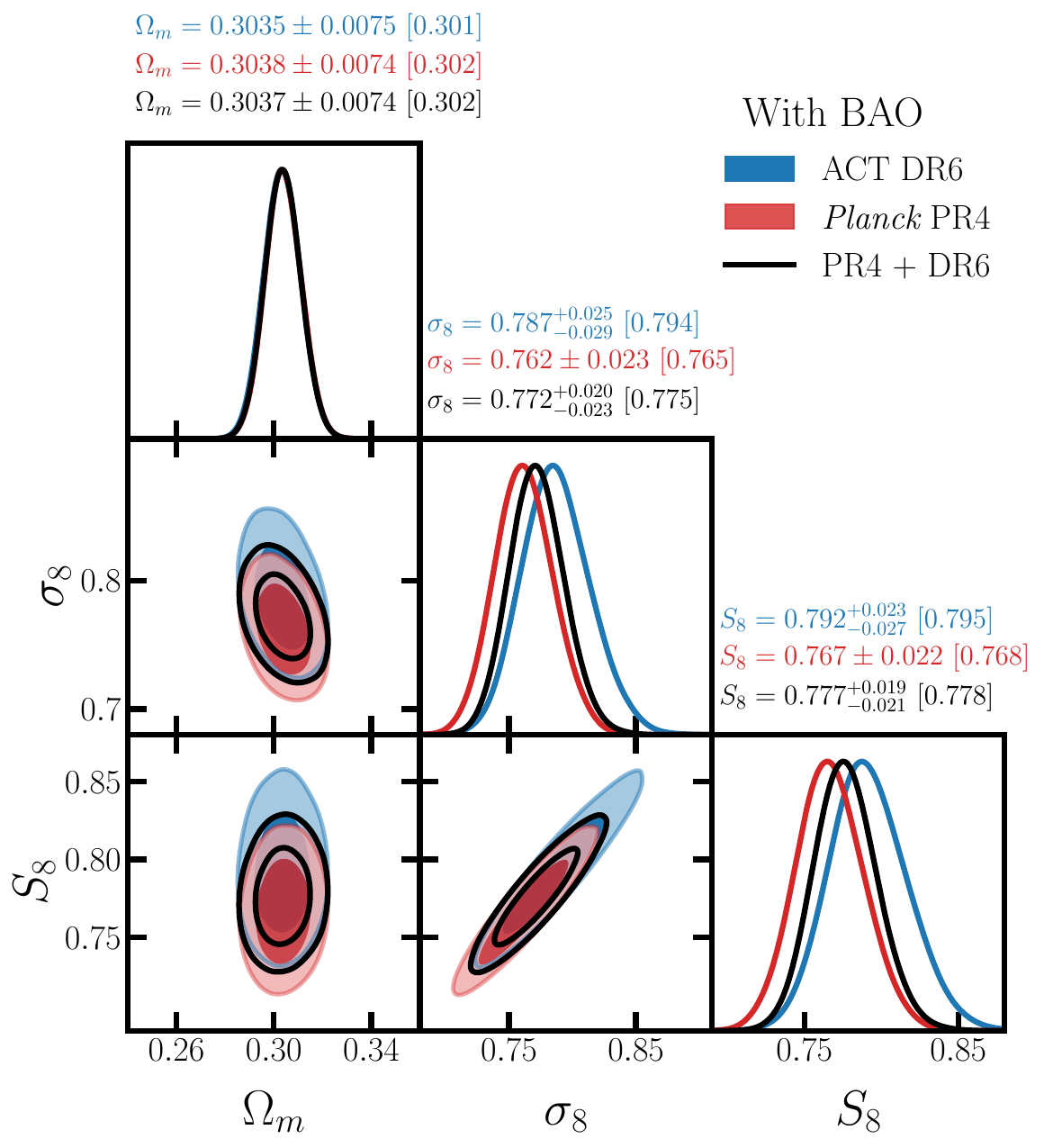}
    \caption{ 
    \textit{Left:} Cosmological constraints when adopting our fiducial analysis choices. We jointly analyze the four redshift bins and their cross-correlations with ACT DR6 (blue), \textit{Planck} PR4 (red) and their combination (black).
    \textit{Right:} Same as before, but now with BAO data included \cite{2011MNRAS.416.3017B,Ross:2014qpa,BOSS:2016wmc}.
    } 
\label{fig:pr4_dr6_triangle}
\end{figure}

We show the PR4 and DR6 $\sigma_8-\Omega_m$ contours (without BAO) in the left panel of Fig.~\ref{fig:pr4_dr6_triangle}.
The difference between the ACT DR6 and $\text{Planck}$ PR4 $S_8$ constraint is $\Delta S_8 = 0.025$. 
As shown in Fig.~6 of the companion paper \cite{Kim2024}, the PR4 and DR6 cross-correlation measurements are correlated by at most $\simeq40\%$ on the scales $(100 \lesssim \ell \lesssim 300)$ where most of the $S_8$ constraint is derived. 
Within linear theory, where the constraint on $S_8$ is schematically driven by the ratio $C^{\kappa g}_\ell/\sqrt{C^{gg}_\ell}$, we expect the correlation coefficient of the PR4 and DR6 $S_8$ measurements to be nearly identical to the correlations of the associated cross-spectra.
Taking $r=0.4$, $\sigma_{\rm PR4} = 0.023$ and $\sigma_{\rm DR6} = 0.0255$ we estimate the error on $\Delta S_8$ to be 
$[\sigma^2_{\rm PR4} +\sigma^2_{\rm DR6} - 2 r \sigma_{\rm PR4} \sigma_{\rm DR6}]^{1/2} \simeq 0.027$. This estimate suggests that the PR4 and DR6 $S_8$ constraints are consistent with one another to within $1\sigma$, as visually suggested by Fig.~\ref{fig:pr4_dr6_triangle}.

Having established the consistency of the \textit{Planck} PR4 and ACT DR6 measurements we consider jointly analyzing the two datasets. 
When analyzing each redshift bin independently we find
\begin{equation}
\begin{aligned}
\label{eq:S8_sig8_individual}
    S_8 &= 
    0.727^{+0.037}_{-0.054}\,\,[0.726],\,
    0.787^{+0.036}_{-0.043}\,\,[0.780],\,
    0.783^{+0.033}_{-0.039}\,\,[0.775],\,
    0.789\pm0.032\,\,[0.789]
    \\
    \sigma_8 &=
    0.717^{+0.039}_{-0.051}\,\,[0.717],\,
    0.776^{+0.035}_{-0.047}\,\,[0.769],\,
    0.781^{+0.032}_{-0.038}\,\,[0.779],\,
    0.790\pm0.030\,\,[0.791]
\end{aligned}
\end{equation}
without and with the addition of BAO data for $S_8$ and $\sigma_8$ respectively.
When jointly fitting the four galaxy auto-spectra, four cross-correlations with PR4, and four cross-correlations with DR6 we find
\begin{equation}
\label{eq:combined_s8}
    \boxed{\text{Combined: } S_8 = 0.775^{+0.019}_{-0.022}\,\, [0.774]}
\end{equation}
without any additional BAO information and 
\begin{equation}
\label{eq:combined_sigma8}
    \boxed{\text{Combined: } \sigma_8 = 0.772^{+0.020}_{-0.023}\,\, [0.775]}
\end{equation}
once BAO data are included. We find $\sim10\%$ improvements to both the $S_8$ and $\sigma_8$ errorbars over \textit{Planck} PR4 alone from the addition of ACT DR6. 

\begin{figure}[!h]
    \centering
    \includegraphics[width=\linewidth]{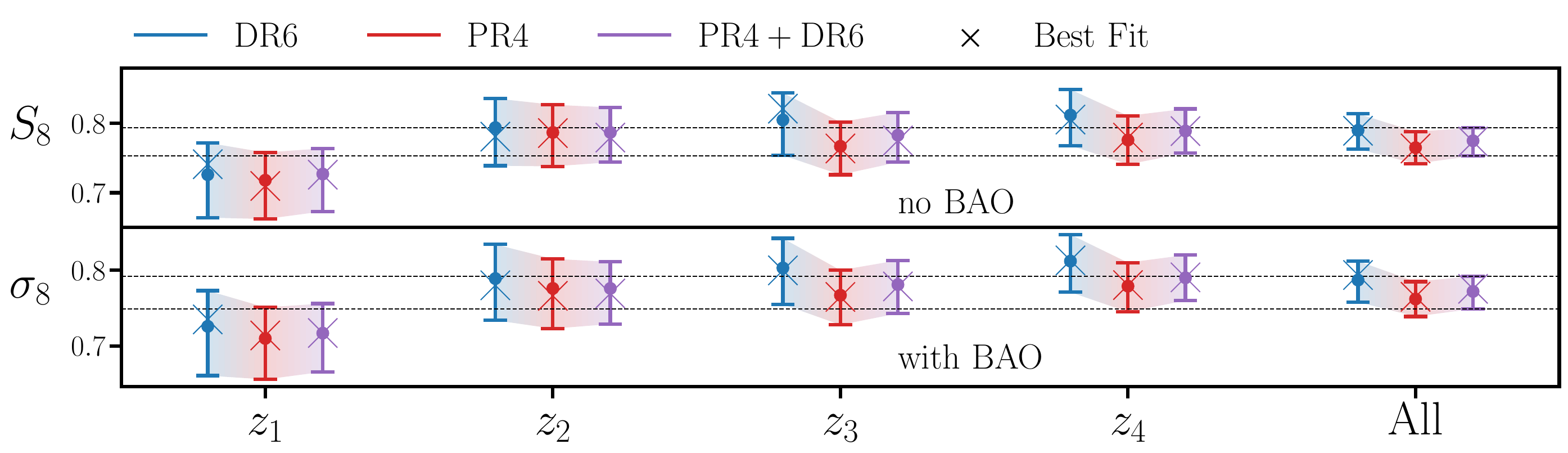}
    \caption{ 
    Marginal $S_8$ (top) and $\sigma_8$ (bottom) constraints when fitting to each redshift bin individually, or jointly fitting to all redshift bins. We show the constraints from PR4 in blue, DR6 in red and their combination in purple. 
    The location of the best-fit value is indicated with $\times$'s. 
    The dashed black curves correspond to the $\pm1\sigma$ interval associated with our joint PR4 + DR6 fit (i.e. Eqs.~\ref{eq:combined_s8} and \ref{eq:combined_sigma8}).
    } 
\label{fig:pr4_dr6_summary}
\end{figure}

We summarize our fiducial results in Fig.~\ref{fig:pr4_dr6_summary}, where we show the marginal $S_8$ (without BAO) and $\sigma_8$ (with BAO) constraints for PR4, DR6 and their combination (in blue, red and purple respectively) for each redshift bin separately, and for the combination of all redshift bins ($x$-axis). 
We note that the DR6 values (blue) are consistently larger than the PR4 (red) results.
This is qualitatively consistent with the results found in the DR6 auto-spectrum analysis \cite{ACT:2023kun, ACT:2023dou} and the cross-correlation with unWISE galaxies \cite{ACT:2023oei}.
As in \cite{White:2021yvw} we find that $z_1$ prefers a slightly lower $S_8$ (and $\sigma_8$) value than the higher redshift samples, however, we have yet to find evidence (see \S\ref{sec:param_based_consistency_tests} for parameter-based tests) that this preference is driven by systematics.

\subsection{Correlations between $S_8(z)$ measurements}
\label{sec:S8_correlations}

To estimate correlations between the $S_8(z)$ constraints listed in Eq.~\eqref{eq:S8_sig8_individual} we reanalyzed our joint PR4+DR6 dataset (including all four redshift bins) with an expanded set of cosmological parameters: $\Omega_c h^2$ and four $A_s(z_i)$'s that set the primordial power spectrum amplitude in each redshift bin. 
We sample each $\ln(A_s(z_i))$ directly with a uniform $\mathcal{U}(2,4)$ prior. We define $S_8(z_i)$ as the $\Lambda$CDM-derived value of $S_8$ given $A_s(z_i)$ and $\Omega_c$ (in addition to the remaining fixed cosmological parameters), serving as a proxy for the value of $S_8$ derived from the $i$'th redshift bin alone. 
Following this procedure we obtain
\begin{equation}
\begin{aligned}
    S_8(z_i) &= 
    0.728^{+0.038}_{-0.051},\,
    0.785^{+0.034}_{-0.048},\,
    0.784^{+0.031}_{-0.038},\,
    0.793\pm0.031,
\end{aligned}
\end{equation}
which is in excellent agreement with the $S_8$ constraints derived from each redshift bin (cf. Eq.~\ref{eq:S8_sig8_individual}).
From the posterior we obtain a covariance matrix for the $S_8(z_i)$'s from which we estimate 9, 13 and 28\% correlations between $S_8(z_1)-S_8(z_2)$, $S_8(z_2)-S_8(z_3)$ and $S_8(z_3)-S_8(z_4)$ respectively. 
The remaining correlations (e.g. $z_1-z_3$) are all $\leq 3\%$. 
In the limit that $S_8$ is derived from the ratio $C^{\kappa g}_\ell/\sqrt{C^{gg}_\ell}$ on linear scales, one would expect the correlations between $S_8(z_i)$ and $S_8(z_j)$ to approximately be equal to the correlation between $C^{\kappa g_i}_\ell$ and $C^{\kappa g_j}_\ell$ at $\ell\sim100$. As illustrated in Fig.~6 of the companion paper \cite{Kim2024} this is indeed the case: there are $\sim$ 15, 15, and 25\% correlations between  $C^{\kappa g_1}_\ell-C^{\kappa g_2}_\ell$, $C^{\kappa g_2}_\ell-C^{\kappa g_3}_\ell$ and $C^{\kappa g_3}_\ell-C^{\kappa g_4}_\ell$ respectively.

To connect the $S_8(z_i)$ constraints to our combined $S_8$ measurement (Eq.~\ref{eq:combined_s8}) we define the following derived quantity, which is an optimal linear combination of the individual amplitudes: $\hat{S}_8 \equiv \sum_{ij} \mathtt{C}^{-1}_{ij} S_8(z_j) / \sum_{ij}\mathtt{C}^{-1}_{ij} $, where $\mathtt{C}_{ij}$ is the covariance matrix of the $S_8(z_i)$'s obtained from the posterior. We find $\hat{S}_8=0.778 \pm 0.021$ which is in excellent agreement with Eq.~\eqref{eq:combined_s8}.

\subsection{Goodness of fit and consistency with \textit{Planck} $\Lambda$CDM}
\label{sec:fit_good}

The best-fit HEFT prediction to the combined (PR4 + DR6) analysis (without BAO) is shown in Fig.~\ref{fig:best_fit}. In each panel we quote the $\chi^2$ value for each individual measurement.
The best-fit $\chi^2 = 53.1$ for 96 data points and 30 free parameters, 16 of which are partially prior-dominated. 
Following Appendix \ref{sec:anamarg} we estimate ${\rm PTE} = 96\%$. 
We note that here and in the companion paper \cite{Kim2024} we approximate the LRGs as Gaussian random fields when estimating covariance matrices. This approximation likely underestimates the covariance between high-$\ell$ bandpowers and hence overestimates the PTE.
We leave an investigation of the trispectrum contribution to the covariance matrix to future work, and note that for our fiducial linear theory analysis (for which these contributions are considerably smaller and likely negligible) we estimate a lower ${\rm PTE} = 83\%$.

We assess the consistency of our results with a \textit{Planck} $\Lambda$CDM cosmology following two approaches. 
First we consider the ``standard" approach of citing tension metrics for a given cosmological parameter, defined [in units of $\sigma$] as the difference between the means divided by the square root of the quadrature sum of the respective errorbars (i.e. ignoring correlations in the datasets). 
Our combined $S_8$ constraint (Eq.~\ref{eq:combined_s8}) is $2.3\sigma$ lower than that preferred by \textit{Planck} 2018 \cite{Planck:2018vyg} primary CMB data (TT,TE,EE+lowE) and $1.8\sigma$ ($2.2\sigma$) lower than the latest PR4 analysis from Tristram et al. \cite{Tristram:2023haj} (Rosenberg et al. + \verb|SRoll2| low $\ell$ EE \cite{Rosenberg:2022sdy,Pagano:2019tci}).
These ``tensions" are less significant when comparing $\sigma_8$ values. 
Our combined $\sigma_8$ constraint (Eq.~\ref{eq:combined_sigma8}) is $1.8\sigma$ lower than PR3 and $1.6\sigma$ ($1.7\sigma$) lower than the aforementioned PR4 analyses. 

Our second approach is to fix the cosmology to \textit{Planck} 2018 (see \S\ref{sec:priors} and Table \ref{tab:priors}) and to compare the resulting best-fit and PTEs to that obtained when $\ln(10^{10}A_s)$ and $\Omega_c h^2$ are allowed to float freely. 
We find $\Delta \chi^2 = 5.0$ between the best-fit with fixed and free cosmology, while with fixed cosmology we estimate ${\rm PTE} = 93\%$ following Appendix \ref{sec:anamarg}. 
We conclude that the \textit{Planck} 2018 cosmology yields a perfectly reasonable fit to our data.

\subsection{Alternative models}
\label{sec:alternative_models}

\begin{figure}[!h]
    \centering
    \includegraphics[width=\linewidth]{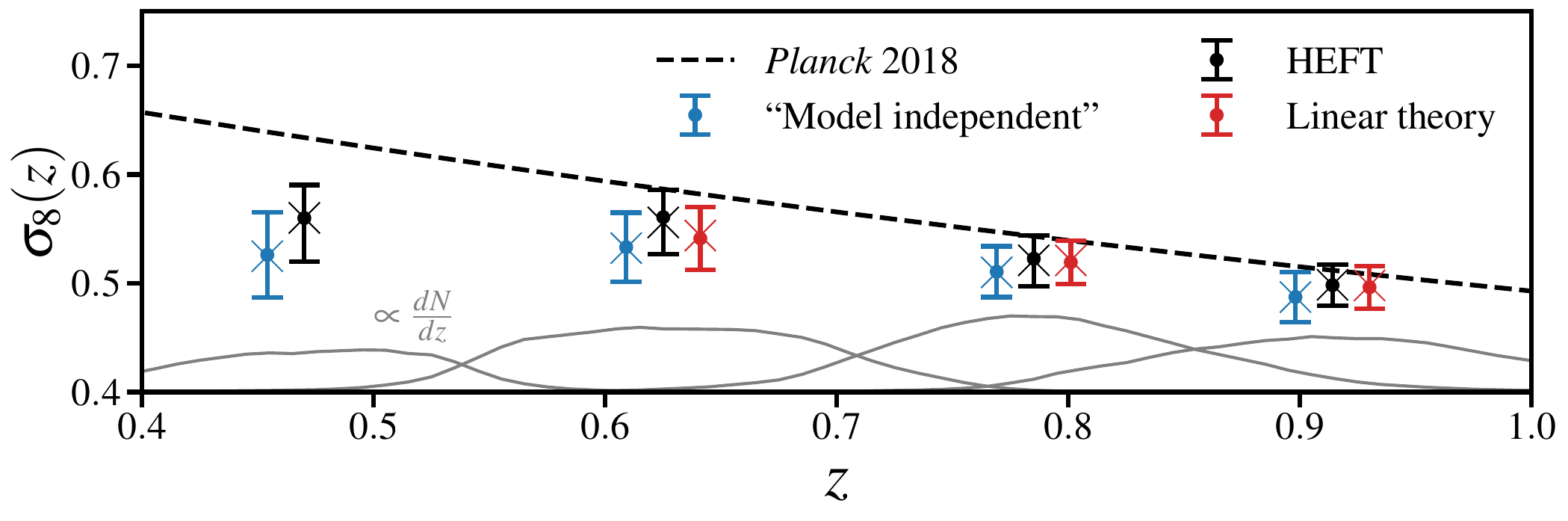}
    \caption{ 
    A comparison of our fiducial HEFT $\sigma_8(z)$ constraints (black) to those obtained from a ``model independent" (blue, \S\ref{sec:modelind}) or linear theory (red, \S\ref{sec:linthy}) approach. 
    We note that there is no linear theory point for $z_1$ as we do not include this bin in our fiducial linear theory analysis.
    In all scenarios we jointly fit to the cross-correlation with \textit{Planck} PR4 and ACT DR6 in each individual redshift bin, and use the fiducial scale cuts and priors listed in Tables \ref{tab:scalecuts} and \ref{tab:priors} respectively. 
    We include a BAO prior for both the HEFT and linear theory constraints, while for the model-independent constraints the cosmology is fixed to the values listed in Table~\ref{tab:priors}.
    We offset the linear theory and model independent constraints by $\Delta z =\pm 0.015$ respectively from the effective redshifts for clarity.
    The black dashed line is the predicted $\Lambda$CDM evolution assuming a \textit{Planck} 2018 cosmology \cite{Planck:2018vyg}. 
    In gray we plot $0.4 + c \,dN/dz$ for each galaxy sample, where $c$ is an arbitrary constant, to illustrate the redshift origin of the signal.
    Following \S\ref{sec:S8_correlations} we estimate 9, 13 and 28\% correlations between the $z_1-z_2$, $z_2-z_3$ and $z_3-z_4$ amplitude measurements respectively.
    } 
\label{fig:sigma8z_model_comparison}
\end{figure}

Here we present cosmological constraints from the alternative models discussed in \S\ref{sec:linthy} and \S\ref{sec:modelind} using our fiducial scale cuts (Table \ref{tab:scalecuts}) and priors (Table \ref{tab:priors}). These results are summarized in Fig.~\ref{fig:sigma8z_model_comparison} and discussed in detail below. 

We first consider ``linear theory" constraints (\S\ref{sec:linthy}). 
When fitting to each redshift bin independently without a BAO prior, we find
\begin{equation}
S_8 = 
0.761\pm 0.042\,\, [0.763],\,\,
0.796\pm 0.034\,\, [0.796],\,\,
0.797\pm 0.032\,\, [0.799]
\end{equation}
for $z_2,\,z_3,\,z_4$ respectively when jointly fitting to the PR4 and DR6 cross-correlations, or
$
S_8 = 0.787\pm 0.024\,\, [0.789]
$ 
from the combination of bins $z_2-z_4$. 
In combination with BAO data we find
\begin{equation}
\sigma_8 = 
0.749\pm 0.040\,\, [0.753],\,\,
0.776\pm 0.030\,\, [0.779],\,\,
0.787\pm 0.031\,\, [0.790]
\end{equation}
from the individual redshift bins, and 
$
\sigma_8 = 0.774\pm 0.024\,\, [0.779]
$
when jointly fitting to $z_2-z_4$. 
We note that the linear theory analysis prefers slightly higher $S_8$ value than our fiducial HEFT analysis (cf. Eq.~\ref{eq:combined_s8}). 
This preference is driven by the exclusion of the lowest redshift bin (by default) in our linear theory fits, which prefers an amplitude that is $\sim1\sigma$ lower than the remaining bins (see Fig.~\ref{fig:pr4_dr6_summary}). When dropping the lowest redshift bin in our combined HEFT analysis, we find 
$S_8 = 0.785\pm 0.023\,\, [0.781]$, which is in better agreement with our linear theory results. Next we consider the model independent approach discussed in \S\ref{sec:modelind}. When fitting to the combination of PR4 and DR6, we find
\begin{equation}
\alpha_8(z_i) = 
0.830\pm 0.062\,\, [0.828],\,\,
0.909\pm 0.054\,\, [0.907],\,\,
0.940\pm 0.043\,\, [0.939],\,\,
0.952\pm 0.045\,\, [0.952]
\end{equation}
for the four redshift bins respectively. In particular, the constraint from the lowest redshift bin is $2.7\sigma$ lower than $1$, while the remaining redshift bins are consistent with $1$ at the $1-2\sigma$ level.

A comparison of the different modeling choices is given in Fig.~\ref{fig:sigma8z_model_comparison}, where we have rescaled the HEFT and linear theory $\sigma_8$ constraints (including BAO) from each redshift bin by $\sigma_8(z_{\text{eff, }i})/\sigma_8(0)$ and the model-independent constraints by $\sigma_8(z_{\text{eff, }i})$, where $\sigma_8(z)$ is computed assuming a \textit{Planck} 2018 cosmology (see \S\ref{sec:priors} and Table \ref{tab:priors}).
We note that the HEFT and linear constraints agree to within $\simeq0.5\sigma$, while the model-independent results are consistently lower than our fiducial HEFT constraints by $\simeq1\sigma$.

\subsection{Parameter based consistency tests}
\label{sec:param_based_consistency_tests}

Here we present variations in our fiducial (\S\ref{sec:pr4_dr6_results}) and  linear theory (\S\ref{sec:alternative_models}) $S_8$ constraints (without BAO) when adopting different analysis choices beyond the alternative models discussed in \S\ref{sec:alternative_models} and when analyzing different data subsets. 

\begin{figure}[!h]
    \centering
    \includegraphics[width=\linewidth]{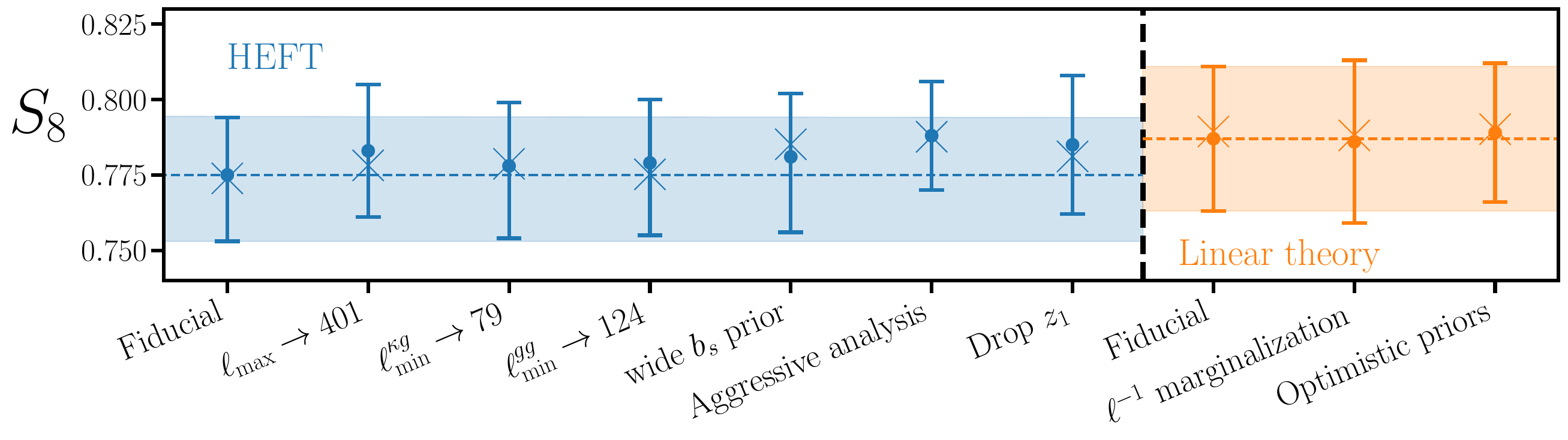}
    \caption{ 
    Variations (or lack thereof) in our baseline PR4+DR6 $S_8$ constraints (without BAO) for both HEFT (blue) and linear theory (orange) with different data subsets, priors and systematic mitigation. Best fit values are indicated with $\times$'s.
    } 
\label{fig:misc_param_consistency}
\end{figure}

First we consider variations in scale cuts and priors in our fiducial PR4+DR6 analysis, which are summarized in the left half of Fig.~\ref{fig:misc_param_consistency}.
When decreasing $\ell_{\rm max}$ for both the galaxy auto- cross-correlation with CMB lensing from 600 to 401, increasing $\ell_{\rm min}$ from 20 (44) to 79 for the cross-correlation with \textit{Planck} (ACT) lensing, increasing the $\ell_{\rm min}$ for the galaxy auto from 79 to 124, or adopting a wider $\mathcal{U}(-5,5)$ prior on $b_s$ in each redshift bin, the mean $S_8$ value shifts by at most $0.4\sigma$ from our fiducial constraint (Eq.~\ref{eq:combined_s8}).
In Fig.~\ref{fig:misc_param_consistency} we also consider a more aggressive analysis where we adopt a more restrictive $\ell_{\rm max} = 401$ but fix the counterterms to zero, reduce the width of the magnification bias prior from 0.1 to 0.05, and adopt highly informative $\mathcal{N}(0,0.6)$ priors on $b_2$ and $b_s$.
Doing so gives $S_8 = 0.788\pm 0.018\,\, [0.788]$, which is $\simeq0.7\sigma$ (in units of the aggressive errorbar) larger than our fiducial result. 
The aggressive constraints are more directly comparable to those found with our fiducial priors but with a reduced $\ell_{\rm max}=401$ (second point in Fig.~\ref{fig:misc_param_consistency}), for which we find $S_8 = 0.783\pm 0.022\,\, [0.778]$, suggesting that our results are fairly insensitive to the assumed priors on higher order nuisance parameters. 

As discussed in \S\ref{sec:alternative_models} our fiducial PR4+DR6 linear theory result (leftmost orange point Fig.~\ref{fig:misc_param_consistency}) is in good agreement with our fiducial HEFT constraint, but is more directly comparable with our HEFT analysis when dropping the first redshift bin (rightmost blue point in Fig.~\ref{fig:misc_param_consistency}), for which we find even better agreement.
Motivated by discrepancies between the inferred and measured galaxy cross-spectra (\S\ref{sec:gxspec} and Fig.~\ref{fig:cgigj}), we consider adding a $\ell^{-1}$ contribution (second orange point in Fig.~\ref{fig:misc_param_consistency}) to $C^{gg}_\ell$ and marginalizing over its amplitude with a $\mathcal{N}(0,10^{-4})$ prior. 
Doing so has a negligible impact on the mean and best fit $S_8$ values of our linear theory fits, which suggests that even if ``systematic contamination" were to blame for the discrepancies observed in the galaxy cross spectra, these contaminants have a negligible impact on our cosmological results. 
Finally, we performed linear theory fits with more optimistic priors on shot noise (Gaussian width decreased from 30\% to 10\%), magnification bias (Gaussian width decreased from 0.1 to 0.05) and counterterms (Gaussian width decreased from $3$ to $0.5$) and found negligible changes.

\begin{figure}[!h]
    \centering
    \includegraphics[width=\linewidth]{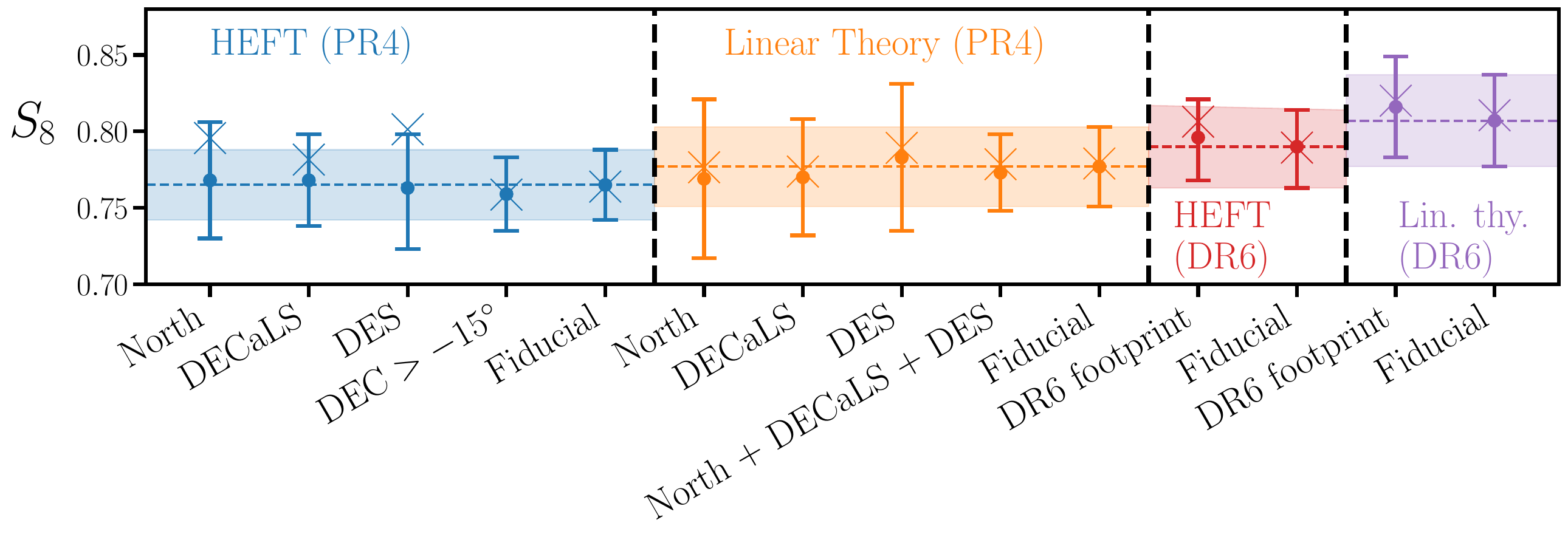}
    \caption{ 
    Variations in $S_8$ constraints (without BAO) when additionally masking the LRG sample with the North, DECaLS, DES, ${\rm DEC}>-15^\degree$, or ACT DR6 lensing mask.
    Best-fit values are indicated with $\times$'s.
    The horizontal lines and shaded bands correspond to mean and $\pm1\sigma$ intervals of the fiducial constraints for each scenario (PR4 or DR6; HEFT or linear theory).
    We suspect that the discrepancies between the mean and best fit $S_8$ values when analyzing the individual imaging footprints with HEFT are the result of noisy minima and volume effects.
    } 
\label{fig:varying_footprints_parameters}
\end{figure}

Motivated by the statistically significant variations in the galaxy auto-spectra on different imaging footprints (\S\ref{sec:imaging_footprint_variations}) we fit to each of these measurements individually as a parameter based test. These result are summarized in Fig.~\ref{fig:varying_footprints_parameters}.
We first consider fitting the galaxy auto- and PR4 cross-correlation measured on the North, DECaLS and DES footprints using our fiducial HEFT model (blue points in Fig.~\ref{fig:varying_footprints_parameters}) with the redshift distributions calibrated on each respective footprint \cite{Zhou:2023gji} (see Fig.~\ref{fig:dNdz_variations}).
For all three footprints we find results that are statistically consistent with our fiducial HEFT constraint.
Next we fit to the data measured on the ${\rm DEC} >-15^\degree$ subset of the LRG footprint, finding $S_8 = 0.759\pm 0.024\,\, [0.759]$ which is in very good agreement with our fiducial PR4 result (cf. Eq.~\ref{eq:pr4_s8}), suggesting that potential variations in the redshift distributions north and south of ${\rm DEC} = -15^\degree$ have a small impact on our final results.

With our fiducial HEFT model we find that the individual footprints prefer a best-fit $S_8$ value that is $\simeq0.5-1\sigma$ larger than the mean. 
When holding $A_s$ fixed to the best-fit of the fiducial PR4 cross-correlation analysis, the best-fit $\chi^2$ for each of the individual footprints increases by at most 1.5.
Given the modest shifts in $\chi^2$ we hypothesize that the difference in best-fit and mean $S_8$ values seen in Fig.~\ref{fig:varying_footprints_parameters} are the result of noisy minima. 
We have repeated the exercise of fitting to each imaging footprint using linear theory (orange points) and find results that are consistent with our fiducial linear theory PR4 constraint: $S_8 = 0.777\pm 0.026\,\, [0.779]$.
Unlike for the HEFT model, here we do not observe significant differences between the best fit and mean $S_8$ values. 
We suspect that these minima are more stable than for the HEFT model, due to fewer nuisance parameters (30 vs $\simeq 14$).
We also performed a joint linear theory fit to the North, DECaLS and DES data (12 galaxy auto- and 12 cross-spectra) where we marginalize over nuisance terms for each sample individually. Doing so yields $S_8 = 0.773\pm 0.025\,\, [0.779]$ which is in extremely good agreement with our fiducial result, suggesting that variations in the LRG linear bias across footprints has a negligible bias on our final results, which is in reasonable agreement with the back-of-the-envelope calculation made in \S\ref{sec:imaging_footprint_variations}.
As a final footprint-variation test we reanalyze the cross-correlation with ACT DR6 when additionally masking the LRGs by the DR6 lensing mask, which ensures that the galaxy auto- and cross-correlation are probing the same effective linear bias (to leading order). 
When using a HEFT (red) or linear theory (purple) model our $S_8$ constraints shift by less than $0.3\sigma$.

\begin{figure}[!h]
    \centering
    \includegraphics[width=\linewidth]{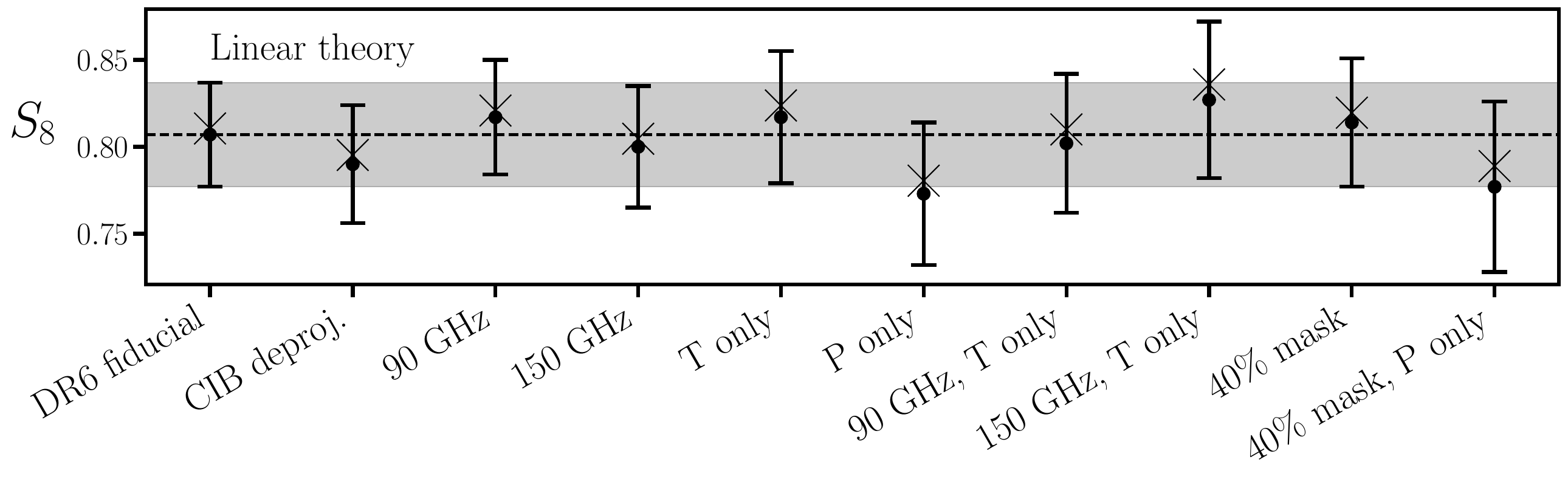}
    \caption{ 
    Linear theory DR6 $S_8$ constraints (without BAO) for alternative lensing reconstruction scenarios. 
    ``T and P only" refer to temperature and polarization only lensing reconstruction respectively; ``CIB deproj." is short for a CIB deprojected internal linear combination of the 90 and 150 GHz channels; while ``40\% mask" is short for a more restrictive Galactic mask that is applied to the primary CMB maps before performing the lensing reconstruction.
    Best-fit values are indicated by $\times$'s. 
    The shaded region corresponds to the $\pm1\sigma$ interval associated with our fiducial linear theory DR6 analysis. 
    We find statistically consistent results across different frequencies, temperature and polarization, different galactic masks, and combinations of the above.
    } 
\label{fig:dr6_alternates}
\end{figure}

The companion paper \cite{Kim2024} performs an extensive set of bandpower-level null tests for the cross-correlation with DR6. 
Here we present parameter-based tests (using linear theory for computational efficiency) for a subset of these checks.
To improve the value of these tests as a cross-check, we have independently measured (following \S\ref{sec:cl_estimation}) the cross-correlation with the \href{https://lambda.gsfc.nasa.gov/product/act/actadv_dr6_lensing_maps_info.html}{publicly available} alternative DR6 lensing maps. 
These results are summarized in Fig.~\ref{fig:dr6_alternates}. 
We find statistically consistent results across different frequencies, temperature and polarization, different galactic masks, and combinations of the above.

\section{Comparison with previous weak lensing analyses}
\label{sec:comparison}

\begin{figure}[!h]
    \centering
    \includegraphics[width=\linewidth]{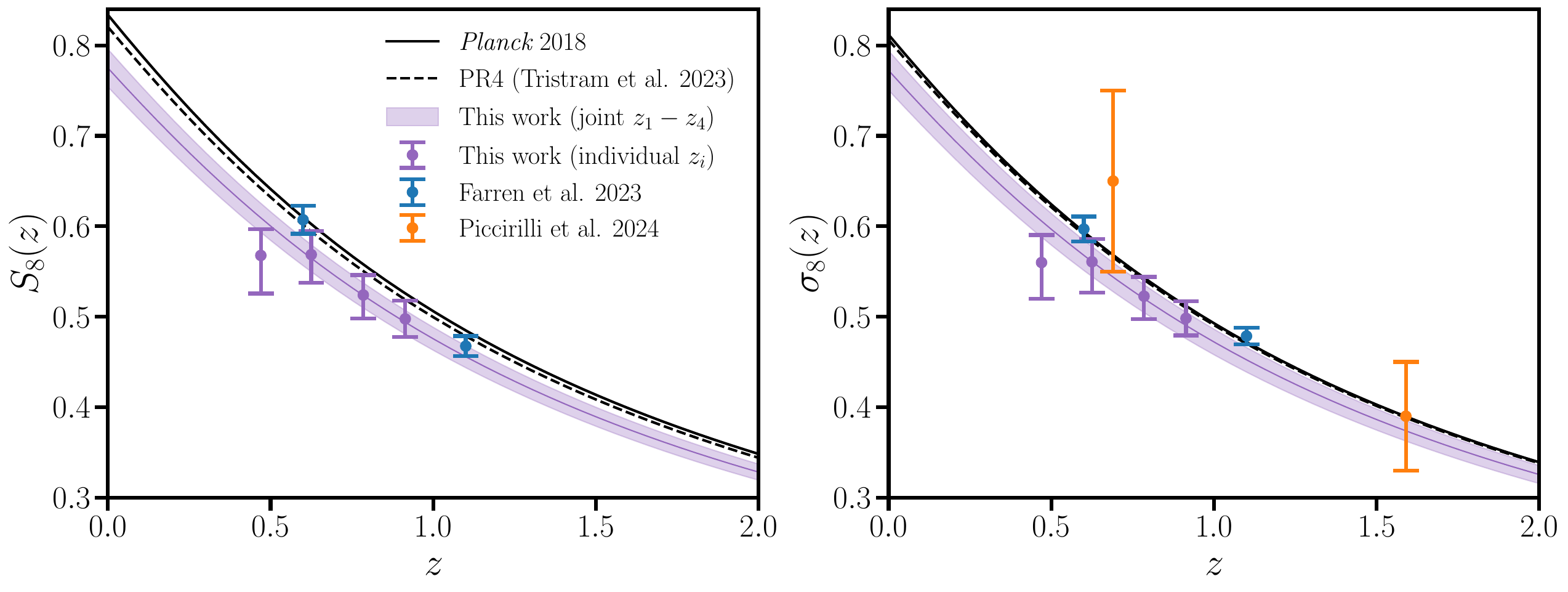}
    \caption{ 
    A comparison of our results with recent tomographic structure growth measurements obtained from cross-correlations of CMB lensing and galaxy positions. In blue we show the results obtained from the blue and green unWISE samples (correlated with PR4 and DR6) \cite{ACT:2023oei}, while in orange we show the results obtained from the lowest two redshift bins of the \textit{Quaia} quasar sample (correlated with PR4) \cite{Piccirilli:2024xgo}.
    Following \S\ref{sec:S8_correlations} we estimate 9, 13 and 28\% correlations between the $z_1-z_2$, $z_2-z_3$ and $z_3-z_4$ amplitude measurements respectively.
    } 
\label{fig:S8z_sigma8z}
\end{figure}

We compare our structure growth measurements with other recent tomographic constraints from correlations of CMB lensing and galaxy positions in Fig.~\ref{fig:S8z_sigma8z}.
In purple we show the evolution of $S_8$ (and $\sigma_8)$ inferred from our baseline PR4+DR6 analysis, where as in \S\ref{sec:alternative_models} we rescale the amplitude constraint from each redshift bin by the factor $\sigma_8(z)/\sigma_8(z=0)$ calculated with a \textit{Planck} 2018 $\Lambda$CDM cosmology. 
We also show the mean (purple line) and $\pm 1\sigma$ (shaded purple region) interval for the $\Lambda$CDM evolution of $S_8(z)\equiv \sigma_8(z)\sqrt{\Omega_m/0.3}$ and $\sigma_8(z)$ as inferred from the PPD of our fiducial PR4+DR6 analysis (including all four redshift bins).
In blue we show the $S_8(z)$ and $\sigma_8(z)$ constraints derived from the blue and green unWISE samples from a recent cross-correlation with the same CMB lensing maps used in this work (Farren et al. \cite{ACT:2023oei}), which have been rescaled to the mean redshifts 0.6 and 1.1 respectively.
In orange we show the $\sigma_8(z)$ constraints obtained from the cross-correlation of the PR4 $\kappa$ map with lowest two redshift bins of \textit{Quaia} quasars (Piccirilli et al. \cite{Piccirilli:2024xgo}). 

\begin{figure}[!h]
    \centering
    \includegraphics[width=\linewidth]{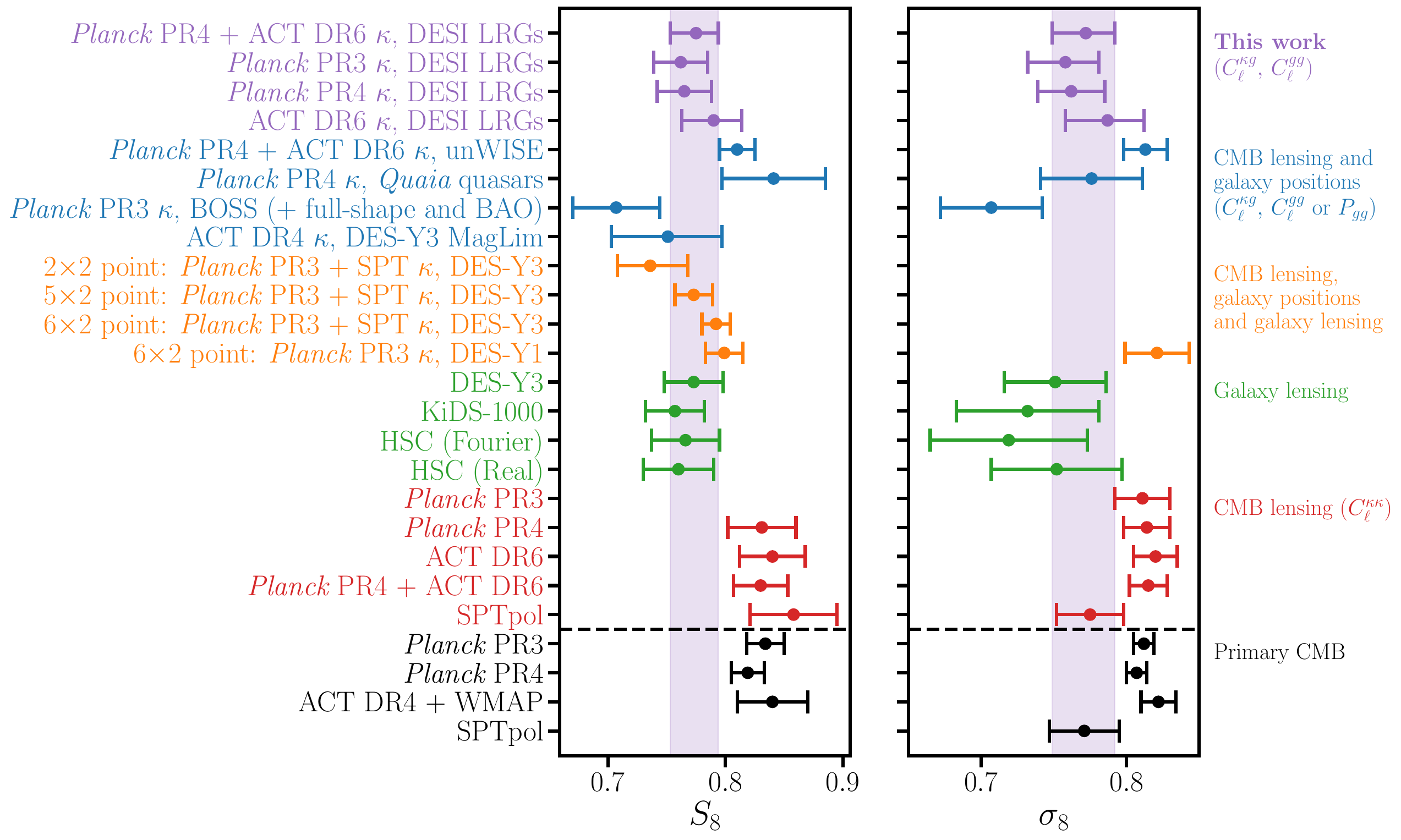}
    \caption{ 
    A compilation of $S_8$ and $\sigma_8$ constraints from cross-correlations of galaxy positions and CMB lensing (purple and blue); cross-correlations of CMB lensing with galaxy lensing and their positions (orange); galaxy lensing (green); the CMB lensing power spectrum (red) and the primary CMB (black).
    We differentiate ``early" and ``late" time measurements with the black dashed line.
    Constraints derived from this work are shown in purple, with the purple band representing the $\pm1\sigma$ contour of our fiducial PR4+DR6 analysis.
    See the text in \S\ref{sec:comparison} for a detailed discussion.
    } 
\label{fig:S8_sig8_compilation_vertical}
\end{figure}

We compare our combined constraints to other weak-lensing based measurements in Fig.~\ref{fig:S8_sig8_compilation_vertical}. As in \S\ref{sec:fit_good} we follow the standard parameter-based approach when quoting ``tensions" with other datasets. 
We quote tensions for the $S_8$ parameter by default, however, for analyses listed in Fig.~\ref{fig:S8_sig8_compilation_vertical} that do not report $S_8$ measurements we instead quote the tension for $\sigma_8$.
We caution the reader that this tension metric neglects correlations between datasets (typically resulting in an underestimation of the tension) and implicitly assumes that the posterior is Gaussian. Beyond this, the assumptions (including priors) made to model these datasets vary considerably with the observable and analyst. 

We first compare our results to those also derived from cross-correlations of CMB lensing and galaxy positions (blue points in Fig.~\ref{fig:S8_sig8_compilation_vertical}).
With the exception of White et al. \cite{White:2021yvw} (see \S\ref{sec:pr3}), our analysis is most directly comparable to Farren et al. \cite{ACT:2023oei}, which analyzes the same CMB lensing maps with a similar (in spirit) EFT-inspired model. From the combination of the blue and green unWISE samples ref.~\cite{ACT:2023oei} finds $S_8 = 0.810\pm 0.015$ (without BAO) which differs from our fiducial result by $1.4\sigma$ (cf. Eq.~\ref{eq:combined_s8}).
We note that our errorbar is $\simeq 37\%$ larger than that obtained from the unWISE sample. This difference is partially driven by the larger number density and broader redshift coverage of the unWISE galaxies, however, we also emphasize that ref.~\cite{ACT:2023oei} adopts highly informative priors on higher order corrections to the bias expansion (in particular, both counterterms were fixed to zero). 
In \S\ref{sec:param_based_consistency_tests} we consider a more ``aggressive analysis" whose scale cuts and priors were chosen to roughly match those used by ref.~\cite{ACT:2023oei}. With these choices we find that our errorbar is $\sim20\%$ larger than ref.~\cite{ACT:2023oei}'s, suggesting that this higher precision is driven just as much by the priors assumed as it is by the larger raw signal-to-noise of the unWISE dataset. 
We find a similar $\sim1.4\sigma$ difference with the cross-correlation of PR4 with \textit{Quaia} quasars \cite{Piccirilli:2024xgo}, which found $S_8 = 0.841 \pm 0.044$.
We find a slightly larger $\sim1.6\sigma$ difference with the $S_8$ measurement derived from cross-correlating PR3 with BOSS galaxies \cite{Chen:2022jzq} (including full-shape and post-reconstruction measurements), however, this analysis did not include a ``normalization" correction to the cross-correlation measurement and is susceptible to volume effects, which when taken into account would reduce the significance of this tension.
Our results are in very good $(\sim0.5\sigma)$ agreement with those obtained from cross-correlating DES-Y3 MagLim galaxies with the ACT DR4 lensing map \cite{ACT:2023ipp}.

Our results are in good agreement with those derived from cross-correlation with CMB lensing and galaxy lensing, with differences ranging from $\sim 0.1-1.1\sigma$. The orange points in Fig.~\ref{fig:S8_sig8_compilation_vertical} show the constraints obtained from the ``$2\times2$ point" analysis \cite{DES:2022xxr} of correlations between PR3+SPT CMB lensing with DES-Y3 galaxy positions and lensing ($\langle \delta_g \kappa\rangle + \langle \gamma \kappa \rangle$), constraints derived from the ``$5\times2$ point" analysis that adds correlations between galaxy lensing and their positions (including each auto-correlation), and the ``$6\times2$ point" analysis that in addition to the above adds the CMB lensing auto-correlation \cite{DES:2022urg}.
We also show constraints from an alternative $6\times2$ point analysis using PR3 and DES-Y1 data that marginalizes over baryonic feedback uncertainties using a principle component-based approach \cite{Xu:2023qmp}, which is in good agreement with the DES-Y3 result.

Our results are in even better agreement (within $\sim0.6\sigma$) with those obtained from galaxy lensing alone. The green points in Fig.~\ref{fig:S8_sig8_compilation_vertical} show the constraints obtained from a reanalysis of DES-Y3 \cite{DES:2021wwk}, KiDS-1000 \cite{Heymans:2020gsg} and HSC (both in Fourier \cite{Dalal:2023olq} and real \cite{Li:2023tui} space) data when adopting the priors on cosmological parameters used in the recent ACT DR6 auto-correlation analysis \cite{ACT:2023kun}. The constraints plotted in Fig.~\ref{fig:S8_sig8_compilation_vertical} are taken from Table 2 of ref.~\cite{ACT:2023kun}.

Our results are consistent to within 1.4, 1.6, 1.9, 1.8, and 2.0$\sigma$ with those obtained from $\textit{Planck}$ PR3 \cite{Planck:2018lbu}, PR4 \cite{Carron:2022eyg}, ACT DR6 \cite{ACT:2023kun}, PR4+DR6 and SPTpol \cite{SPT:2019fqo,SPT:2023jql} analyses of the CMB lensing power spectrum respectively, indicated by the red points in Fig.~\ref{fig:S8_sig8_compilation_vertical}.
We emphasize that the CMB lensing power spectrum probes different redshifts (with the signal peaking around $z\sim2$) and scales than our tomographic analysis, leaving open the possibility that non-standard structure growth for $z<1$ could consistently reconcile these (mildly significant) differences.
We defer the reader to \S\ref{sec:fit_good} for a discussion of the consistency of our results with \textit{Planck} primary CMB measurements, which are also shown in Fig.~\ref{fig:S8_sig8_compilation_vertical} (black points). We find a similar level of agreement ($\sim1.8\sigma$) with primary CMB measurements from ACT DR4 + WMAP \cite{ACT:2020gnv}, and even better agreement with measurements from SPTpol \cite{SPT:2017jdf}.

\section{Discussion and conclusions}
\label{sec:discussion_and_conclusions}

We infer the amplitude of late time large scale structure from the cross-correlation of DESI legacy survey LRGs with the latest CMB lensing maps from \textit{Planck} and ACT.
The code base used to perform this analysis is publicly available (\verb|MaPar| \href{https://github.com/NoahSailer/MaPar/tree/main}{\faGithub}) and discussed in more detail in Appendix~\ref{sec:mapar}.
From the joint analysis of all four LRG redshift bins we obtain a 2.6\% constraint of $S_8 = 0.775^{+0.019}_{-0.022}\,\, [0.774]$, while with the addition of BAO data we obtain a 2.7\% constraint of $\sigma_8 = 0.772^{+0.020}_{-0.023}\,\, [0.775]$ 
(Eqs.~\ref{eq:combined_s8} and \ref{eq:combined_sigma8}). 
These results are less discrepant with primary CMB measurements than was found in a previous LRG cross-correlation analysis \cite{White:2021yvw} (see \S\ref{sec:pr3} for a discussion).
Our $S_8$ measurement is 5\% (7\%) lower than obtained from \textit{Planck} PR4 (PR3) primary CMB data with a statistical significance ranging from $1.8-2.3\sigma$ (\S\ref{sec:fit_good}). 
Our results are consistent with the decade-long claims of a low $S_8$ from weak galaxy shear measurements, however, unlike for shear our data's preference for a low $S_8$ cannot be explained by baryonic feedback \cite{Amon:2022azi} as our signal is derived primarily from the linear regime.
When jointly analyzed with primary CMB measurements, our data would have a mild preference for beyond-$\Lambda$CDM physics (see e.g. \cite{Abdalla:2022yfr}) that suppresses structure evolution on linear scales and at late times.

Through individually analyzing four photometric-redshift bins we constrain the evolution of structure growth across $0.4 \lesssim z \lesssim 1$. As indicated by Fig.~\ref{fig:S8z_sigma8z} we find a rate of growth that is largely consistent with a $\Lambda$CDM prediction conditioned on primary CMB observations with an overall normalization that is $\simeq 5-7\%$ lower. 
As was found in White et al. \cite{White:2021yvw}, the constraints derived from the lowest redshift bin are most discrepant with a \textit{Planck} $\Lambda$CDM prediction. 
The lowest redshift bin is also the least statistically constraining, and as such it is inconclusive if this deviation is the result of a statistical fluctuation or a preference for a slower rate of structure growth for $z\lesssim 0.5$. 
These scenarios could potentially be distinguished through the cross-correlation of CMB lensing with lower-redshift samples (albeit with modest precision due to the limited dynamic range of linear scales and limited overlap with the CMB lensing kernel at these redshifts), such as the DESI Bright Galaxy Survey \cite{Hahn:2022dnf}, which we leave to future work.

In our fiducial analysis we adopt a Hybrid EFT (HEFT) model (\S\ref{sec:heft}) that approximates the early-time galaxy distribution as a perturbative bias expansion that is advected to the present day using the non-linear CDM displacements measured from the \verb|Aemulus| $\nu$ simulations \cite{DeRose:2023dmk}. 
In \S\ref{sec:buzzard_fits} we verify that our likelihood recovers unbiased cosmological results from a set of realistic mock measurements for both our fiducial HEFT model and alternatives presented in \S\ref{sec:alternative_models}.
We marginalize over both quadratic and lowest-order derivative corrections to the bias expansion with relatively wide priors. 
Doing so robustly regulates the cosmological information that can be obtained from smaller scales, such that our fiducial cosmological constraints are derived primarily from the linear regime ($k < 0.1\,h\,{\rm Mpc}^{-1}$).
We explicitly demonstrate this point by performing a linear theory analysis (\S\ref{sec:alternative_models}) whose scale cuts were chosen such that only modes with $k < 0.1\,h\,{\rm Mpc}^{-1}$ efficiently contribute to the signal, and find excellent agreement with our fiducial results with a comparable statistical precision. 

Our data are largely immune to systematic uncertainties faced by photometric surveys. CMB lensing has a well-characterized source distribution and is measured using very well-understood statistical properties of primordial fluctuations, while the redshift distributions of the four LRG photometric redshift bins were spectroscopically calibrated by ref.~\cite{Zhou:2023gji}.
To further ensure the robustness of our measurements we performed an extensive set of systematics, null and parameter-based consistency tests. In a companion paper \cite{Kim2024} we present a suite of bandpower-level null tests for the ACT cross-correlation measurement and confirm with simulations that the expected bias from extragalactic foregrounds is negligible. 
In this work we thoroughly inspect the uniformity of the LRG sample's physical properties by measuring their power spectra on different footprints (see sections \ref{sec:imaging_footprint_variations} to \ref{sec:dec_pm15}), confirm that our results are robust to stricter cuts on extinction and stellar contamination (\S\ref{sec:strict_ebv_star}) and quantify the (negligible) bias arising from residual correlations between the systematic weights and (systematics-corrected) LRG maps (\S\ref{sec:sysweights}).
We verify that our results are robust to variations in the scale cuts and priors assumed in \S\ref{sec:param_based_consistency_tests}.
While the vast majority of these tests find no significant evidence for systematic contamination, there are two notable exceptions: (1) we find statistically significant variations in the high-$z$ LRG auto-correlations on the DES footprint vs elsewhere (\S\ref{sec:imaging_footprint_variations}) and (2) the measured galaxy cross-spectra (\S\ref{sec:gxspec}) of neighboring redshift bins are slightly smaller than predicted from fits to the galaxy auto- and cross-correlation with CMB lensing.
In passing, we detect a non-zero magnification signal by cross-correlating the lowest and highest LRG bins (which have negligible overlap in their redshift distributions) at $8\sigma$ significance.
In \S\ref{sec:param_based_consistency_tests} we perform a set of parameter-based consistency tests which suggest that neither of these potential sources of systematic contamination significantly bias our cosmological results. 
We encourage future work to further investigate these findings to improve the fidelity of the LRG sample.

In the near future DESI will obtain redshifts for the majority of LRGs used in this work, improving both the calibration of the redshift distribution and enabling more accurate treatments of systematic contaminants.
These data will also be vital to improving the systematic characterization (e.g. through indirect redshift distribution calibration) of other photometric samples, which in particular may prove useful for diagnosing the (mild) discrepancy with the latest unWISE analysis \cite{ACT:2023oei}.
In addition, these data will enable ``full-shape" analyses of the 3D galaxy-power spectrum that can (somewhat) independently constrain $\sigma_8(z)$ from redshift-space distortions.
Full-shape analyses will serve as a valuable consistency check of our results, and can be used to assess the dynamical consistency of relativistic and non-relativistic tracers within General Relativity when combined with CMB lensing cross-correlations.
A joint analysis including LRG full-shape and BAO data will be presented in future work.

Future galaxy surveys such as Rubin Observatory’s Legacy Survey of Space and Time (LSST) \cite{2019ApJ...873..111I}, Roman \cite{Roman}, Euclid \cite{EUCLID18} and other next-generation imaging and spectroscopic surveys will further refine these cross-correlation measurements due to their greater number densities and wider sky areas. At the same time, these cross-correlations will benefit significantly from lower noise CMB lensing maps produced from upcoming high-resolution telescopes such as Simons Observatory \cite{2019JCAP...02..056A} and CMB-S4 \cite{CMBS4}. The methods presented here can be readily extended to these datasets, enabling even more precise constraints on low redshift structure growth while maintaining an acceptable level of accuracy.

\acknowledgments

We would like to thank 
Alex Krowleski,
Ant\'{o}n Baleato Lizancos,
Niall MacCrann,
Bruce Partridge,
John Peacock, 
David Spergel,
and Noah Weaverdyck
for helpful discussions during the preparation of this manuscript.
NS is supported by the Office of Science Graduate Student Research (SCGSR) program administered by the Oak Ridge Institute for Science and Education for the DOE under contract number DE‐SC0014664. 
JK acknowledges support from NSF grants AST-2307727 and AST-2153201.
SF is supported by Lawrence Berkeley National Laboratory and the Director, Office of Science, Office of High Energy Physics of the U.S. Department of Energy under Contract No.\ DE-AC02-05CH11231.
MM acknowledges support from NSF grants AST-2307727 and  AST-2153201 and NASA grant 21-ATP21-0145. 

IAC acknowledges support from Fundaci\'on Mauricio y Carlota Botton and the Cambridge International Trust.
EC acknowledges support from the European Research Council (ERC) under the European Union’s Horizon 2020 research and innovation programme (Grant agreement No. 849169).
JD acknowledges support from NSF award AST-2108126.
CEV received the support of a fellowship from “la Caixa” Foundation (ID 100010434). The fellowship code is LCF/BQ/EU22/11930099.
GSF acknowledges support through the Isaac Newton Studentship and the Helen Stone Scholarship at the University of Cambridge. GSF furthermore acknowledges support from the European Research Council (ERC) under the European Union’s Horizon 2020 research and innovation program (Grant agreement No. 851274).
KM acknowledges support from the National Research Foundation of South Africa.
CS acknowledges support from the Agencia Nacional de Investigaci\'on y Desarrollo (ANID) through Basal project FB210003.
This research has made use of NASA's Astrophysics Data System and the arXiv preprint server.

This material is based upon work supported by the U.S. Department of Energy (DOE), Office of Science, Office of High-Energy Physics, under Contract No. DE-AC02-05CH11231, and by the National Energy Research Scientific Computing Center, a DOE Office of Science User Facility under the same contract. Additional support for DESI was provided by the U.S. National Science Foundation (NSF), Division of Astronomical Sciences under Contract No. AST-0950945 to the NSF's National Optical-Infrared Astronomy Research Laboratory; the Science and Technologies Facilities Council of the United Kingdom; the Gordon and Betty Moore Foundation; the Heising-Simons Foundation; the French Alternative Energies and Atomic Energy Commission (CEA); the National Council of Science and Technology of Mexico (CONACYT); the Ministry of Science and Innovation of Spain (MICINN), and by the DESI Member Institutions: \url{https://www.desi.lbl.gov/collaborating-institutions}.

The DESI Legacy Imaging Surveys consist of three individual and complementary projects: the Dark Energy Camera Legacy Survey (DECaLS), the Beijing-Arizona Sky Survey (BASS), and the Mayall $z$-band Legacy Survey (MzLS). DECaLS, BASS and MzLS together include data obtained, respectively, at the Blanco telescope, Cerro Tololo Inter-American Observatory, NSF's NOIRLab; the Bok telescope, Steward Observatory, University of Arizona; and the Mayall telescope, Kitt Peak National Observatory, NOIRLab. NOIRLab is operated by the Association of Universities for Research in Astronomy (AURA) under a cooperative agreement with the National Science Foundation. Pipeline processing and analyses of the data were supported by NOIRLab and the Lawrence Berkeley National Laboratory. Legacy Surveys also uses data products from the Near-Earth Object Wide-field Infrared Survey Explorer (NEOWISE), a project of the Jet Propulsion Laboratory/California Institute of Technology, funded by the National Aeronautics and Space Administration. Legacy Surveys was supported by: the Director, Office of Science, Office of High Energy Physics of the U.S. Department of Energy; the National Energy Research Scientific Computing Center, a DOE Office of Science User Facility; the U.S. National Science Foundation, Division of Astronomical Sciences; the National Astronomical Observatories of China, the Chinese Academy of Sciences and the Chinese National Natural Science Foundation. LBNL is managed by the Regents of the University of California under contract to the U.S. Department of Energy. The complete acknowledgments can be found at \url{https://www.legacysurvey.org/}.

Any opinions, findings, and conclusions or recommendations expressed in this material are those of the author(s) and do not necessarily reflect the views of the U. S. National Science Foundation, the U. S. Department of Energy, or any of the listed funding agencies.

The authors are honored to be permitted to conduct scientific research on Iolkam Du'ag (Kitt Peak), a mountain with particular significance to the Tohono O'odham Nation.

Support for ACT was through the U.S.~National Science Foundation through awards AST-0408698, AST-0965625, and AST-1440226 for the ACT project, as well as awards PHY-0355328, PHY-0855887 and PHY-1214379. Funding was also provided by Princeton University, the University of Pennsylvania, and a Canada Foundation for Innovation (CFI) award to UBC. ACT operated in the Parque Astron\'omico Atacama in northern Chile under the auspices of the Agencia Nacional de Investigaci\'on y Desarrollo (ANID). The development of multichroic detectors and lenses was supported by NASA grants NNX13AE56G and NNX14AB58G. Detector research at NIST was supported by the NIST Innovations in Measurement Science program. Computing for ACT was performed using the Princeton Research Computing resources at Princeton University, the National Energy Research Scientific Computing Center (NERSC), and the Niagara supercomputer at the SciNet HPC Consortium. SciNet is funded by the CFI under the auspices of Compute Canada, the Government of Ontario, the Ontario Research Fund–Research Excellence, and the University of Toronto. We thank the Republic of Chile for hosting ACT in the northern Atacama, and the local indigenous Licanantay communities whom we follow in observing and learning from the night sky.

\appendix

\section{Data and code availability}
\label{sec:mapar}

The code base used to estimate ancillary power spectra and covariances, perform mock tests and cosmological analyses is publicly available (\verb|MaPar| \href{https://github.com/NoahSailer/MaPar/tree/main}{\faGithub}).
The fiducial bandpowers, window functions and covariance used in our fiducial analysis are available at this \href{https://zenodo.org/records/12613408}{URL}, along with the chains and best-fit prediction for a subset of the analyses presented here.
The \verb|Aemulus| emulator weights will be made available upon the publication of an upcoming DES-DESI cross-correlation analysis \cite{DeRose24b}.
The configuration for our baseline analysis is given by \verb|fiducial_noBAO_pr4-dr6.yaml|, located in the \verb|MaPar/yamls/| directory along with several other alternative analysis configurations.
The fiducial redshift distributions are located in the \verb|MaPar/data/dNdzs/| directory, while the fiducial power spectra used for covariance estimation can be found in \verb|MaPar/spectra/fiducial/|. See the \verb|README| for further information.

\section{Signal to noise}
\label{sec:snr}

For a power spectrum $C^{ab}_\ell$ with measurement error $\sigma(C^{ab}_\ell)$ the signal-to-noise ratio can be expressed as 
\begin{equation}
\begin{aligned}
\label{eq:snr}
    {\rm SNR}^2 
    &=
    \sum_\ell 
    \frac{(C^{ab}_\ell)^2}{\sigma^2(C^{ab}_\ell)}
    \\
    &=
    \int dz \,d\ln k \left[\frac{W^a(\chi) W^b(\chi)}{\chi^2 H(z)} kP_{ab}(k)   \sum_\ell \frac{1}{\ell\,\sigma^2(C^{ab}_\ell)} W^a(\ell k^{-1}) \, W^b(\ell k^{-1}) \, P_{ab}(\ell\chi^{-1}) \right].
\end{aligned}
\end{equation}
where we used
\begin{equation}
    C^{ab}_\ell
    \simeq
    \int d\chi \frac{W^a(\chi) W^b(\chi)}{\chi^2} P_{ab}(\ell\chi^{-1})
    =
    \ell^{-1}\int dk \, W^a(k^{-1}l) \, W^b(k^{-1}l) \, P_{ab}(k)
\end{equation}
to go from the first to the second line.
In Fig.~\ref{fig:snr_per_dzdk} we plot ${\rm SNR}(k,z)$, which is defined as the square root of the quantity in brackets in Eq.~\eqref{eq:snr}.

\section{Parameter-based comparison of analysis choices}
\label{sec:joshuas_vs_noah_measurements}

Our methodology for estimating ancillary power spectra and covariances is summarized in \S\ref{sec:cl_estimation}. There are several differences between the methods adopted here and in the companion paper: (1) the companion paper \cite{Kim2024} adopts a slightly different binning for high-$\ell$ bandpowers ($\ell > 971$) that has minor impacts on the numerical computation of the window functions and bandpowers for $\ell<600$, (2) the covariance matrices computed in \cite{Kim2024} are either estimated numerically or estimated as an analytic-numerical hybrid, whereas the ancillary covariance matrices used in this work are all estimated analytically, (3) we always remask the CMB lensing map when computing pseudo-$C_\ell$'s whereas \cite{Kim2024} remasks the PR4 map but not the DR6 map, and (4) we adopt a slightly different approach for computing the ``normalization" correction, which we now describe. We refer the reader to the companion paper \cite{Kim2024} for a description of the transfer function for computing this correction.

\begin{figure}[!h]
    \centering
    \includegraphics[width=\linewidth,valign=c]{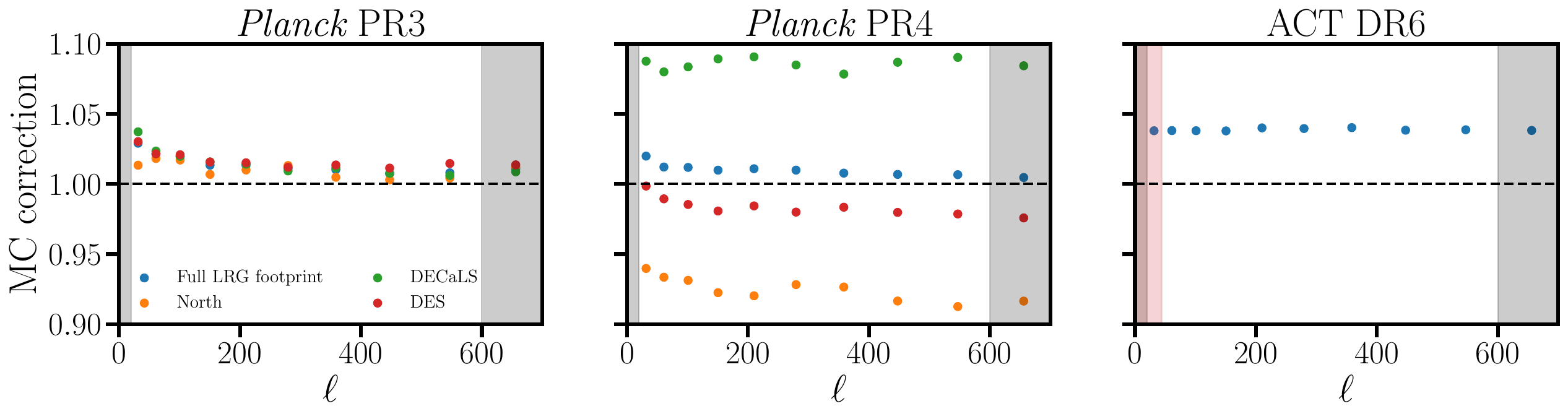}
    \caption{ 
    Monte-Carlo normalization corrections for the baseline PR3 (left), PR4 (middle) and DR6 (right) CMB lensing reconstructions for different LRG masks. For PR3 and PR4 we show the normalization corrections for the full LRG footprint (blue) and the intersection of the LRG footprint with the Northern (orange), DECaLS (green) and DES (red) imaging regions. 
    Note in particular the large $\mathcal{O}(10\%)$ variations for PR4, which arise from the optimal anisotropic filtering adopted by \cite{Carron:2022eyg}.
    } 
\label{fig:mccorr}
\end{figure}

Mode-couplings arising from e.g. masking or anisotropic filtering of CMB maps are not properly forward modeled in CMB lensing cross-correlations measurements (which are integrals of squeezed limit bispectra) by the MASTER algorithm. To approximately account for these effects, the current state-of-the-art (see \cite{Sailer2024} for an improved approximation) is to multiply a CMB lensing cross-correlation measurement by a Monte-Carlo ``normalization correction" estimated from a set of simulated CMB lensing reconstructions, which we compute as
\begin{equation}
    (\text{MC correction})_L
    =
    \frac{
    \sum_{\ell} W_{L\ell}
    \sum_{i=1}^{N_\text{sim}}
    \sum_{m=-\ell}^\ell
    \{M^\kappa \kappa^i\}_{\ell m}\{M^g \kappa^i\}^*_{\ell m}
    }{
    \sum_{\ell} W_{L\ell}
    \sum_{i=1}^{N_\text{sim}}
    \sum_{m=-\ell}^\ell
    \{M^\kappa \hat{\kappa}^i\}_{\ell m}\{M^g \kappa^i\}^*_{\ell m}
    }
\end{equation}
where $M^\kappa$ ($M^g$) is the CMB lensing (galaxy) mask, $\hat{\kappa}^i$ ($\kappa^i$) is the reconstructed (input) CMB lensing convergence from the $i$'th CMB lensing reconstruction simulation, $\{AB\}_{\ell m} \equiv \int d^2\hat{\bm{n}}Y^*_{\ell m}(\hat{\bm{n}}) A(\hat{\bm{n}}) B(\hat{\bm{n}})$, and $W_{L\ell}$ is the window function corresponding to the bandpower $L$. We show the normalization corrections used in our ancillary cross-correlation measurements in Fig.~\ref{fig:mccorr}. 

\begin{figure}[!h]
    \centering
    \includegraphics[width=0.5\linewidth,valign=c]{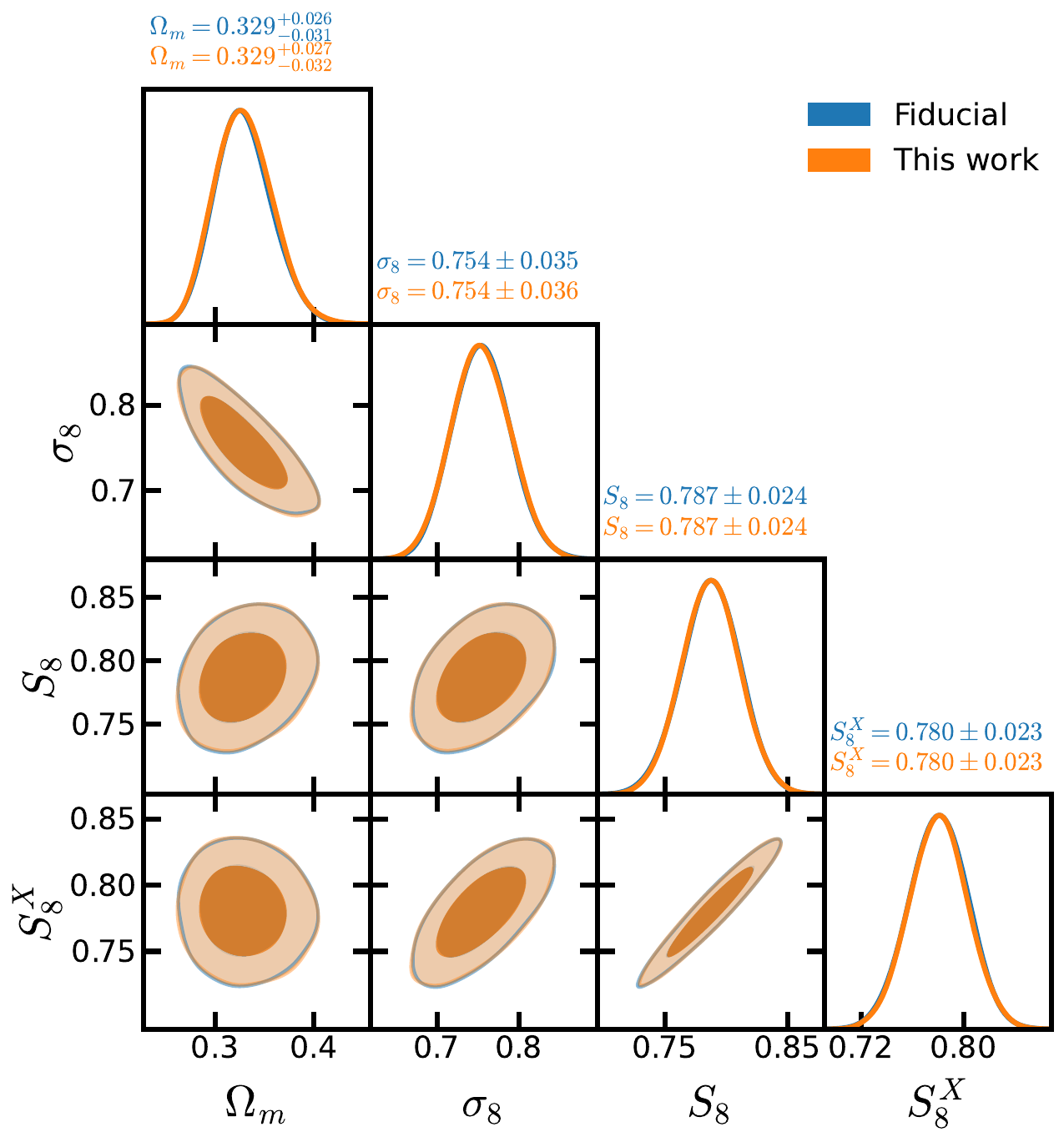}
    \caption{ 
    Linear theory constraints from PR4+DR6 using our fiducial measurements, window functions, and hybrid covariance vs those obtained following \S\ref{sec:cl_estimation} and Appendix \ref{sec:joshuas_vs_noah_measurements}.
    } 
\label{fig:joshua_vs_noah}
\end{figure}

The ancillary power spectra and covariances are used in the main text to perform systematics checks and to reanalyze the cross-correlation with \textit{Planck} PR3, while for our fiducial analysis we use the power spectra measurements and hybrid covariance from the companion paper \cite{Kim2024}. 
As a cross-check, we independently measured the galaxy auto-correlations, the PR4 and DR6 cross-correlations and estimated their associated covariances following \S\ref{sec:cl_estimation} and the discussion above. 
In Fig.~\ref{fig:joshua_vs_noah} we verify at the parameter-level the consistency of these measurements for the linear theory PR4+DR6 analysis. In particular, we find find that the $S_8$ constraints are identical to three decimal places.

\section{Analytic marginalization and Bayesian PTEs}
\label{sec:anamarg}

\textbf{Analytic marginalization:} Suppose that our theory prediction $\bm{t}(\bm{\theta},\bm{\phi}) \equiv \bm{A}(\bm{\theta}) + \sum_i\phi_i \bm{B}_i(\bm{\theta})$ depends generically on a set of parameters $\bm{\theta}$ and linearly on a set of parameters $\bm{\phi}$. We assume that the likelihood and priors on linear parameters are Gaussian, in which case the posterior is
\begin{equation}
\label{eq:post}
    P(\bm{\theta},\bm{\phi}|\bm{d})
    \propto
    e^{-\frac{1}{2}(\bm{t}-\bm{d})^T\bm{C}_d^{-1}(\bm{t}-\bm{d})}
    \,\,
    e^{-\frac{1}{2}(\bm{\phi}-\bm{\mu})^T\bm{C}_\phi^{-1}(\bm{\phi}-\bm{\mu})}
    \,\,
    P(\bm{\theta})
    ,
\end{equation}
where $\bm{d}$ is the data, $\bm{C}_d$ is the data covariance and $\bm{\mu}$ ($\bm{C}_\phi$) is the mean (covariance) of the prior on the linear parameters. Defining $\bm{\delta} \equiv \bm{A}-\bm{d}$, $(\bm{M}^{-1})_{ij} \equiv \bm{B}_i^T \bm{C}^{-1}_d\bm{B}_j + (\bm{C}^{-1}_\phi)_{ij}$, $V_i \equiv \bm{B}^T_i \bm{C}^{-1}_d \bm{\delta} - (\bm{C}_\phi^{-1} \bm{\mu})_i$ and $\bar{\bm{\phi}} \equiv -\bm{M}\bm{V}$, we may re-express the posterior as
\begin{equation}
\label{eq:simple_post}
    P(\bm{\theta},\bm{\phi}|\bm{d}) 
    \propto
    e^{-\frac{1}{2}\bm{\delta}^T \bm{C}^{-1}_d \bm{\delta}}
    \,\,
    e^{\frac{1}{2} \bm{V}^{T} \bm{M}\bm{V}}
    \,\,
    e^{- \frac{1}{2} \big(\bm{\phi}-\bar{\bm{\phi}}\big)^T \bm{M}^{-1} \big(\bm{\phi}-\bar{\bm{\phi}}\big)}
    \,\,
    P(\bm{\theta}),
\end{equation}
where we have absorbed all terms independent of $(\bm{\theta},\bm{\phi})$ into a proportionality constant. We perform the (Gaussian) integral over the linear parameters analytically to get the marginal posterior
\begin{equation}
\label{eq:marg_post}
    P_m(\bm{\theta}|\bm{d})
    \equiv
    \int d\bm{\phi}\,\,P(\bm{\theta},\bm{\phi}|\bm{d})
    \propto
    e^{-\frac{1}{2}\bm{\delta}^T \bm{C}^{-1}_d \bm{\delta}}
    \,\,
    e^{\frac{1}{2} \bm{V}^{T} \bm{M}\bm{V}}
    \,\,
    \sqrt{{\rm det}(\bm{M})}
    \,\,
    P(\bm{\theta}).
\end{equation}

\begin{figure}[!h]
    \centering
    \includegraphics[width=\linewidth,valign=c]{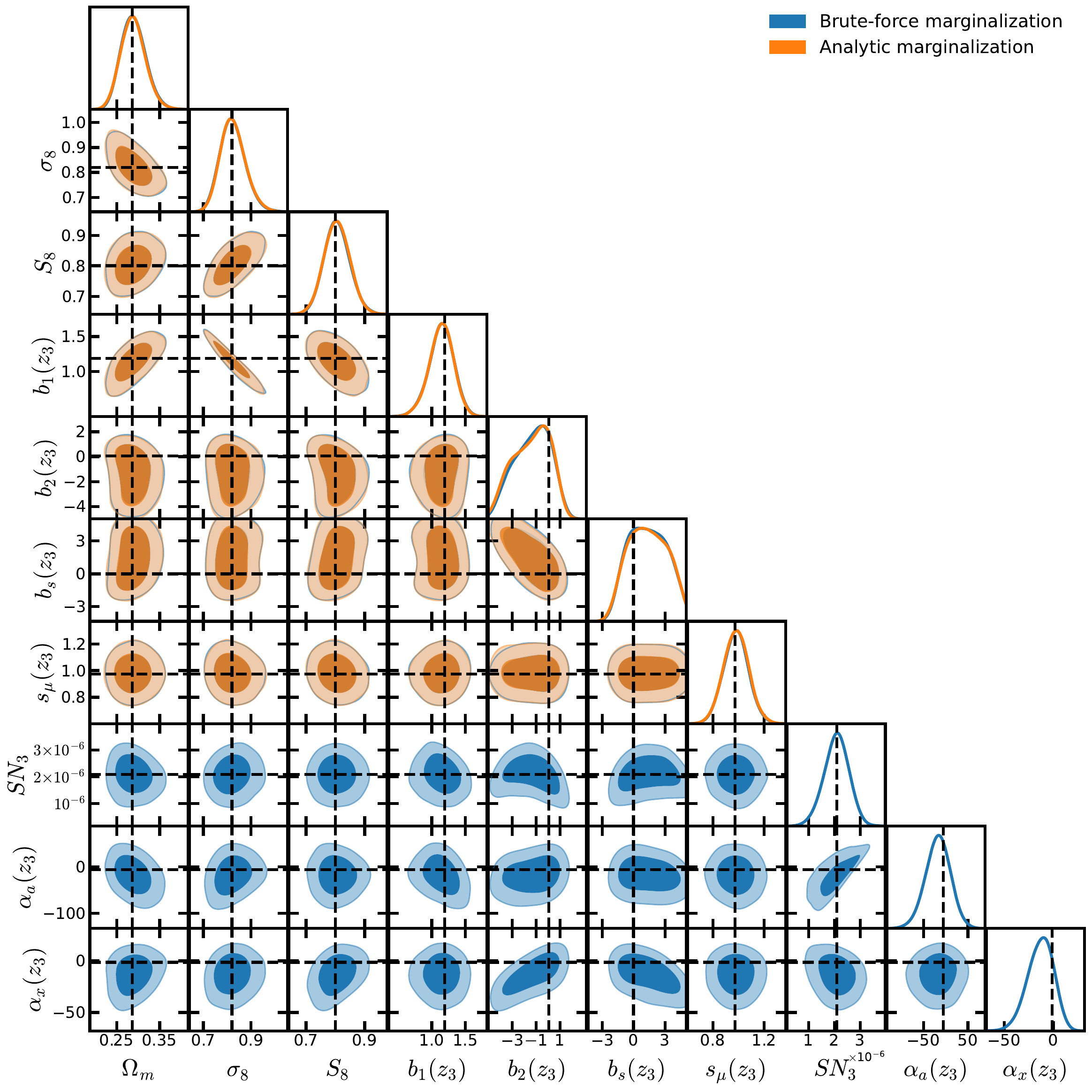}
    \caption{ 
    A comparison of analytic (blue) and brute-force marginalization (orange) over shot noise and counterterms. The true values are indicated by black dashed lines. We fit to the $z_3$ model prediction (see \S\ref{sec:verification_and_volume}) using our fiducial HEFT model and a PR4-like covariance. In this figure we have adopted wider priors for $b_s$ and $\alpha_x$ than used in our fiducial analysis (see text of \S\ref{sec:anamarg}).
    } 
\label{fig:anamarg}
\end{figure}

The linear parameters in our fiducial analysis include shot noise and counterterms, which we marginalize over analytically using Eq.~\eqref{eq:marg_post}. When maximizing the posterior we set the linear parameters to their best fit values $(\bm{\phi}_{\rm best-fit} = \bar{\bm{\phi}} = -\bm{M}V)$ for each set of generic (non-linear) parameters $\bm{\theta}$. As an explicit check of our numerical implementation of Eq.~\eqref{eq:marg_post} we consider fitting to the $z_3$ model prediction (see \S\ref{sec:verification_and_volume}) with and without analytic marginalization, where for the latter we directly sample shot noise and counterterms. We adopt the fiducial HEFT model (\S\ref{sec:heft}) with a PR4-like covariance using the priors listed in Table~\ref{tab:priors} with the following exceptions: we place a uniform $\mathcal{U}(-5,5)$ prior on $b_s$ and sample $\alpha_x$ directly (instead of $\epsilon$) with a wide $\mathcal{N}(0,50)$ prior. These results are summarized in Fig.~\ref{fig:anamarg}. We find excellent agreement between the posteriors of non-linear parameters.

\textbf{Bayesian PTE:} The Bayesian probability to exceed (PTE) is defined as the probability that a random data realization $(\bm{d}^{\rm rand})$ will have a larger average\footnote{Bayesian PTEs can be defined for any test-statistic (e.g. $\chi^4$). We take the test statistic to be the ``standard" $\chi^2$.} $\chi^2$ than the observed data $(\bm{d})$ \cite{reason:GelCarSteRub95}:
\begin{equation}
\begin{aligned}
    {\rm PTE}
    &\equiv
    {\rm Prob}\big(
    \chi^2(\bm{\theta},\bm{\phi},\bm{d}^{\rm rand})
    \geq 
    \chi^2(\bm{\theta},\bm{\phi},\bm{d})
    \,|\,
    \bm{d}
    \big)
    \\
    &=
    \int 
    d\bm{\theta}
    \,\,
    d\bm{\phi}
    \,\,
    d\bm{d}^{\rm rand}
    \,\,
    H\big(\chi^2(\bm{\theta},\bm{\phi},\bm{d}^{\rm rand})
    - 
    \chi^2(\bm{\theta},\bm{\phi},\bm{d})\big)
    P(\bm{\theta},\bm{\phi},\bm{d}^{\rm rand}|\bm{d})
    \\
    &
    {\rm with}
    \,\,\,\,
    \chi^2(\bm{\theta},\bm{\phi},\bm{d}) = \big(\bm{t}(\bm{\theta},\bm{\phi})-\bm{d}\big)^T\bm{C}^{-1}_d(\bm{t}\big(\bm{\theta},\bm{\phi})-\bm{d}\big)
\end{aligned}
\end{equation}
where $H(x)$ is the Heaviside step function and $P(\bm{\theta},\bm{\phi},\bm{d}^{\rm rand}|\bm{d}) = P(\bm{d}^{\rm rand}|\bm{\theta},\bm{\phi})P(\bm{\theta},\bm{\phi}|\bm{d}) $ is the joint distribution of the random data realization and model parameters $(\bm{\theta},\bm{\phi})$ given a measurement of the data $(\bm{d})$.

Given that the likelihood $P(\bm{d}^{\rm rand}|\bm{\theta},\bm{\phi})$ is a Gaussian with mean $\bm{t}(\bm{\theta},\bm{\phi})$ and (presumed known) covariance $\bm{C}_d$, the quantity $\chi^2(\bm{\theta},\bm{\phi},\bm{d}^{\rm rand})$ is drawn from a $\chi^2$ distribution with $N_d$ (number of data points) degrees of freedom for a fixed set of $(\bm{\theta},\bm{\phi})$. The probability then that this quantity exceeds the measured value $\chi^2(\bm{\theta},\bm{\phi},\bm{d})$ is given by the PTE ($=1-{\rm CDF}$) of a $\chi^2$ distribution with $N_d$ degrees of freedom:
\begin{equation}
\label{eq:pteb}
    \int 
    d\bm{d}^{\rm rand}
    \,\,
    H\big(\chi^2(\bm{\theta},\bm{\phi},\bm{d}^{\rm rand})
    - 
    \chi^2(\bm{\theta},\bm{\phi},\bm{d})\big)
    P(\bm{d}^{\rm rand}|\bm{\theta},\bm{\phi})
    =
    1
    -
    \frac{1}{\Gamma(N_d/2)}
    \gamma
    \left(
    \frac{N_d}{2},
    \frac{\chi^2(\bm{\theta},\bm{\phi},\bm{d})}{2}
    \right)
\end{equation}
where $\gamma(s,x) = \int_0^x t^{s-1} e^{-t} dt$ is the incomplete gamma function and $\Gamma(s) = \lim_{x\to\infty} \gamma(s,x)$. 

Plugging this result in to the definition of the Bayesian PTE we have:
\begin{equation}
\begin{aligned}
    {\rm PTE}
    &=
    \int 
    d\bm{\theta}
    \,\,
    d\bm{\phi}
    \left[
    1
    -
    \frac{1}{\Gamma(N_d/2)}
    \gamma
    \left(
    \frac{N_d}{2},
    \frac{\chi^2(\bm{\theta},\bm{\phi},\bm{d})}{2}
    \right)
    \right]
    P(\bm{\theta},\bm{\phi}|\bm{d}).
\end{aligned}
\end{equation}
The remaining integrals over the model parameters are performed numerically. Analytic marginalization complicates the numerical integration slightly. For each step in the chain (fixed $\bm{\theta}$) we integrate over the linear parameters $\bm{\phi}$ via Monte-Carlo, where each $\bm{\phi}$ is drawn from a Gaussian distribution with mean $\bar{\bm{\phi}}$ and covariance $\bm{M}$ (see Eq.~\ref{eq:simple_post}). We integrate over non-linear parameters by averaging the PTEs (averaged over $\bm{\phi}$) from random elements of the chain.

\section{Supplemental constraints}
\label{sec:big_triangle}

\begin{figure}[!h]
    \centering
    \includegraphics[width=\linewidth,valign=c]{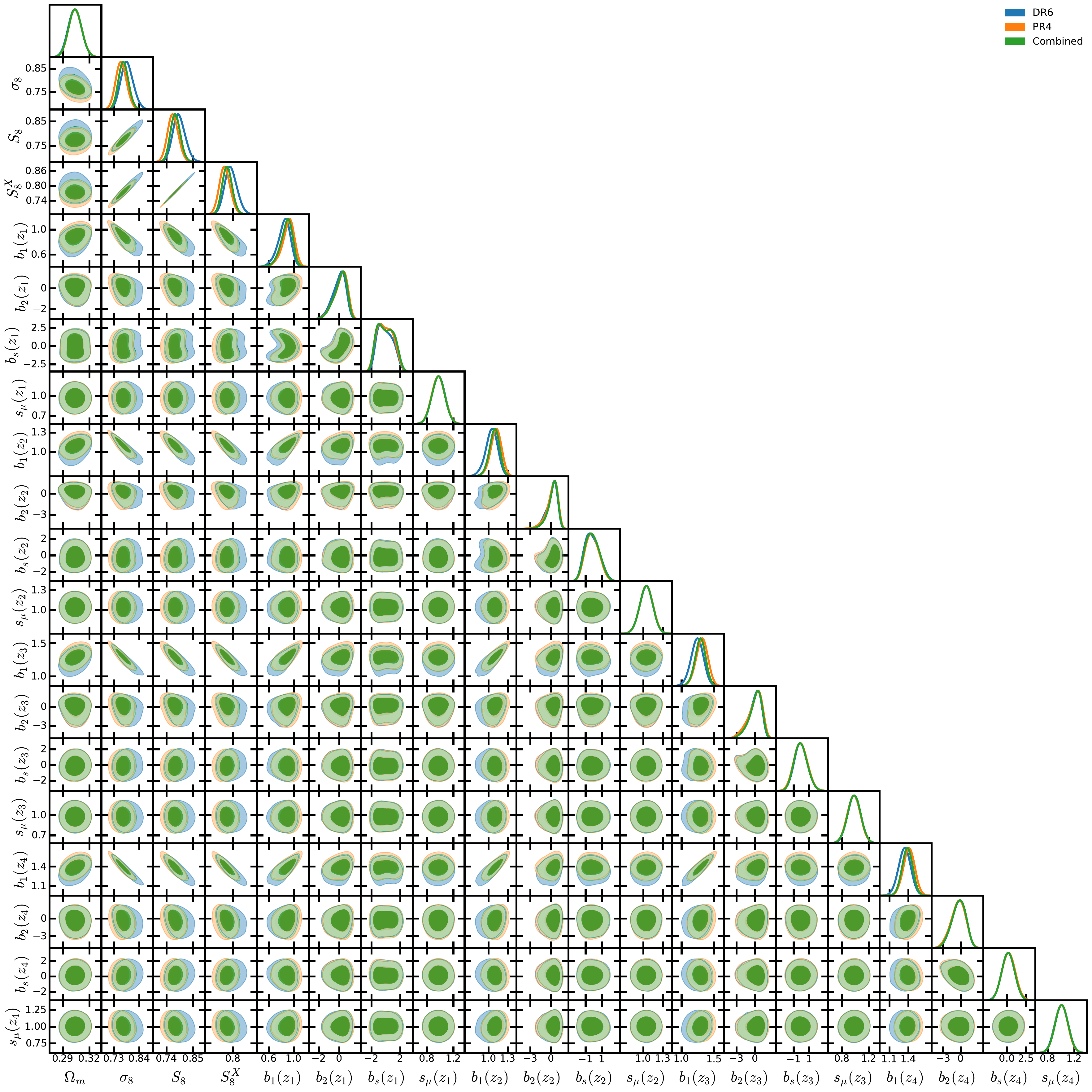}
    \caption{ 
    Parameter constraints from the baseline PR4+DR6 fit.
    } 
\label{fig:giant_triangle}
\end{figure}

We show the full set of parameter constraints from our baseline PR4+DR6 fit in Fig.~\ref{fig:giant_triangle}.

\bibliographystyle{JHEP}
\bibliography{main}
\end{document}